\documentclass[12pt, letterpaper]{article}

\usepackage[usenames,dvipsnames,svgnames,table]{xcolor}
\usepackage{dsfont}
\usepackage{tabularx}
\usepackage{svg}
\usepackage{caption}
\usepackage{titlesec}
\usepackage{siunitx}
\usepackage{makecell}
\usepackage{adjustbox}
\usepackage[flushleft]{threeparttable}
\usepackage{booktabs}
\usepackage{helvet}
\usepackage{times}
\usepackage{placeins}
\usepackage{multirow}
\usepackage{subcaption}
\usepackage{arydshln}
\usepackage{longtable}
\usepackage{float}
\usepackage{multicol}
\usepackage{supertabular}
\usepackage[framemethod=TikZ]{mdframed}
\usepackage{fullpage}
\usepackage[switch]{lineno}
\usepackage{amssymb}
\usepackage{amsmath}
\usepackage{rotating}
\usepackage{array}
\usepackage{mathtools}
\usepackage[ruled]{algorithm2e}
\usepackage{algorithmic}
\usepackage{bm}
\usepackage{comment}
\usepackage{enumitem}
\usepackage{graphics}
\usepackage{latexsym}
\usepackage{mathrsfs}
\usepackage{morefloats}
\usepackage{nicefrac}
\usepackage{authblk}
\usepackage[english]{babel}
\usepackage{blindtext}
\usepackage{graphicx}
\usepackage{hyperref}
\usepackage{xurl} % allows URLs to break anywhere
\hypersetup{breaklinks=true}
\usepackage{xr-hyper}
\usepackage{dcolumn}
\usepackage{todonotes}
\usepackage{subfiles} % Best loaded last in the preamble
\usepackage[normalem]{ulem}
\makeatletter
\newcommand*{\addFileDependency}[1]{
  \typeout{(#1)}
  \@addtofilelist{#1}
  \IfFileExists{#1}{}{\typeout{No file #1.}}
}
\makeatother

\newcommand*{\myexternaldocument}[1]{
    \externaldocument{#1}
    \addFileDependency{#1.tex}
    \addFileDependency{#1.aux}
}

\graphicspath{{main_figures/}{../main_figures/}}

\myexternaldocument{supplementary}

\newcolumntype{d}[1]{D{.}{.}{#1}}

\title{Media Coverage of War Victims: Journalistic Biases in Reporting on Israel and Gaza}

\author[1*]{Bedoor AlShebli}
\author[2]{ Bruno Gabriel Salvador Casara}
\author[2*]{Anne Maass}

\affil[1]{\normalsize Social Science Division, New York University Abu Dhabi, UAE}

\affil[2]{\normalsize Psychology, Science Division, New York University Abu Dhabi, UAE}

\affil[*]{\footnotesize Corresponding author e-mails:\ bedoor@nyu.edu, anne.maass@nyu.edu}

\date{}

\begin{document} 
\nolinenumbers
\maketitle

\begin{abstract}

\noindent {\color{black} October 7, 2023 marked the start of a war against Gaza, one of the most devastating conflicts in modern history, which quickly produced a stark global attitudinal divide. To examine the role of media bias in shaping public understanding of this asymmetrical war, we analyzed more than 14,000 news articles published during its first year across three major Western outlets (The New York Times, BBC, CNN) and one non-Western English-language outlet (Al Jazeera English). Focusing on media narratives surrounding Israeli and Palestinian victims, we identify three systematic biases in Western coverage: (1) Identifiable Victim Reporting: Israeli victims were substantially more likely to be depicted as identifiable individuals, whereas Palestinian victims were predominantly represented as undifferentiated collectives. (2) Equalization Bias: Despite the profound asymmetry in casualties, displacement, and other forms of suffering, Western reporting repeatedly invoked the October 7 attacks to equalize Israeli and Palestinian victimhood, even in the absence of new Israeli-casualty events. (3) One-sided Doubt Casting: Journalists disproportionately used language that casts doubt on the credibility of casualty figures and the reliability of sources when reporting Palestinian (vs. Israeli) victim counts, selectively undermining trust in information about Palestinian suffering. Across all three phenomena, these patterns were either absent or greatly attenuated in Al Jazeera English. Taken together, our analysis uncovers a coherent set of systematic biases in high-profile Western media coverage of the Gaza war, with implications for how global audiences come to understand and morally evaluate the conflict.}

\end{abstract}

\section*{Introduction}
{\color{black} More than other geopolitical events, the ongoing war on Gaza has produced an attitudinal divide between Western and Non-Western countries, which is reflected at the level of governmental decision making, support for UN resolutions, and public opinion. In the present research we investigate the role of media bias that may reflect, and contribute to, this attitudinal gap.

%Background information on the 2023/24 war on Gaza
The ongoing U.S.-backed war by Israel against Gaza began on October 7\textsuperscript{th}, following an attack by Hamas' military wing on Israel. The attack resulted in over 1,200 deaths, including 38 children, the taking of approximately 250 hostages, and nearly 5,500 injured~\cite{unnews2024}. The attack is considered the deadliest attack in Israel’s history. This was followed by an extensive war of Israel against Gaza that - already by the end of the first year - had produced an unprecedented humanitarian crisis~\cite{who2024}, the destruction of over half of Gaza’s homes~\cite{npr2024}, the devastation of the school, university, and health systems, extreme deprivation, and the forcible displacement of almost the entire population~\cite{unnews2024}. After one year of war, the death toll among Palestinians in Gaza amounted to over 42,000 people~\cite{ocha_website}, the majority of whom women and children. However, independent estimates suggest that the true death toll may greatly exceed these official statistics, both for violent mortality and when non-violent excess death rates are taken into account~\cite{khatib2024, Spagat2025LancetGlobHealth}

Over 97,000 people were injured~\cite{ocha_website}, one fourth of whom with life-changing injuries such as amputations~\cite{who2024rehab}. Different from other conflicts (such as Ukraine), the population was trapped in Gaza due to the 17-year-old air, sea and land blockade. The war has been particularly dire on children. In the first twelve months alone, over 14,000 had been killed~\cite{unicef2024oneyear}, at least 17,000 were unaccompanied~\cite{unicef2024statement}, and 90\% suffered acute malnutrition~\cite{unicef2024foodpoverty}. According to Save the Children, more than 10 children each day had one or both legs amputated~\cite{savethechildren2024}, and by the time of the writing, the humanitarian crisis has tragically deteriorated and multiple organizations, including the United Nations~\cite{UNSpecialCommittee2024}, Human Rights Watch~\cite{HumanRightsWatch2024}, and Amnesty International~\cite{AmnestyInternational2024}, have declared that a genocide is taking place.

This war is also considered the deadliest conflict in history for aid workers with over 300 killed and for journalists with 128 killed within the first 12 months~\cite{unocha2024} and many others missing, injured or arrested~\cite{unocha2024, ifj2024}. As a unique case in the history of journalism, news reports have relied almost exclusively on local journalists, given that international reporters were prevented from reaching the war zone, except for sporadic visits as embedded journalists under the strict control of the Israeli military~\cite{tobitt2024}. Since the beginning of the war, foreign media such as \textit{The New York Times, BBC, CNN, ABC, NBC}, and \textit{Fox News}, have very limited access to Gaza, can operate only in the presence of Israeli soldiers, and ``\textit{have to submit all materials and footage to the IDF for review before publication}''~\cite{vox2024}. Thus, international reporters are prevented from baring witness to the tragedy unfolding in Gaza. Beyond the impact in the Middle East, the war also had global effects on intergroup relations, as reflected in a considerable increase in both anti-Jewish and anti-Muslim hate crimes since October 2023~\cite{itv2023, bbc2024}. 

Against this backdrop, the present study examines how such a humanitarian disaster and information bottleneck are reflected in international media coverage. The current analysis, informed by psychological research and theory, focuses on subtle media biases in war journalism. Specifically, we examine three potential journalistic biases that may have played a role in creating or sustaining attitudinal divides:

\begin{enumerate}
    \item \textbf{Identifiable victim reporting:} War journalism has been criticized for its primary focus on frontline reporting, often neglecting the human consequences for the civilian population, and for children in particular~\cite{ferguson2023life}. The media may either quantitatively give little space to civilian victims or report on them in a way that conceals their full human qualities (dehumanization), including their intellect, their culture, their morality, their feelings, and their individuality.  One way in which journalists may downplay the civilian experience of war is by reporting group-level experiences rather than individual stories. Social scientists such as Slovic~\cite{slovic2007psychic} and Schelling~\cite{schelling1968life} have long observed that people often remain indifferent to mass atrocities such as war, mass murder, and genocide, yet are profoundly moved by individual stories. This psychological tendency is captured by what scholars call \textit{identifiable victim effect}~\cite{lee2016identifiable}: people feel more compassion toward single victims and are more likely to take action (for instance, through donations) when text or images describe the suffering of a single individual rather than that of larger groups. Compassion quickly fades as the number of victims increases~\cite{slovic2007psychic}, with the exception of very cohesive and interconnected small groups such as families~\cite{vastfjall2014compassion}. For instance, in a classical study by Kogut and Ritov~\cite{kogut2005identified}, a single child in need of a costly life-saving treatment elicited greater distress and received greater donations than a group of sick children with identical needs. This phenomenon, mostly investigated in experimental work, is driven by different mechanisms,  most notably that identifiable victims attract greater attention and evoke greater emotional reactions (such as sympathy or distress) than statistical victims or large groups~\cite{slovic2007psychic}. Also, by providing personalized information (such as age or name), as is typically done for single identifiable victims, narratives tend to be more vivid, which in turn induces a sense of familiarity and closeness. To examine the relative humanization of victims through individual stories, we therefore analyzed the degree to which Israeli and Palestinian civilian victims were portrayed through individualized (rather than collective) narratives. In the absence of bias, one would expect that the ratio of individualized to group-based narratives would be approximately the same for both sides. To the best of our knowledge, this is the first large-scale media analysis of individualizing vs. group-level reporting in the context of war (for a small-scale qualitative analysis on identifiable victim reporting, see~\cite{grimm2025hierarchies}).
    
    \item \textbf{Equalizing bias:} The second journalistic bias investigated here refers to the tendency to deny hierarchies in human suffering. In asymmetrical conflict situations, in which victims on one side greatly outnumber those on the other, one might expect media coverage to roughly mirror this difference, whereas stark deviations from this baseline may suggest that not all lives are equally news-worthy. We coined the term Equalization Bias to describe the tendency to dedicate equal media attention or space to two groups that, in reality, show very different levels of human suffering. For instance, giving the same coverage to Bosnian and Serbian victims in the Yugoslav war would constitute a case of equalizing bias. The equalization bias resembles, and was inspired by, the concept of \textit{false balance}~\cite{terzian2025epistemic}, an established journalistic bias, where two positions are presented as epistemically equivalent and receive equal space or attention (e.g. airtime) despite differences in actual validity and trustworthiness, such as climate skeptics and climate scientists or pro- and anti-vaccination proponents~\cite{terzian2025epistemic,boykoff2004balance,salvador2019viral}. Both, the false balance and the equalization bias create a false equivalence between non-comparable objects, but whereas false balance revolves around the question of validity of information, the equalization bias proposed here alludes to the allocation of narrative resources to different groups (e.g., number of stories are told, space devoted, etc) that is not proportionate of the actual scale of human suffering. Furthermore, given the novelty of the proposed bias, we complement our large-scale media analysis with a controlled experiment designed to identify the psychological mechanism that may give rise to this bias.

    \item \textbf{One-sided doubt-casting:} The third bias refers to linguistic devices (including hedges such as ``allegedly'') that signal that a reported fact is unconfirmed and, therefore, that the claim is uncertain (e.g., Bailey et al., 2014~\cite{bailey2014grammatical}). Such statements imply a lack of verifiability or consensus and, hence, are likely to induce doubt in the reader's mind. In the context of war or natural disasters, where actual victim numbers can not easily be established or independently verified, journalists necessarily rely on external sources, typically governmental agencies such as the police or the health ministries of the countries involved. Whereas doubt-casting language is generally considered a sign of journalistic integrity and may reflect high ethical standards in situations of uncertainty, it becomes problematic when applied in a one-sided manner. To uncover potential bias, we therefore investigate the degree to which doubt-casting is used to either undermine the credibility of the source of information (e.g., according to the Hamas-controlled health ministry) or to question the reliability of the victim numbers reported in an article (such as allegedly;  reportedly; appear). 
\end{enumerate}

Having outlined the three forms of journalistic bias, we next clarify how they are explored in the present study. Importantly, throughout this work we use the term ``bias'' to describe systematic asymmetries or deviations in media coverage relative to objective baselines, such as actual casualty rates. The concept is applied to all three phenomena discussed above, without implying any form of intentional distortion or malevolent intent on the part of journalists. Possible explanations for the emergence of such patterns are discussed in later sections.

Our analyses focus on the human consequences of war for the civilian population (including children), that constitute the overwhelming majority of victims on both, the October 7 attack (69\% of all victims) and in the subsequent war on Gaza (estimated to be above 80\% among Palestinian victims~\cite{ayoub2024comparative}). To investigate the three proposed biases, we employ two complementary analytical approaches: one centered on \textit{narrative representation} and the other on \textit{quantitative reporting}.

\begin{itemize}
    \item Narrative representation examines how victims are portrayed in media stories, whether as individualized human beings or as undifferentiated collectives, capturing degrees of (de-)humanization in reporting.
    \item Quantitative reporting investigates how numbers are used to convey the human costs of war, assessing whether the distribution of attention and reported casualties reflects the actual scale, the asymmetry of suffering across groups and whether such numbers are questioned.
\end{itemize}

Together, these two analytical dimensions allow us to systematically assess how subtle reporting tendencies may shape perceptions of conflict and contribute to the broader attitudinal divides observed across the international media landscape.
}

\section*{Results}

In this study, we analyze over 14,000 news articles published between October 7, 2023 and October 7, 2024 by four major international media outlets: The New York Times (NYT), BBC, CNN, and Al Jazeera English (AJE). {\color{black} To ensure analytical consistency and cross-outlet comparability, we restrict our corpus to textual news content and exclude videos, podcasts, and transcripts. A detailed description of the outlet selection criteria, temporal scope, and data collection procedures is provided in the Materials and Methods (Data Selection and Sample Construction subsection).}

\begin{figure}[H]
    \centering    \includegraphics[width=.95\textwidth,height=0.86\textheight,keepaspectratio]{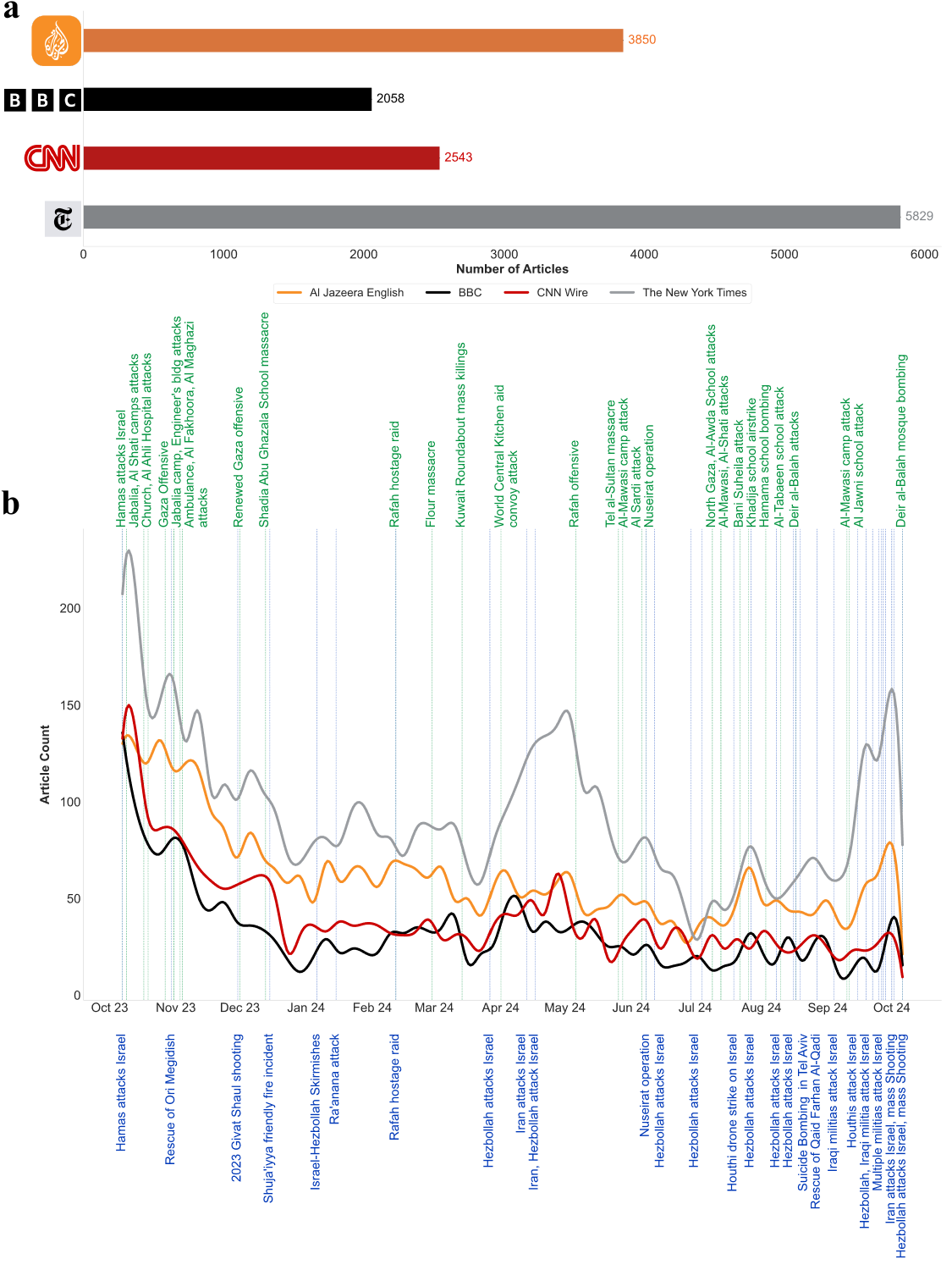}
    \caption{
    \textbf{Total Number of Relevant Articles per News Source.} \textbf{a)} The bar plot shows the overall count of relevant articles per each of the four media sources. \textbf{b)} the timeline plot below shows the same number of articles but spread out over time along with major war events that affected Palestinian and Israeli sides displayed in green and blue respectively. Both plots cover the first 12 months of the conflict.}
    \label{fig:Figure 1}
\end{figure}

We start our analysis by examining the data landscape presented in Figure~\ref{fig:Figure 1}, which illustrates the total article counts by source (Figure~\ref{fig:Figure 1}a) alongside the temporal distribution of article counts across sources (Figure~\ref{fig:Figure 1}b). The data reveal that NYT published the most articles, followed by AJE and CNN, with BBC contributing the least. The temporal distribution highlights significant spikes in response to key events, such as the October 7th attacks, the World Central Kitchen massacre in April 2024, attacks on Israel in both April 2024 and Sep 2024, as well as the Israeli attacks on schools and refugee buildings during Sep 2024; for a full list of the events and their exact dates please refer to Supplementary Tables~1 and 2. Furthermore, each outlet exhibited distinct sensitivity patterns: NYT consistently responded more than the others. CNN seems to adopt a similar overall publishing profile to that of NYT with publication spikes after relevant events, which, however, abate faster than NYT's, suggesting they do not dwell on the event as long as NYT. The BBC maintained the lowest publishing profile and demonstrated a selective sensitivity to events, spiking its publications for some events, while remaining unresponsive to others. AJE, by contrast, maintained a relatively high publication rate, which remained comparably steady even after major events.

{\color {black} Using this dataset, we examined the three biases outlined in the Introduction. We began with narrative representation, focusing on the first two biases, identifiable victim reporting and equalizing bias, by asking whose stories are individualized, whose remain collective, and how casualty-related reporting reflects underlying asymmetries in victimization.

\subsection*{Narrative Representation as an Indicator of Identifiable Victim and Equalizing Bias}

Narrative representation concerns how civilian victims are made visible (or remain obscured) in news coverage. Individualized stories tend to humanize victims and elicit empathy, whereas collective or statistical references do so to a much lesser extent. Examining these narrative choices enables us to identify two forms of bias: whether some victims are more likely to be individualized (identifiable victim bias) and whether narrative space is distributed in ways that obscure large disparities in suffering (equalizing bias). 

\subsubsection*{Identifiable Victim Bias: Individualized vs Collective Reporting}
We began by examining the identifiable victim bias, that is, the tendency to portray some victims as individualized persons while others are presented as undifferentiated collectives.} To do this, we employed a large language model (LLM) to extract every reference to civilians, whether individual or collective, and classify each instance as pertaining to either the Palestinian or Israeli side. Specifically, we instructed the model to identify individualized civilian stories that highlight personal experiences of hardship, such as: \textit{``{Noralin `Nataly' Babadilla, 60, was visiting Kibbutz Nirim, to celebrate the community's 70th anniversary. Her husband was murdered....''}} \cite{bbc2023_article}. These types of narratives humanize the conflict by centering on a specific, emotionally resonant character. In addition, we asked the model to categorize the most severe type of hardship being described, enabling us to analyze not just whether individuals are mentioned, but also how their suffering is framed. Group-level references, which are typically broader and less personalized, included generalized hardships, such as \textit{``With many Israelis traumatized by the attack...''} \cite{Kingsley_2024}, statistical accounts, such as \textit{``More than 40,738 people have been killed in Gaza''} \cite{bbc2024hostages}, and event descriptions, like \textit{``Gazans are dying of starvation''} \cite{Ahmad_Kershner_Bashir_Alghorra_2025}. A detailed explanation of how these classifications were extracted and validated is provided in Supplementary Note~1.

\begin{figure}[H]
    \centering      \includegraphics[width=.95\textwidth,height=0.9\textheight,keepaspectratio]{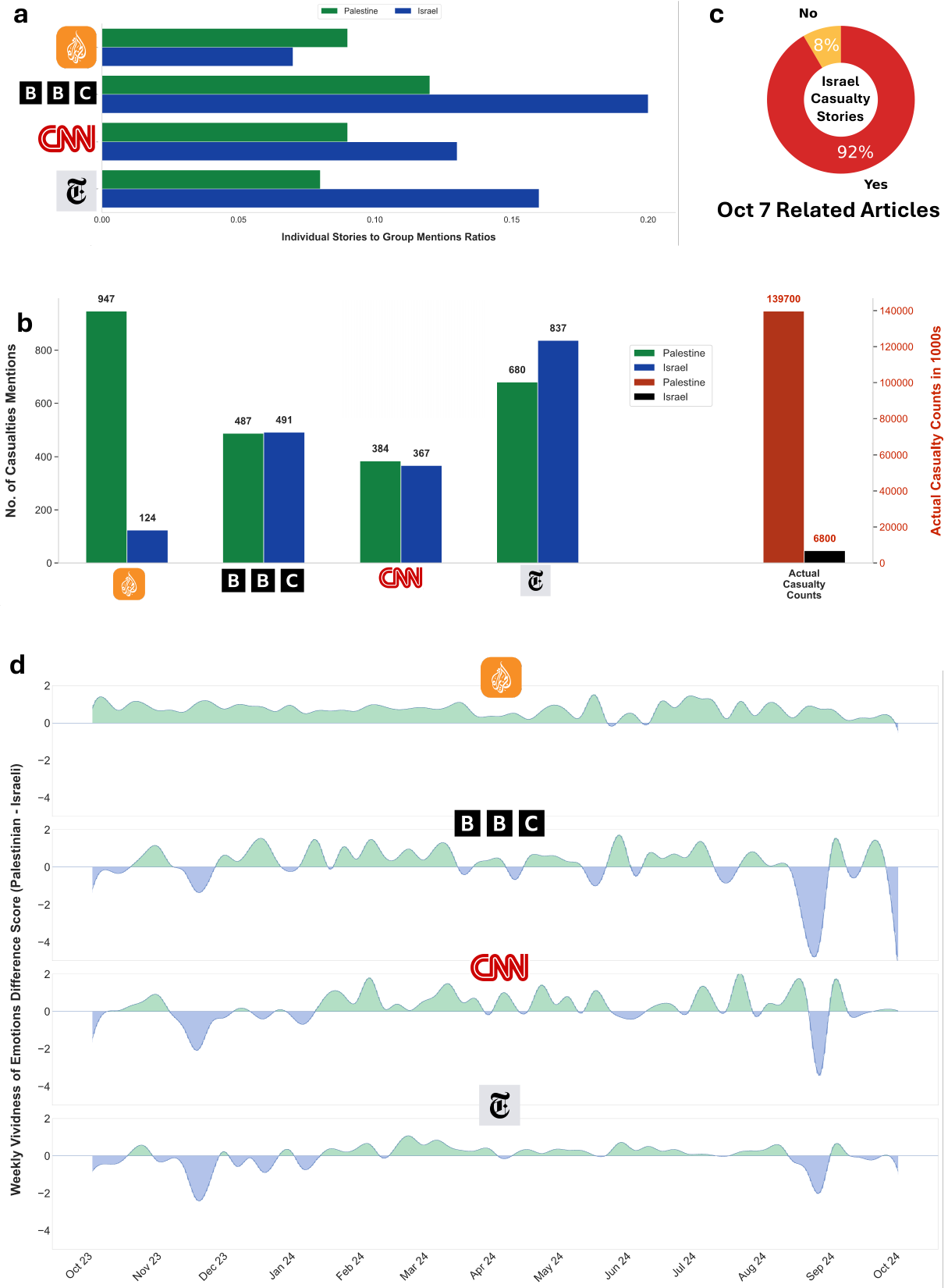}
    \caption{\textbf{Results of Individualized vs. Category-based Reporting Analysis.} \textbf{a)} Ratio of Individualized to Grouped mentions per side for each media source. \textbf{b)} Left: Individualized casualty-related story counts per side for each media source. Right: Actual casualty counts for both sides in the first 12 months. \textbf{c)} Proportion of Western media’s Israeli casualty stories mentioning Oct 7 or hostages. \textbf{d)} Weekly average difference in Vividness of Emotions scores across media source during the first 12 months.
    \label{fig:Figure 2}
    }
\end{figure}

Next, we examined the extent to which each side is represented through individualized versus group narratives. As shown in Figure~\ref{fig:Figure 2}a, Western media outlets, BBC, CNN, and NYT, consistently featured higher ratios of individualized to group mentions for the Israeli compared to the Palestinian side. This asymmetry was most pronounced in NYT, where Israeli individuals were referenced once for every six group mentions (ratio = 0.16), whereas Palestinians were individualized only once for every twelve group mentions (ratio = 0.08). Similar patterns emerged for BBC (0.20 vs. 0.12) and CNN (0.13 vs. 0.09), though the disparity was less stark (see Supplementary Table~3). In contrast, AJE showed a more balanced pattern, with similar ratios for both sides (0.09 for Palestinians vs. 0.07 for Israelis). This suggests that Israeli individuals were more frequently singled out in Western media coverage, while Palestinians were more often portrayed as part of an undifferentiated collective. 

We formalized this asymmetry using a Generalized Estimating Equations (GEE) regression analysis (see Materials and Methods, subsection ``Testing Individualization Bias with GEE Models''), which confirmed that Western outlets were significantly more likely to individualize the Israeli side in their reporting. Specifically, we find that the odds of a reported story being individualized are \textcolor{black}{14.7\%} lower for Palestinian instances compared to Israeli ones in the case of BBC, \textcolor{black}{28.7\%} lower in the case of CNN, and \textcolor{black}{41.3\%} lower in the case of NYT. No statistically significant difference was observed for AJE. {\color {black} Taken together, these results provide clear evidence of an identifiable victim bias in Western coverage, whereby Israeli civilians are more often presented as fully individuated persons, whereas Palestinians are more frequently portrayed as part of an anonymous collective. \\

\subsubsection*{Equalizing Bias in Casualty Reporting}}
We then turned our attention to the content of the stories themselves, categorizing each instance based on the most severe type of hardship described. These included categories such as `casualties', `displacement and refugees', `imprisonment and detention', among others. For a complete list of hardship categories and their distribution across media outlets, see Supplementary Figure~3 and Supplementary Note~1. Among these, we opted to concentrate on stories labeled as ``casualties’’ (i.e., deaths and injuries), as this category was the most frequent across all sources, relevant to both sides, and supported by reliable statistics from international organizations such as the UN, allowing for systematic comparison.

{\color{black} After identifying casualty-related stories, we next assessed the presence of equalizing bias by comparing how frequently outlets reported on Palestinian versus Israeli casualties.} As shown on the left side of Figure~\ref{fig:Figure 2}b, a stark contrast was revealed between AJE and Western outlets. AJE published roughly eight Palestinian casualty stories for every Israeli story, while BBC, CNN, and NYT maintained a close to one-to-one ratio. NYT even showed a slight tilt in the opposite direction, reporting eight Palestinian stories for every ten Israeli ones. At first glance, this might suggest a pro-Palestinian bias in AJE’s reporting. However, when we compared these ratios to the actual casualty figures, used here as a crude baseline for contextualizing media attention, a more complex picture emerged. As shown on the right side of Figure~\ref{fig:Figure 2}b, for every Israeli casualty, there were roughly 20 Palestinian casualties by the end of the first year of war. This stark disparity in actual casualties reflects the marked power differential between the two groups, and contrasts with the seemingly ``balanced'' reporting in Western media. In theory, if all events were covered equally and without selection bias, one would expect a roughly proportional discrepancy in the number of stories reflecting the underlying reality. Instead, the reporting style of Western media and, to a lesser degree of AJE, {\color{black} is coherent with our phenomenon of equalizing bias, namely the tendency to create the impression that two sides are equivalent when, in reality, they are not.  In particular, Western outlets create a false equivalence by giving equal narrative weight to Israeli experiences, despite the vastly different realities of suffering on the ground}. 

It is important to note, however, that one would not necessarily expect casualty reporting to mirror casualty rates perfectly. In fact, no outlet, including AJE, displayed a fully proportional relationship. Some degree of over-reporting of the events of October 7th relative to the subsequent months is psychologically plausible. First, the October 7th massacre was extremely salient, involving a large number of victims in a short time span. Research on the availability heuristic shows that single-mass casualty events tend to attract disproportionate attention and are remembered more vividly than larger, but gradually accumulating, death tolls \cite{gigerenzer2004dread}. Second, events that occur first in time or space often receive greater attention than those that follow; a pattern observed in phenomena such as the ballot-order effect \cite{meredith2013causes}. Thus, if October 7 is perceived as the ``starting point'' of the war, some degree of disproportionate coverage would be understandable. Nevertheless, such cognitive and temporal biases cannot account for the extent to which media reporting equated casualties on both sides. For instance, while the 38 Israeli children killed on October 7th might be expected to draw more attention than 38 Palestinian children killed, on average, each day over the subsequent year, it would be implausible for the cumulative 14,000 Palestinian child deaths over the course of the year to warrant narrative attention equivalent to those 38 Israeli deaths. {\color{black} This pattern is consistent with equalizing bias: a structural tendency to allocate roughly equal narrative space to two sides despite profound asymmetries in underlying human suffering. Such patterns may be shaped, in part, by asymmetries in identification with conflict parties, which we next attempted to test experimentally.}\\

{\color{black}\subsubsection*{Experimental Evidence on Identification as a Mechanism of Equalizing Bias}

To examine whether equalizing bias can arise from social identification alone, i.e. the degree to which individuals perceive a group as part of their social self and feel attached to it, we conducted a controlled experiment in which participants in the US and UK were presented with a fictitious asymmetric geopolitical conflict between two countries differing in military strength and civilian death tolls (10 vs. 40). Participants were randomly assigned to adopt the role of a journalist aligned with the stronger country, the weaker country, or an independent international outlet (see Supplementary Note~2 for full details). Participants reviewed brief profiles of 50 civilian casualties (10 from the stronger country; 40 from the weaker country) and were asked to select 10 victims whose stories they would feature in their news articles. In the absence of bias (random selection), proportional coverage would yield, on average, two strong-country victims and eight weak-country victims.

Consistent with the equalizing bias observed in our media analysis, participants assigned to the stronger-country condition substantially over-selected strong-country victims (mean of 3.7 rather than 2) relative to the proportional benchmark. Participants assigned to the weaker-country condition did not show such over-selection (mean of 1.8 rather than 2), while independent participants occupied an intermediate position.

Critically, identification predicted selection behavior: the more participants identified with the strong country, the more strong-country victims they selected ($r = .43$, $p < .001$). Further supporting the critical role of identification, a mediation model (see Supplementary Note~2) demonstrated that the equalizing bias is driven by journalists' identification with the country to which they were assigned. Together, these findings provide causal evidence that equalizing bias can arise from asymmetries in social identification, lending support to the interpretation that the patterns observed in BBC, CNN, and NYT coverage are shaped by stronger identification with Israel.}

{\color{black}\subsubsection*{Sustaining Equalizing Bias in the Face of Discrepancy} }

This led us to a critical line of inquiry: How do Western media outlets sustain parity in reporting casualty-related stories between both sides when the actual number of casualties is so drastically imbalanced?  To investigate this, we analyzed the content of Israeli casualty-related stories published by Western media (generated using \textit{Prompt 2} in Supplementary Note~1). Strikingly, we found that 92\% of these stories were related to the events of October 7, as shown in Figure~\ref{fig:Figure 2}c. This concentration persisted even in stories published several months after the attack. In other words, a single moment in time, the October 7 event, was repeatedly revisited to maintain a steady volume of Israeli casualty narratives, thereby flattening the temporal and numerical asymmetry in actual deaths {\color {black} to sustain equalizing bias}.

To explore this pattern further, we examined how Western media reported on the Israeli side during moments of significant Palestinian civilian casualties. We began by testing whether spikes in Palestinian deaths were associated with renewed publication of October 7–related Israeli stories. Indeed, a strong positive correlation emerged (Pearson’s r = 0.8; see Supplementary Figure~4), suggesting that Israeli narratives were often recycled precisely when Palestinian deaths were peaking. To better understand how this dynamic played out in practice, we selected four equidistant dates of high Palestinian casualties across the year and reviewed the stories that appeared on those dates and the day after.

Across these cases, a clear pattern emerged: while Palestinian tragedies were reported, Israeli stories received more narrative depth, emotional detail, and prominence, even on days when no new Israeli casualties occurred. For instance, on 1 December 2023, nearly 184 Palestinians were killed \cite{Jazeera_2023} as Israel resumed its assault following a ceasefire, yet NYT ran stories quoting hostage families: \textit{``Mentally, she’s definitely doing better. She’s more engaged with people and becoming more independent.''}\cite{Rosman_2023}, while CNN reminded readers that \textit{``around 240 people, from infants to octogenarians, were taken hostage during Hamas' attack on October 7''}~\cite{Clarke_2023}. Across two days, BBC, CNN, and NYT published 19, 21, and 24 Israeli stories respectively, versus just 17 Palestinian stories combined. A similar imbalance appeared during the ``Flour Massacre'' on 29 February 2024, when 117 Palestinians were killed and 750 injured \cite{aljazeera2024un} while receiving food aid. Despite the scale of the tragedy, Western outlets again devoted substantial space to Israeli narratives, collectively running 10 Israeli-focused stories versus 16 on Palestinians, a ratio that muted the scale of the Palestinian tragedy. CNN even described the incident as the ``worst tragedy'' yet, but in the same news cycle published an opinion piece urging readers: \textit{``When you kiss them goodnight — please think of Kfir and Ariel Bibas. Think of their mother and father''}~\cite{levy2024forgotten}.

The third date explored was 8 June 2024, which recorded the highest number of Palestinian casualties among the dates we examined: 274 killed and nearly 700 injured \cite{Amon_2024}. Given this extraordinary scale, and the tentative shift toward balance observed on 29 February, one might have expected coverage to tilt more heavily toward Palestinians. This was not the case, as the date also coincided with a major Israeli development: the rescue of four hostages from Gaza’s Nusseirat area. Coverage of the rescues was celebratory,\textit{“Saturday was an emotional and happy day for the state of Israel and the IDF”}~\cite{McKenzie2024FourIsraeli}, while Palestinian casualties were reported with skepticism: \textit{“CNN has no way of verifying casualty numbers reported by Palestinian officials in Gaza”}~\cite{McKenzie2024FourIsraeli}. Across that day and the next, BBC, CNN, and NYT ran a total of 40 Israeli stories, more than double the 19 on Palestinians, raising the possibility that when Israeli and Palestinian events coincided, Israeli narratives overshadowed even the deadliest Palestinian tragedies.

By the fourth date explored, 10 September 2024, coverage appeared to move back toward a more balanced pattern. An Israeli strike on a humanitarian zone killed 19 Palestinians~\cite{Tawfeeq_Nasser_2024}, and both BBC and CNN reported only on Palestinian casualties. The New York Times still ran hostage-focused stories, re-emphasizing their suffering: \textit{“the hostages suffered from significant malnutrition, severe weight loss and long-term physical neglect”} \cite{Livni_2024}, yet overall published more on Palestinians than Israelis (7 vs 4). This suggested that, with time, the narrative might have been shifting toward greater balance. However, the 8 June example, where Israeli and Palestinian events coincided, raised the question of whether this balance only held when Israeli developments were absent. To probe this further, we examined two additional dates where both sides experienced major events: 15 January 2024, marking 100 days since October 7, and 1 September 2024, following the recovery of six Israeli hostage bodies \cite{Rasgon_Sobelman_Shankar_Fuller_2024}.

On 15 January 2024, Israeli airstrikes on Gaza’s Al Thalatheni Street killed 22 Palestinians and injured many more \cite{aljazeera2024_gaza_day101}. Yet the day was largely framed in Western media as the “100 days since October 7” milestone. Coverage focused heavily on hostage testimonies and recollections, such as NYT's headline \textit{“They thought they knew death but that didn’t prepare them for Oct 7”} \cite{Bergman2024Their} and personal accounts like \textit{“Israeli teenager recounts her time as a hostage in Gaza”} \cite{nyt2024hostage}. BBC emphasized military perspectives with stories such as \textit{“They were Israel’s eyes on the border but their Hamas warnings went unheard”}~\cite{Cuddy_2024}. Palestinian casualties, by contrast, received comparatively little attention. Across two days, BBC, CNN, and New York Times ran 13, 10, and 12 Israeli stories, compared to just 6 on Palestinians combined, with NYT publishing none at all. This imbalance suggests that symbolic Israeli milestones may have drawn coverage away from concurrent Palestinian losses.

On 1 September 2024, an Israeli strike on a school-turned-refugee camp killed 11 Palestinians, including several asleep in their tents \cite{aa2024gaza}. Yet coverage that day was dominated by the recovery of six Israeli hostage bodies \cite{Rasgon_Sobelman_Shankar_Fuller_2024} the previous evening. Western outlets devoted extensive space to personal tributes: the BBC described one victim as \textit{“among the innocents brutally attacked while attending a music festival for peace… He had just turned 23. He planned to travel the world”}~\cite{bbc2024hostages}. NYT highlighted family memories: \textit{“Hersh Goldberg-Polin loved soccer and music. He was curious, respectful and passionate about geography and travel”}~\cite{nyt2024polin}. CNN portrayed him as \textit{“one of the most recognizable faces of the enduring hostage crisis… Banners and murals demanding his return were often displayed in Jerusalem and around the world”}~\cite{CNN2024IsraelGazaHostages}. In contrast, Palestinian victims were mentioned only briefly. Across 1–2 September, BBC, CNN, and New York Times published 86, 38, and 57 Israeli stories, compared to just 5 Palestinian ones combined, with CNN publishing none. Here again, the overlap of major events may help explain why Israeli narratives overshadowed Palestinian casualties, despite the ongoing humanitarian toll.

Taken together, these cases show that although Western media did cover major events affecting both Palestinians and Israelis, at times even including vivid depictions of Gaza’s devastation, such as describing Al-Ahli Hospital as \textit{“like a horror movie”} \cite{Elbursh_Davies_2023} or reporting that \textit{“children [were] torn apart and scattered in the streets”} \cite{McKenzie2024FourIsraeli}, the overall volume, depth, and emotional richness of coverage skewed heavily toward the Israeli side. Coverage appeared more balanced only when no major Israeli events occurred. When significant events on both sides coincided, however, Israeli narratives were amplified to such a degree that they overshadowed even the most severe Palestinian tragedies (see Supplementary Table~4). Israeli stories were also far more frequently personalized, with names, ages, aspirations, and family histories highlighted \cite{nyt2024polin, CNN2024IsraelGazaHostages, bbc2024hostages}, and often revisited the same individuals over time. This suggests that the imbalance was not merely the result of coverage decay but was actively reinforced whenever Israeli stories could be interwoven with Palestinian casualty events, thereby sustaining a structural asymmetry in how loss and victimhood were represented.

A more detailed explanation of our date-selection procedure and findings, including additional quotes, is provided in Supplementary Note~3. The number of Israeli stories, related to October 7, published by each outlet on the dates examined can be found in Supplementary Table~4.\\

{\color {black}
\subsubsection*{Empathy-Related Narrative Elements in Individualized Stories} 
Given that Israeli victims are more likely to be portrayed as identifiable individuals than Palestinian victims, the question arises whether the individualized stories about the former are also told in a more empathic narrative style. To explore this}, we quantified differences in narrative elements that can influence reader empathy, using the validated Human Empathy and Narrative Taxonomy (HEART) framework \cite{shen-etal-2024-heart}. Specifically, we applied two HEART metrics: (i) Vividness of Emotions and (ii) Plot Volume (the main predictors of perceived empathy in the HEART model) to all hardship-related individualized stories in our corpus. While these metrics are not direct measures of empathy, prior research shows they are key stylistic features that contribute to how much empathy a narrative may elicit. Scores for each metric were first calculated at the story level and then aggregated by side within each article, given that articles often addressed both sides. For each metric, we then calculated a difference score by subtracting the Israeli aggregate score from the Palestinian one, with positive values indicating that these empathy-related narrative elements were more prominent in Palestinian stories, and negative values indicating the reverse. For full details on the scoring procedure, see ``Measuring Narrative Elements Associated with Empathy in News Texts'' in the Materials and Methods.

Weekly averages of the difference scores for each outlet are shown in Figure~\ref{fig:Figure 2}d (Vividness of Emotions) and Supplementary Figure~5 (Plot Volume). In these plots, values above 0 indicate that, on average, stories published during that period were more vivid or detailed for the Palestinian side; values below 0 indicate the reverse, and 0 represents parity.

AJE consistently leaned toward more vivid Palestinian narratives, though without dramatic fluctuations, mostly hovering between 0 and 2. In contrast, all three Western outlets showed two clear spikes toward the Israeli side: one in the final two weeks of November 2023, coinciding with the hostages–prisoners swap, and another in the last two weeks of August 2024, when hostage negotiations collapsed and six Israeli hostages were found dead in Gaza. In both cases, the spike in vividness for Israeli stories exceeded even AJE’s highest levels for Palestinians.

The Plot Volume results show the same pattern, including the two spikes, with Israeli-focused stories in those periods receiving even higher relative scores than in the Vividness of Emotions metric. At the same time, it is worth noting that, on average, stories about Palestinians were written with greater emotional vividness across outlets (albeit less so for NYT), which is not surprising given the scale of Palestinian suffering. {\color{black} Taken together, the above findings suggest that the asymmetry lies less in the narrative tone than in the relative frequency with which individualized stories are told in the first place: In Western media, Israeli suffering is more likely to be told through individual stories, whereas collective statements abound for Palestinian victims. Yet, once a journalist decides to report individualized stories about Palestinians, these stories are, on average, no less empathic towards the victims than those told about Israelis.

\subsubsection*{Children as a test case of Equalizing Bias} 

Children offer a particularly revealing test case for equalizing bias. Given the high vulnerability of children and the fact that they cannot be held responsible for armed conflict nor for the election of leaders who promote armed conflicts, it is not surprising that children have a special status in international humanitarian law (among which the UN Convention on the Rights of the Child~\cite{icrc2024children}). Similarly, lay people value the life of the young higher than that of adults~\cite{goodwin2014valuing}, feel a greater moral obligation to help children~\cite{erlandsson2024beneficiary}, and are more likely to assist them~\cite{moche2021helping, erlandsson2020moral}. 
Children (including older children) are prioritized in helping decisions because they evoke greater nurturing concern and compassion~\cite{oveis2010compassion}, because they are generally not held responsible for their fate, and because they have, on average, a greater number of years ahead, which makes helping them a rational choice. As such, one might expect coverage to track the underlying disparity in child casualties especially closely. We therefore extended our analyses of individualized reporting to specifically track mentions of children.}

On average, the four media outlets dedicated roughly 46.5\% of their individualized stories to children (AJE 47\%, BBC 45\%, CNN 50\%, NYT 44\%; Supplementary Table~5a). Taken at face value, this looks like responsible reporting: the media gave children more prominence than their actual share of casualties (about 32\%), which seems appropriate given children’s heightened vulnerability and moral salience. However, when the data are broken down by side, the picture changes. Palestinian children do take up a commendable share of stories—roughly in line with their 34\% share of deaths—but Israeli children receive far more attention than their casualty numbers would suggest. Although they make up only 3\% of Israeli deaths, their stories account for as much as 44\% of child-related individualized reporting in CNN, with similarly elevated shares in BBC (36\%) and NYT (40\%). This disproportionate emphasis mirrors the {\color{black} equalization bias} documented in our earlier analyses, where the relative weight given to Israeli and Palestinian suffering diverged sharply from the underlying reality.

The imbalance becomes even clearer when looking at the distribution of child-related stories within each outlet (Supplementary Table~5b). If each child’s life were given equal weight, nearly all individualized stories about children (over 99\%) would focus on Palestinian victims. Instead, Western outlets devoted a strikingly large share of their child coverage to Israeli children: BBC 44\%, CNN 43\%, and NYT 51\%. This pattern reflects a form of {\color{black} equalization bias}: by the end of the first year of war, 38 Israeli~\cite{wikipedia2024oct7} and 14,000 Palestinian children~\cite{unicef2024oneyear} had been killed, a disparity likely even greater for other hardships such as injuries, displacement, and famine. If casualty numbers alone were used as a baseline, individualized stories on Israeli children would be expected to account for less than 1\% of all child references. AJE came closest to reflecting the underlying disparity, with 91\% of its child stories about Palestinians. The most extreme deviation is observed in the New York Times, where Israeli and Palestinian children were given nearly equal attention, despite the enormous difference in the actual number of victims.

{\color{black}
\subsection*{Quantitative Reporting as an Indicator of Equalizing Bias and One-Sided Doubt-Casting}

Quantitative reporting provides a second lens through which media bias may manifest. Whereas narrative representation concerns whose suffering is made visible through individualized storytelling, quantitative reporting examines how numerical information is allocated, sourced, and qualified across the two sides. This dimension allows us to assess two forms of bias: (i) \textit{equalizing bias}, reflected in the distribution of casualty-related numbers that do not mirror the underlying asymmetry in victimization, and (ii) \textit{one-sided doubt-casting}, reflected in selective source attribution or linguistic uncertainty applied disproportionately to one side’s figures. Having analyzed biases in narrative representation, we next assess how they manifest in the numerical reporting of the four news outlets.

Because numerical information plays a central role in shaping how audiences understand the scale and distribution of harm, disparities in how numbers are presented offer an important window into these biases.} Reporting of victim numbers in news articles is crucial for conveying the magnitude of casualties and the impact of violence on communities. Quantitative data not only provides readers with a clear understanding of the scale of incidents but also shapes public perception and policy responses to crime and victimization. Research indicates that the inclusion of victim statistics can significantly influence how audiences interpret the severity of an issue, thereby fostering a sense of urgency and empathy towards affected populations~\cite{johns2019civilian}. For instance, when media outlets report on violent crimes, the number of victims can serve as a stark reminder of the human cost associated with such events, prompting discussions on prevention and intervention strategies~\cite{kraitzman2025civilian}. Furthermore, the reporting of victim numbers can also reflect broader societal attitudes towards certain groups, as disparities in reporting are likely to lead to the devaluation of victims from marginalized communities~\cite{white2021whose}. This selective emphasis on victim statistics not only affects public discourse but can also influence legislative and law enforcement priorities, underscoring the media's role in shaping narratives around crime and victimhood~\cite{bouchard2020differential}. Thus, accurate and comprehensive reporting of victim numbers is essential for fostering informed public dialogue and effective policy-making.

Recognizing the importance of reporting numbers in conflict scenarios, we employed an LLM to analyze all articles and extract instances containing numerical references to civilians being harmed in any capacity, such as ``100 civilians were killed in that strike.'' For clarity and convenience, these instances will be referred to as Civilian Victim Numbers (CVNs). A detailed explanation of how this information was extracted and validated can be found in Supplementary Note~4. {\color{black} These extracted CVNs allow us to examine whether numerical attention is allocated proportionally across the two sides (equalizing bias) and whether numerical claims are treated differently in terms of source attribution or skepticism (doubt-casting bias).} We begin by examining the total number of articles containing at least one CVN.

\begin{figure}[H]
    \centering    
    \includegraphics[width=.95\textwidth,height=0.9\textheight,keepaspectratio]{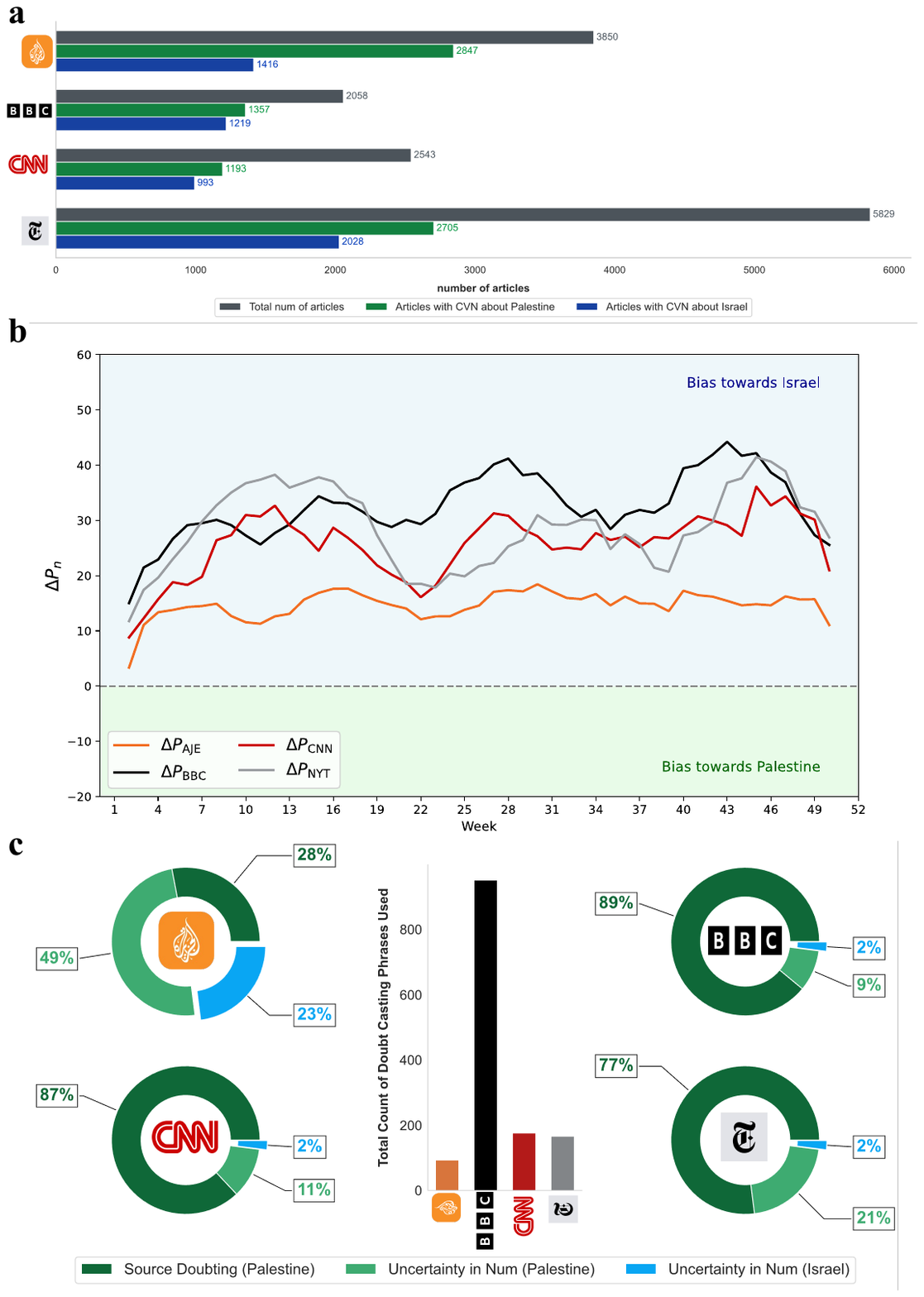}
    \caption{\textbf{Results of Quantifying the Human Cost of War Analysis.} \textbf{a)} A comparison of the total number of articles published by each news outlet compared to the subsets containing at least one CVN about Palestine and those with at least one CVN about Israel. \textbf{b)} The weekly proportion of expected mentions allocated to the Palestinian side ($\Delta P_n$) based on the baseline model, shown for each news outlet. \textbf{c)} A bar plot displaying the total number of doubt-casting phrases per source, alongside additional charts breaking down the counts by the type of doubt-casting technique used by each outlet.
    \label{fig:Figure 3}
    }
\end{figure}

Figure~\ref{fig:Figure 3}a presents a bar plot displaying the number of articles for each of the four media sources. Each source is represented by three bars: one showing the total number of articles (grey), one for articles mentioning CVNs related to the Palestinian side (green) and one for the Israeli side (blue). Notable differences in reporting styles emerge when comparing AJE with the Western news outlets. AJE reports twice as many articles with Palestinian CVNs than Israeli, while BBC and CNN report on both sides almost equally. Furthermore, NYT presents an interesting case: although the absolute number of its articles mentioning Palestinian CVNs (2,705) is comparable to AJE’s (2,847), the proportion is markedly lower (46\% for NYT versus 74\% for AJE). At the same time, NYT reports more on Israeli CVNs in absolute terms than any other outlet, whereas the BBC devotes the highest proportion of its coverage to Israeli CVNs (59\%).

Given the significant disparity in civilian casualties reported by October 7, 2024 (over 41,689 Palestinians versus 1,200 Israelis killed, and 96,625 Palestinians versus 5,432 Israelis injured, as documented by the United Nations~\cite{ocha_website}), this imbalance should logically be reflected in media reporting, with greater attention directed toward the side experiencing more severe and prolonged impacts. To evaluate this, we assessed the proportion of CVNs specifically related to casualties (for a breakdown of the different types of CVNs, refer to Supplementary Table~6). Our findings confirm that the majority of CVNs were indeed linked to casualty-related statistics. Moreover, we observe that, across all outlets, Palestinian CVNs outnumber Israeli CVNs. However, in Western media specifically, Israeli CVNs still amount to more than half the number of Palestinian CVNs, despite the vast disparity in actual casualties. Once again, this points to a form of {\color{black} equalizing bias}, where Israeli suffering is elevated to a level of prominence in coverage that does not reflect the underlying reality.

{\color{black} \subsubsection*{Equalizing Bias in Coverage: Baseline Comparison of Casualty-Related Numbers}}

To assess whether the observed imbalance is limited to the early stages of the war following the October~7 attack, we analyzed the aggregate article counts from each news source over time. Supplementary Figure~6 illustrates the difference in the percentage of articles mentioning Palestinian casualty-related CVNs versus Israeli casualty-related CVNs. Data points above zero indicate more articles reporting Palestinian CVNs than Israeli ones, and data points at zero indicate equal reporting for both sides. The data reveal that AJE consistently reported more articles on Palestinian casualties, with an average percentage difference of approximately 41\%, whereas BBC, CNN, and NYT hovered closer to zero, averaging 10\%, 11\%, and 14\%, respectively. Given the substantial disparity in casualty numbers between the two sides, a higher proportion of Palestinian CVNs is expected. Moreover, while Israeli casualties were concentrated largely around October~7 and its immediate aftermath, Palestinian casualties were sustained throughout the year, which would further suggest more frequent reporting on Palestinians to reflect this temporal distribution. However, the question arises: how large should that difference be? Does a line far above zero indicate bias toward Palestinians, while one close to zero would reflect bias toward Israelis? Does AJE's reporting reflect fairness, while other outlets lean toward bias, or is the opposite true? Determining at what point the percentage difference transitions from fair reporting to bias requires establishing a baseline to objectively evaluate the reporting styles of the four media sources.

To this end, we propose a baseline model to predict the expected number of mentions of casualty-related CVNs in news outlets based on weekly casualty numbers reported by the United Nations Office for Coordination of Humanitarian Affairs (OCHA) during the first year of the conflict~\cite{ocha_website}. The model calculates the expected share of mentions for Palestine and Israel, relative to their casualty numbers. This aligns with findings by Miller and Albert (2015), who observed that media coverage is strongly influenced by casualty figures, following the adage ``If it bleeds, it leads,'' wherein fatal incidents dominate narratives and amplify their emotional and psychological impact~\cite{miller2015media}. The model also accounts for sudden increases in media coverage, or ``spikes,'' caused by sharp rises in casualties. These spikes, reflecting significant shifts in violence, temporarily alter media dynamics, as shown by previous research on media reactions to crises~\cite{hansen2011media, eisensee2007news}. Additionally, the model incorporates a decay factor to capture the natural decline in coverage over time. This decay, often referred to as ``agenda-decay,'' is consistent with findings by Pfefferbaum et al. (2014) and Downing et al. (2004), which highlight that media attention diminishes without continuous stimuli, reflecting the temporal dynamics of waning public interest as the news cycle progresses~\cite{pfefferbaum2014media, downing2004sage}. These components enable the model to account for the fluctuating patterns of media attention throughout the conflict. For a detailed description, kindly refer to the subsection ``Baseline Model of Expected Weekly Casualty Reporting'' in Materials and Methods.

{\color{black} By comparing expected versus observed numerical attention, the model allows us to quantify the degree of equalizing bias for each outlet.} The comparison between the expected number of casualty-related CVN mentions, based on the baseline model, and the actual number of casualty-related CVN mentions for each side is visualized over time for each news source in Supplementary Figure~7. A consistent trend emerges across all outlets: Israeli CVNs appear to be overreported, while Palestinian CVNs seem to be underreported. However, comparing the magnitude of these gaps between expected and actual mentions across different news sources is challenging. Just how much overreporting or underreporting—and thus potential bias—exists relative to the baseline? To address this, we calculate the weekly proportion of expected mentions allocated to each side based on the baseline model for each news outlet.

For example, suppose that in a given week $t$, the baseline model predicts that, in news source $n$, 90\% of all casualty-related CVN mentions should concern Palestinians and 10\% should concern Israelis, based on the actual number of deaths that week. In reality, however, news source $n$ might report 60\% of its casualty-related CVN mentions about Palestinians and 40\% about Israelis. This would indicate a 30-percentage-point shortfall in Palestinian coverage (\( \Delta P_n^t = +30\% \)) and an equivalent overrepresentation of Israeli coverage for that week (\( \Delta I_n^t = -30\% \)). This ensures the differences are symmetric in absolute terms, reflecting the same magnitude of deviation for both sides. We display the Palestinian gap (\( \Delta P_n^t \)) to visualize the discrepancy between actual reporting and the baseline expectation, allowing us to assess the bias of each news outlet. These results are illustrated in Figure~\ref{fig:Figure 3}b. The figure reveals that NYT, CNN, and BBC exhibit a greater bias toward Israel, with the US-based sources (NYT and CNN) frequently interchanging over time. BBC stands out as the most biased news source overall, with an average gap of approximately 31\%. In contrast, AJE consistently demonstrates the least bias, maintaining an average gap of just 13\% over time in favor of Israel.\\

\subsubsection*{Selective Citation of Sources}
Our analysis thus far has focused on whether one side's CVNs are being under-reported, operating under the assumption that these numbers are fact-checked and accurate, thereby granting media sources the benefit of the doubt. However, the reporting practices of mainstream media often reveal biases that can profoundly influence public perception and understanding of events. One prominent aspect of this bias is the selective citation of sources and facts, which serves to reinforce particular narratives while marginalizing alternative viewpoints. Research indicates that such selective reporting can create a skewed representation of reality, as news sources may prioritize information that aligns with their ideological leanings, thereby engaging in what has been termed ``partisan coverage filtering''~\cite{broockman2025consuming}. Furthermore, the phenomenon of confirmation bias suggests that media organizations may favor facts that support their existing narratives, while casting doubt on those that do not~\cite{elejalde2019understanding,edgerly2024speaking}. This can manifest in the use of language that introduces uncertainty around facts that contradict the media's preferred storyline, thereby shaping audience perceptions in a way that aligns with the outlet's biases~\cite{jomini2007media,salvador2019viral}. The implications of these practices are profound, as they not only distort the truth but also contribute to a polarized media landscape where audiences are increasingly exposed to biased information that aligns with their pre-existing beliefs~\cite{stroud2010polarization,kim2015does}.

Building on the previous observations, we conducted an analysis to investigate the possibility of selective citation of sources for CVNs. We utilized an LLM to evaluate each sentence containing a CVN, prompting it to determine whether a source was cited (for details, see Supplementary Note~5). Subsequently, we examined the frequency of cited sources for CVNs on each side and compared patterns across different news outlets (examples of the most popular ways to cite sources for each side can be found in Supplementary Tables~7 and 8). The results, detailed in Supplementary Table~9, highlight distinct differences in reporting practices. AJE demonstrated a balanced approach, citing sources with almost equal frequency for both Palestinian and Israeli CVNs, indicating a consistent standard for source attribution. In contrast, Western news outlets were significantly more prone to cite sources when reporting on Palestinian CVNs than Israeli ones. U.S.-based outlets, in particular, referenced sources for Palestinian data nearly twice as often as for Israeli figures. BBC, on the other hand, reported Israeli CVNs with the least source attribution, at only 16.9\% of instances, while attributing sources to Palestinian CVNs three times more frequently. These disparities raise questions about potential biases in validating casualty figures based on the context.

\subsubsection*{Creating Skepticism {\color{black} through Doubt-Casting}}
Next, we analyzed how referenced CVNs were phrased to determine whether they were presented in a ``casting doubt'' style (details can be found in Supplementary Note~6). Using an LLM, we examined full sentences containing CVNs for phrases implying skepticism or questioning the source or the reported numbers. These doubt-casting phrases were categorized into two main types: (1) Source Doubting, i.e. phrases that question the credibility of the reporting source, thereby indirectly doubting the reported numbers, such as: ``Israel’s relentless airstrikes have killed more than 8,000 people, \textit{according to the Hamas-run health ministry}''~\cite{Fichera_2023} and ``The health ministry in Gaza, \textit{which is controlled by Hamas}, estimates that 11,000 civilians have been killed there over the last month,''~\cite{Guo_2023}
and (2) Uncertainty in Numbers, i.e. phrases that cast doubt specifically on the reported numbers, such as: ``...dozens of civilians were \textit{reportedly} killed when jets struck buildings...''~\cite{Rabin_2024}. 
For a complete list of doubt-casting phrases and their categories, see Supplementary Table~10. Notably, our analysis found no source-doubting phrases on Israeli-reported numbers, even after using a second independent LLM prompt designed specifically to detect such instances. This suggests that doubt-casting techniques are disproportionately applied to Palestinian-reported numbers. While this study does not aim to verify the accuracy of these numbers, its purpose is to evaluate, through numerical analysis, whether both sides are presented equally in news coverage. We explore this further below.

After identifying doubt-casting phrases, we analyzed their occurrences for each side (Palestine vs. Israel) across four news sources. The results, summarized in Figure~\ref{fig:Figure 3}c, reveal that BBC overwhelmingly leads in using doubt-casting phrases, with 951 CVN mentions, 98\% of which target Palestinians, averaging over two CVNs doubted daily during the first year. This far exceeds CNN (176), NYT (166), and AJE (93). Common Source Doubting phrases like ``Hamas-run health ministry'' were heavily used by BBC, CNN, and NYT, whereas analogous expressions for Israel (such as ``Knesset-run health ministry'') were absent. CNN and NYT also favored phrases like ``Hamas-controlled Gaza.'' For Uncertainty in Numbers, terms such as ``reportedly'' , ``allegedly'', and ``claim'' were predominantly used to describe Palestinian casualties by Western outlets, whereas AJE applied them more evenly across both sides (see Supplementary Tables~11-14 for detailed lists of phrases by outlet). Western sources attributed 77-89\% of their cast-doubting phrases to doubting the sources of Palestinian CVNs, compared to 28\% for AJE. Additionally, AJE was the only outlet to use Uncertainty in Numbers phrases in a balanced manner: 49\% referred to Palestinians and 23\% to Israelis, a proportion consistent with the fact that AJE published roughly twice as many articles about Palestinian CVNs as about Israeli ones. In contrast, the other outlets applied such language almost exclusively to Palestinians, with 2\% or less referring to Israelis. These findings underscore in particular BBC’s significant reliance on doubt-casting phrases, particularly when reporting on Palestinian casualties, compared to other sources. 

Doubting Palestinian sources is certainly justified, given the inherent difficulties in providing precise victim estimates during wartime. Statistics released by Palestinian authorities in Gaza were indeed frequently corrected and updated as new information became available. However, the same holds true for victim estimates on the Israeli side, where official statistics were revised repeatedly for up to six months after the events. Notably, skepticism toward Palestinian figures was often amplified by the practice of explicitly associating them with Hamas, an organization formally designated as terrorist group in the United States~\cite{us_state_department_terrorist_orgs} and the United Kingdom~\cite{uk_home_office_proscribed_groups}, thereby highlighting their perceived lack of credibility. This rhetorical framing tends to obscure the fact that similar or identical casualty estimates were also reported and validated by independent researchers, international institutions, and humanitarian organizations (e.g.,~\cite{huynh2024no}).

\subsubsection*{Children in the Narrative: Imbalances in CVN Reporting}

To conclude this section of our analysis, we examined how frequently media outlets referenced children in their reporting of civilian victim numbers (CVNs) for both Palestinians and Israelis (Supplementary Figures~8 and 9, and Supplementary Table~15). To contextualize these proportions, we benchmark them in Supplementary Table~15 against the approximate share of children among civilian deaths on each side: around 5\% for Israeli civilians on October~7 (and roughly 3\% when including soldiers, based on public compilation)~\cite{ayoub2024comparative, wikipedia2024oct7}, compared to 40–44\% for Palestinian civilians over the first year of the war (October~7, 2023–October~7, 2024)~\cite{OHCHR2024GazaUpdate, SaveTheChildren2024GazaChildren}.

Relative to these baselines, Al Jazeera English (AJE) provides the most proportionate coverage of Palestinian children, with casualty-related child mentions at 29.3\% (closest to the 40–44\% reference) but reports the least on Israeli children, at only 2\% compared to the 5\% baseline. The New York Times (NYT) exhibits the reverse pattern: it comes closest to the Israeli baseline (4.2\% vs.~5\%) but falls furthest below the Palestinian reference (15.7\% vs.~40–44\%). BBC and CNN occupy intermediate positions, with 20.8\% of casualty-related Palestinian child mentions for both, and 2.6–3.4\% for Israeli children. Overall, these patterns highlight how outlets differ not only in their quantitative attention to children but also in how proportionally, or selectively, they mirror the actual distribution of child victims across the two populations.

\section*{Discussion}
Three main conclusions can be drawn from our media analyses of over 14.000 articles published by four international news outlets during the first year of the 2023 war on Gaza. 

First, unlike AJE, Western media (BBC, CNN, and NYT) tend to emphasize the individuality of Israeli victims experiencing hardship, while treating Palestinian victims more frequently as a collective. This is evident in the higher ratio of individualized to group-level reporting for the Israeli compared to the Palestinian side. The only outlet showing a more balanced pattern is AJE; here the prevalence of individualized (vs. group-level) narratives is very similar for the two sides. These findings suggest that even highly professional (and relatively “liberal”) Western media tend to downplay the individuality of Palestinian victims experiencing the traumatic (and often fatal) consequences of war, while acknowledging, to a greater extent, the personal suffering of Israeli victims. {\color{black} This finding confirms what Grimm et al. (2025)~\cite{grimm2025hierarchies} observed in a small-scale qualitative discourse analysis of 40 articles published in four German newspapers during the first 7 weeks of war, namely that Israeli victims tended to receive more personalized coverage than Palestinian victims.} Based on previous literature on the identifiable victim effect~\cite{slovic2007psychic}, the Western reporting pattern is likely to elicit greater empathy and compassion for Israeli than for Palestinian victims. Although testing the impact of this reporting bias on the readership was beyond the aims of the present study, this appears to be a worthwhile endeavor for future research. 

Looking at the empathic style of reporting, we see a more complex pattern. The two main indices of the Human Empathy and Narrative Taxonomy~\cite{shen-etal-2024-heart} show a small but consistent tendency of AJE journalists to report in greater detail (plot volume) and with more vivid emotions on individual Palestinian than Israeli victims. In contrast, Western media appear more reactive to specific events, showing modest pro-Palestinian spikes at some moments and stark pro-Israeli spikes in others. Taken together, these findings suggest that it is not as much the style of writing than the under-reporting of individual stories that contributes to the dehumanization of Palestinians in Western media.

The second systematic bias in Western media (and to a lesser degree in AJE) was the tendency to equate Israeli to Palestinian suffering despite the stark difference in all forms of hardship (casualties, displacement, arbitrary detention and kidnapping, destruction of the health, school, and university systems, etc.){\color{black}, We termed this pattern ``equalizing bias''}. In the absence of bias, news reports should reflect the actual gaps in human suffering. To give only one example, if Israeli child deaths are less than 1\% and Palestinian children more than 99\% of the total number of children killed during the first year of war, news reports would roughly reflect this discrepancy. In contrast to this assumption, we find a systematic deviation from actual discrepancies that emerges consistently across observations: (a) In Western media, 47\% of individual narratives on children referred to Israeli children who constitutes less than 1\% of the victims; (b) Including all age groups, the ratio of Palestinian to Israeli casualty-related individual stories was approximately 1:1 in Western media, despite an actual casualty ratio of 20:1; (c) Especially in Western media, Israeli Civilian Victim Numbers are over-reported and Palestinian Numbers under-reported compared to what would be expected based on the actual number of victims and the distribution over time. Together, these reporting patterns insinuate a false premise of symmetry, where, in reality, the entity and prevalence of human suffering is blatantly asymmetrical. {\color{black} This raises the question of whether the equalizing bias is unique to the Israeli-Palestinian conflict or whether it constitutes a more general bias likely to occur whenever journalists (and possibly laypeople) identify with the dominant side of an asymmetrical conflict. What argues in favor of the latter interpretation is our experimental study replicating the equalizing bias among lay people acting as journalists in an imaginary, greatly tilted conflict. }

How did BBC, CNN, and the NYT reporting convey an impression of (false) balance despite the extreme disparities in military power and human suffering? One contributing factor was the frequent return to the events of October 7th, even months later and in the absence of new incidents involving Israeli victims. Our analysis suggests that reiterations of October 7 were particularly pronounced during periods when significant events occurred on both sides, often overshadowing coverage of Palestinian suffering and diverting attention from the toll in Gaza. This practice, which contradicts the common ``agenda decay'' phenomenon, not only creates an illusion of balance, but also reiterates the idea that the 7th of October was the starting point and ``cause'' of the subsequent events. Importantly, we do not intend to suggest that the reiteration of October 7th was part of a deliberate strategy pursued by the analyzed media, however, even such unintentional biases may have contributed to the neglect of Palestians' victimhood in public perception. Biases can arise from common and often automatic cognitive processes. As previously discussed, a single mass-casualty event tends to attract disproportionate attention and to be remembered more vividly than a higher but gradually accumulating death toll. In addition, events that occur first in time or space typically receive greater attention than those that follow. Together, these mechanisms may partly explain why the October 7th massacre remained a persistent reference point in subsequent reporting. At the same time, perfect parity in coverage should not be expected, given the profound asymmetry in casualties between the two sides and the unequal number of conflict-related events that could be reported. {\color{black} Yet these cognitive mechanisms alone can hardly explain the almost perfect parity in coverage we observe, particularly in light of the profound asymmetry in casualties and the greatly unequal number of conflict-related events that could be reported.

We interpret the reporting pattern observed in Western media as an instance of what we have termed the equalizing bias, namely the tendency to portray two groups as equally affected by a conflict despite stark differences in suffering. Although this is beyond the scope of the present media analysis, the equalizing of unequal parties may not only obscure asymmetries in suffering, but also in power and in culpability (such as violations of international law), leading audiences to (mis-)perceive the conflict as an intractable clash of opposing but equivalent sides. As one of the main consequences of false equivalence is the production of uncertainty~\cite{dixon2013heightening}, it is plausible that, in the context of war journalism, equalizing portrayals will reduce public recognition of victimization and aggression, thereby dampening calls for accountability and action. Whether the equalizing bias does, indeed, have these effects remains an open question for future research. 

Furthermore, why would Western journalists portray Israeli suffering at par with Palestinian suffering? We can envisage at least three, not mutually exclusive reasons: First, the equalization bias may reflect the journalistic norm of pursuing balance and impartiality (although the same desire for balance was not visible in the choice of individuating and doubt-casting reporting). This interpretation, however, was not supported by our experimental simulation study in which the endorsement of journalistic norms failed to predict the magnitude of the equalizing bias. Second, to the degree that Western reporters may identify more with the Israeli than the Palestinian side~\cite{terzian2025epistemic}, they may engage in what is known as ``competitive victimhood'', namely the tendency to claim that one’s own group has suffered more than a relevant outgroup~\cite{young2016competitive}. In particular, competitive victimhood provides a tool for regaining moral ground whenever a group is accused of inflicting harm to the outgroup (Sullivan et al., 2012~\cite{sullivan2012competitive}). This interpretation is supported by our experimental findings, which show that identification with the stronger country significantly predicted the equalizing bias and fully mediated the effect of experimental condition on victim selection, suggesting that group identification alone is sufficient to generate competitive victimhood dynamics. The third and most disquieting possibility is that the equalizing bias reflects deliberate journalistic malpractice involving editorial censorship and/or self-censorship~\cite{guardian2024cnn}. Regardless of which underlying mechanism drives the phenomenon and whether it occurs in an intentional or unconscious manner, the equalizing bias is likely to muddy public understanding of asymmetrical power relations and lead people to suspend judgement and action~\cite{terzian2025epistemic}.
}

Finally, our third finding concerns the presumed reliability of the reported information and the trustworthiness of the source providing it. Unlike AJE, Western journalists were found to engage in strategies aimed at planting seeds of doubt in their readership when reporting on Palestinian victims. When quantifying the human cost of war for civilian victims, Western journalists tended to cite sources more frequently for Palestinian than for Israeli victims, as if Israeli (but not Palestinian) statistics can be taken for granted. This interpretation is further corroborated by a systematic and asymmetrical ``casting doubt'' style that questions the veridicality of data predominantly when referring to Palestinian victims (such as ``reportedly'', ``claims'', ``according to the Hamas-run health ministry'', and the like). By discrediting sources in a subtle way, journalists undermine the reader's trust in the information provided and in the reputation of the source.

Together, the present media analysis identifies multiple biases in Western media that may have contributed to the global divide in public opinion and to the delayed public outrage in the Western world in response to the unfolding genocide. Somewhat surprisingly, we did not find a complementary pro-Palestinian bias in AJE, which---with a few exceptions, such as the underreporting of Israeli children’s CVNs---exhibited a relatively balanced reporting pattern.

{\color{black} \subsubsection*{Scope and Generalizability} 

Beyond the present case, the study contributes a theory-driven analytic framework for examining how media bias manifests in the coverage of asymmetrical warfare. This framework can be extended to compare media outlets, conflicts (e.g., Israel–Gaza vs. Serbia–Bosnia), or temporal phases within the same conflict (e.g., Gaza wars between 2008 and 2023). To exemplify the latter, we conducted a follow-up replication for our main bias measures to analyze how the same media (AJE, BBC, CNN, NYT) reported on the previous 3 wars on Gaza; for a detailed analysis see Supplementary Note 7. Just like the most recent (2023) war, these three wars, which took place in 2008, 2012, and 2014, showed a strong asymmetry in power and victimhood with Palestinians representing over 90\% of all casualties.  Importantly, during the 2008 and 2012 wars, Israel was accused of war crimes, but not genocide. In 2014, some organizations voiced genocide accusations, but these did not lead to formal court rulings, unlike the recent 2023 war, where such accusations were brought before the International Court of Justice. 

If accusations of ingroup wrong-doing increase the motivation to restore the lost moral ground through competitive victimhood (Noor et al., 2012~\cite{noor2012suffering}; Sullivan et al., 2012\href{applewebdata://7037B8A0-E962-4AD4-AFC7-9531E87F1767\#_msocom_2}{[2]} , Young \& Sullivan, 2016\href{applewebdata://7037B8A0-E962-4AD4-AFC7-9531E87F1767\#_msocom_3}{[3]} ) and if the equalizing bias offers a way to achieve this, then one would expect a greater equalizing bias the more serious the accusation of wrong-doing. 
A comparison across the four Gaza wars is broadly consistent with this reasoning. Equalization in casualty reporting appears to have been present in earlier wars and to have intensified over successive conflicts, becoming significantly most pronounced in 2023 (see Supplementary Figure 13), with the notable exception of AJE. It is also in line with another time trend observed in our analysis, namely the tendency to individualize Israeli (vs. Palestinian) victims more clearly in the more recent wars (2014 and 2023) where accusations became more serious (see Supplementary Figure 12). 
While this interpretation is post-hoc, emphasizing the victimhood of the more powerful group (Israel) remains relevant. As Noor et al.~\cite{noor2012suffering} argue, competitive victimhood serves several functions, including inflating the perceived threat from the out-group, justifying in-group violence, and mobilizing moral and material support from third parties. For our last analysis concerning doubt-casting language when reporting casualty statistics, both source-doubting (such as “Hamas-run health ministry”) and doubt-casting expressions (such as “reportedly”) were rare before 2023. They saw a steep increase concerning Palestinian casualty figures in 2023, specifically for BBC. Together, this timeline suggests an increase in Western media bias over time, which is most pronounced for the individualizing and the equalizing biases. This additional analysis illustrates that our framework lends itself to multiple comparisons, including historical trends.}

\subsubsection*{Limits and future directions}
Despite its extensive coverage, the present study has several limitations that should be acknowledged and could serve as a starting point for future research. First of all, we took a conservative approach and included only highly respected and relatively ``liberal'' media in our analysis. This has almost certainly led to an underestimate of actual media bias. Thus, future studies should broaden the scope to include more conservative media (such as Fox News or Daily Mail).

Another important future development regards the extension of the analysis beyond the first year of war. It is conceivable that the pro-Israeli / anti-Palestinian bias in Western media may have declined during the second year of the ongoing genocide and that this may have contributed to greater awareness and the shift in public opinion in the West. Investigating whether the media biases identified here did indeed shape public opinion and to what extent remains another important issue to be addressed in future research. Only longitudinal or experimental methods will be able to address this question and allow inferences about the actual impact of the biases observed here on public opinion and collective action.

\newpage
%TC:ignore
\section*{Materials and Methods}

\subsection* {Data Selection and Sample Construction}

{\color{black} \noindent \textbf{Media Outlet Selection.}
Our analysis focuses on four major international news organizations: The New York Times (NYT), BBC, CNN, and Al Jazeera English (AJE). Outlet selection was guided by four criteria:
\begin{itemize}
    \item publication in English,
    \item digital accessibility of archival content,
    \item the presence of independent reporting infrastructure (including field correspondents), and
    \item substantial global audience reach.
\end{itemize}

Audience reach and influence rankings consistently identify BBC, NYT, and CNN among the most widely consumed English-language news organizations worldwide. According to PressGazette (2025)~\cite{PressGazette2025}, BBC, NYT, and CNN rank first, second, and fourth, respectively, in global visits among the top 50 English-language news websites. MSN, which ranks third, was excluded because it functions as a news aggregator without its own reporting staff or field reporters. Among non-Western outlets, several Indian news websites (including India Times, Hindustan Times, The Indian Express, India Today, and One India) appear high in the rankings, but their readership is overwhelmingly domestic or diaspora-based. The highest-ranked truly international non-Western outlet is AJE (46th place). This pattern is consistent with rankings produced by PRlab~\cite{PRlab2025}, which list CNN, BBC, and NYT as the first, second, and third most influential outlets for news coverage, while AJE—ranked sixth—is the leading non-Western organization. The prominence of these outlets is further supported by audience research such as the 2024 Digital News Report~\cite{DigitalNewsReport2024}. Using the share of respondents who reported seeing content from each brand in the past month as a measure of global reach, CNN (25\%), BBC (22\%), and NYT (12\%) emerge as the top three news brands worldwide, while AJE (10\%; ranked fifth) is again the highest among non-Western outlets. The audience scale is substantial: the three Western outlets alone attract an estimated average of 1.906 billion visits per month, including approximately 421 million unique visitors.\\

\noindent \textbf{Temporal Scope.}
We restrict our analysis to the first year of the war (October 7, 2023 - October 6, 2024), as this period captured both the overwhelming majority of global public attention and the phase of maximal media influence. Independent indicators of information demand, including Google Trends and Wikipedia pageview statistics (Supplementary Figures 1–2), show a sharp spike in interest immediately following October 7, followed by sustained levels of attention that remained substantially elevated throughout the first year relative to the subsequent period. 

This early attention surge marks the interval in which news organizations allocated exceptional resources, public reliance on media was greatest, and core interpretive frames crystallized. It is also the period containing most of the conflict’s humanitarian and military variance, when reporting was most intense and consequential for public understanding. 

%By contrast, coverage in the second year is structurally different: audience attention is far lower, reporting becomes routinized and sporadic, and editorial resources contract. Such low-salience, lower-frequency coverage is not comparable to the high-intensity environment of Year 1 and offers limited diagnostic value for identifying the origins of systematic media bias. Thus, restricting the analysis to the first year provides a theoretically conservative and methodologically coherent test: if systematic asymmetries emerge during the phase of maximal scrutiny, they cannot be attributed to marginal coverage or declining newsworthiness, but instead reflect structural features of early war reporting.

By contrast, coverage beyond the first year occurs under conditions of markedly lower public attention and reduced reporting intensity. As audience demand stabilizes at lower levels, coverage becomes more routinized and episodic, and editorial investment contracts. Because media effects and framing dynamics are strongest during periods of high salience and interpretive uncertainty, the first year provides the most analytically informative window for assessing systematic reporting asymmetries. Examining this interval therefore allows us to evaluate patterns of representation under conditions of maximal scrutiny, when media influence and agenda-setting capacity are at their peak.}\\

\noindent \textbf{Data Retrieval and Filtering.}
The New York Times (NYT), CNN, and Al Jazeera English (AJE) news articles were accessed via the LexisNexis Academic Web Services API (WSApi)\footnote{LexisNexis NYU guide, https://guides.nyu.edu/lexisnexis-rest-api}\footnote{LexisNexis WSApi https://solutions.nexis.com/wsapi/, accessed through October 2024}. These articles cover the period from October 7, 2023, to October 7, 2024. To ensure the inclusion of only highly relevant news articles, we excluded specialized content such as podcasts and audio transcripts. We also removed summary articles that condensed multiple topics into brief overviews. Additionally, we filtered the articles based on country and subject tag, retaining only those with content relevant to Palestine and Israel, specifically tagged with ``Hamas,'' ``Israel,'' ``Gaza,''  ``West Bank," and ``Palestin.'' (the term Palestin was used to account for the variations of the word, such as Palestinian(s) and Palestine).

For BBC news articles, which were not part of the LexisNexis dataset, we retrieved them directly from the BBC website. Similar to the LexisNexis filtering, we focused on articles categorized under tags pertinent to our research, including ``Israel,'' ``Israel and the Palestinians,'' ``Gaza,'' ``Palestinian territories,'' ``Hamas,'' ``West Bank,'' and ``Israel-Gaza war.'' Non-news media, such as live feeds and audio transcripts, were also excluded. Further, we used Google News Search to find relevant articles, restricting our query to return results that belong to the BBC website and we only kept articles with titles containing the following terms: ``Hamas,'' ``Israel,'' ``Gaza,''  ``West Bank," ``IDF," and ``Palestine.'' This process resulted in a dataset comprising 14,280 news articles, distributed as follows: 5,829 (40.8\%) from NYT, 3,850 (27\%) from AJE, 2,543 (17.8\%) from CNN, and 2,058 (14.4\%) from the BBC.

\subsection*{Testing Individualization Bias with GEE Models}

To test whether one side was more likely to be individualized in news coverage, we conducted a series of Generalized Estimating Equations (GEE) logistic regression analyses. This approach accounts for potential correlation among observations clustered within the same article, recognizing that instances within an article may be shaped by shared context, author style, or topical focus. Each LLM-identified mention (instance) was coded using two binary variables:
\begin{enumerate}
\item \textbf{Instance Type}, where 1 = Individualized and 0 = Grouped.
\item \textbf{Side}, where 1 = Palestinian and 0 = Israeli.
\end{enumerate}

To account for clustering, each instance was linked to a unique Article ID, ensuring that multiple mentions from the same article were recognized as non-independent.

We estimated four separate GEE logistic regression models, one for each media outlet (AJE, BBC, CNN, and NYT). %using the \texttt{statsmodels} package in Python (version 3.12.2). 
In each model, the dependent variable was the binary \textit{Instance Type}, and the independent variable was \textit{Side}. We specified an exchangeable working correlation structure and used robust (Huber-White) standard errors to account for within-article correlation.

The estimated coefficients represent the log odds of an instance being individualized rather than grouped, comparing Palestinian to Israeli mentions. To aid interpretation, we also report odds ratios (exponentiated coefficients). A positive and statistically significant coefficient (or odds ratio $> 1$) indicates that mentions of the Palestinian side are more likely to be individualized relative to Israeli mentions, while a negative coefficient (or odds ratio $< 1$) indicates the reverse.

\subsection*{Measuring Narrative Elements Associated with Empathy in News Texts}

To systematically assess the level of empathy conveyed in story narratives, we adopted a validated framework from Shen et al. (2024) \cite{shen-etal-2024-heart}. In their study, the authors identified a set of narrative style elements that contribute to eliciting empathy and integrated them into the Human Empathy and Narrative Taxonomy (HEART). Each element in the taxonomy represents a distinct dimension of empathy, operationalized through targeted prompts for large language models (LLMs). For example, under the Vividness of Emotions metric, the LLM is instructed to assign a score of 1, 2, or 3 to indicate whether the emotional expressions in a text are not vivid at all, somewhat vivid, or extremely vivid.

Shen et al. (2024) found that three metrics contributed significantly to empathy:
\begin{enumerate}
\item \textbf{Vividness of Emotions:} the degree to which emotions are expressed vividly in stories.
\item \textbf{Plot Volume:} the frequency and significance of events in a story.
\item \textbf{Character Development:} the depth and evolution of characters throughout a narrative.
\end{enumerate}

Of these, we retained only the first two. Character Development was excluded, as it is generally ill-suited to news reporting, where stories are concise and rarely allow for sustained character arcs, unlike the longer-form social media and podcast narratives used in the original study.

Given that the HEART prompts had already been validated against human-coded responses and shown to achieve ``reasonable human-level'' agreement, we adopted them with minimal adaptation. Specifically, we modified the in-prompt examples to better reflect the kinds of content the LLM would encounter in our corpus while preserving the conceptual intent of the originals. For instance, an example rated as 3 under Vividness of Emotions in Shen et al.: 

\begin{quote}
    \textit{``The pain of losing someone is like being stabbed in the chest. I was devastated when I lost her''}   
\end{quote}
 
was replaced with:

\begin{quote}
    \textit{``And you lived there in a horrific sense of fear. Every second that you live with this feeling is a terrible feeling, that you don't really know if you're going to wake up in the morning, or in a minute, if a missile is going to fall on you, if they're going to come in with a Kalashnikov and start spraying us with bullets. The conditions are very, very difficult there.''}
\end{quote}

Similarly, an example rated as 1 for lack of vividness:

\begin{quote}
    \textit{``I didn’t feel great about the situation''}
\end{quote}

was replaced with:

\begin{quote}
    \textit{``Farhan al-Qadi, 52, a member of Israel’s Bedouin minority who was working at a kibbutz when he was seized in the Hamas raid of Oct 7.''}  
\end{quote}

We also added a brief introductory statement in the prompt to orient the LLM to the type of text it would process. Other than these targeted adjustments, the structure, instructions, and scoring criteria from the original HEART prompts were preserved.

After scoring each story individually using these prompts, we addressed the fact that a single news article could contain multiple stories, often from both Israeli and Palestinian perspectives. To account for this, we aggregated the story-level HEART scores to create an article-level empathy metric, enabling us to capture the relative balance of empathy elicited for each side within the same piece of reporting.

For each of the two retained HEART metrics (Vividness of Emotions and Plot Volume), the calculation proceeded as follows:

\begin{enumerate}
\item \textbf{Aggregate story-level ratings by side}: For each article, we summed the ratings for all Palestinian stories to obtain a Palestinian aggregate score, and likewise summed the ratings for all Israeli stories to obtain an Israeli aggregate score.
\item \textbf{Compute the difference}: We subtracted the Israeli aggregate score from the Palestinian aggregate score. This yielded a single article-level empathy difference score for that metric, where positive values indicated higher empathy toward Palestinian stories, and negative values indicated higher empathy toward Israeli stories.
\end{enumerate}

For example, consider an article containing three Palestinian stories with Vividness of Emotions ratings of $3$,$2$, and $2$ (aggregate $= 7$), and two Israeli stories with ratings of $1$ and $2$ (aggregate $= 3$). The article-level empathy difference score for Vividness of Emotions would be: $7-3=4$. A score of $+4$ in this case suggests that, on balance, the Palestinian narratives in the article were described with greater emotional vividness than the Israeli narratives. The same procedure was applied independently for Plot Volume, producing two article-level empathy difference scores, one per metric.

\subsection*{Baseline Model of Expected Weekly Casualty Reporting}

To estimate the expected weekly reporting of casualty-related CVNs for each side across four news sources during the first year, we propose a baseline model. This model preserves the total weekly mentions for each media outlet, adjusts for significant events, and incorporates a decay factor to account for the natural decline in coverage over time.\\

\noindent The model consists of the following components:

\begin{enumerate}
    \item \textbf{Reporting Ratio}: The reporting ratio \( r_n(t) \) captures the overall tendency of news source \(n\) to report casualty-related CVNs relative to actual casualty numbers in week \(t\). It is defined as:

   \[
   r_n(t) = \frac{M_{p,n}(t) + M_{i,n}(t)}{C_p(t) + C_i(t)}
   \]

   where:
   \begin{itemize}
   \item        \( M_{s,n}(t) \): the actual number of casualty-related CVN mentions in week \( t \) by news source \( n \) for side \( s \), where \( s \in \{p, i\} \), representing Palestine (\( p \)) and Israel (\( i \)) 
       \item \( C_s(t) \): the actual number of casualties reported in week \( t \) for side \( s \)

   \end{itemize}

   \item \textbf{Spike Factor}: The spike factor \( S_s(t) \) accounts for abrupt increases in casualty numbers. It is defined as: 

   \[
   S_s(t) = \begin{cases} 
   2, & \text{if } C_s(t) \geq 2 \times C_s(t-1) \\
   1, & \text{otherwise}
   \end{cases}
   \]

   \item \textbf{Initial Mentions}: The initial mentions \(G_{s,n}(t)\) generated in week \(t\) for side \(s\) by news source \(n\) are given by:

   \[
   G_{s,n}(t) = r_n(t) \times M_{s,n}(t) \times S_s(t)
   \]

    \item \textbf{Cumulative Expected Mentions with Decay}: The expected number of mentions \( M'_{s,m}(t) \) for side \( s \) in week \( t \) by news source \( n \) incorporate contributions from initial mentions in the current and prior weeks, adjusted by side-specific decay factors \( \delta_{p}=0.6\) (Palestine) and \( \delta_{i}=0.8\) (Israel); see Side Note 1 below for the rationale behind the choice of side-specific decay factors:

   \[
   M'_{s,n}(t) = \sum_{k=0}^{t-1} G_{s,n}(k) \cdot \delta_s^{t-k} + G_{s,n}(t)
   \]

   where:
   \begin{itemize}
       \item \( \delta_{s}\): the decay factor, where \( \delta_{p}=0.6\) for \(s=p\) (Palestine) and \( \delta_{i}=0.8\) for \(s=i\) (Israel)
       \item \( G_{s,n}(k) \cdot \delta_s^{t-k} \): the decayed mentions from previous week \( k \). 
       \item \( G_{s,n}(t) \):the initial mentions for the current week, which do not decay until subsequent weeks.
    \end{itemize}
   
    \item \textbf{Normalization of Weekly Mentions:} Finally, to normalize the expected mentions and reflect the actual number of mentions for each week, we define the normalized values for each side as follows:

    The normalized reporting ratio \(w_s(t)\) for each side \(s\) in week \(t\) is calculated by:
    
    \[
    w_s(t) = \frac{M'_{s,n}(t)}{M'_{p,n}(t) + M'_{i,n}(t)}
    \]

    Then, the final normalized expected mentions for side \(s\) is recalculated as:

     \[
     M'_{s,n}(t) = (M_{p,n}(t) + M_{i,n}(t)) \times w_s(t) 
    \]

    This normalization ensures that the final expected mentions for both Palestine and Israel are reflective of the actual number of mentions for each side, adjusted for the different decay rates. The model effectively accounts for variations in reporting tendencies, the impact of major events, and the natural decay of coverage over time.
\end{enumerate}

\noindent \textbf{Side Note 1: Rationale for Side-Specific Decay Factors}

The choice of decay factors ($\delta_p=0.6$ for Palestine and $\delta_i=0.8$ for Israel) reflects differences in the frequency and timing of occurring casualties, as well as how media coverage responds to these dynamics. The Palestinian side experienced a continuous stream of casualties throughout the first year, causing media coverage to decay more rapidly as fresh incidents emerged weekly. This aligns with Pfefferbaum et al.~\cite{pfefferbaum2014media}, who observed that ongoing crises lead to shifting media focus and faster declines in coverage for earlier reports. Additionally, Friedman \& Sutton~\cite{friedman2013selling} suggest that the continuous nature of such events accelerates the turnover of media attention.

In contrast, Israeli casualties were concentrated in the initial week, leading to a slower decay factor. The media's prolonged interest in these early events stems from their novelty and dramatic impact, consistent with research by Downing et al. and Rohner \& Frey~\cite{downing2004sage, rohner2007blood}, which shows that significant initial events can sustain attention longer in the absence of ongoing crises. Thus, the differing decay factors capture these dynamics and are supported by literature on media responses to crises and the temporal aspects of news reporting.

%\section*{Acknowledgements}

%\section*{Data Availability}

% Bibliography
%\bibliography{bib}
%\bibliographystyle{naturemag}

\pagebreak

\section*{Figures}

\pagebreak

\pagebreak

%TC:endignore

\end{document}

\nolinenumbers
\baselineskip24pt
\maketitle

\pagebreak

\begin{figure}[htp]
    \centering    \includegraphics[width=1\textwidth,height=0.9\textheight, keepaspectratio]{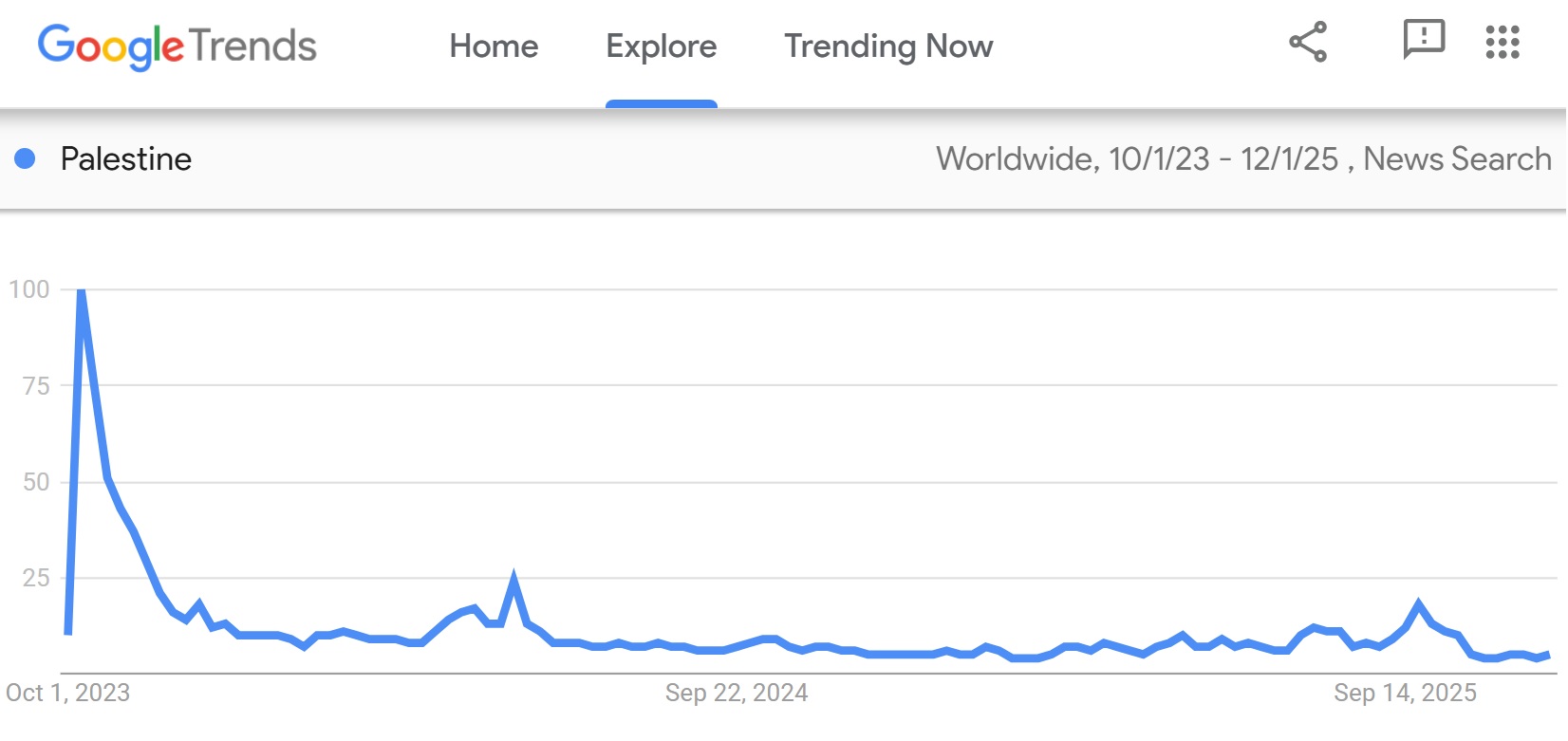}
    \caption{\textbf{Google Trends search interest for the term ``Palestine'' (News Search), October 1, 2023–December 1, 2025.} Values represent normalized relative search volume (0–100), with 100 indicating the peak level of global attention during the observation period.
    }
    \label{suppfig:Google_Trends}
\end{figure}

\begin{figure}[htp]
    \centering    \includegraphics[width=1\textwidth,height=0.9\textheight, keepaspectratio]{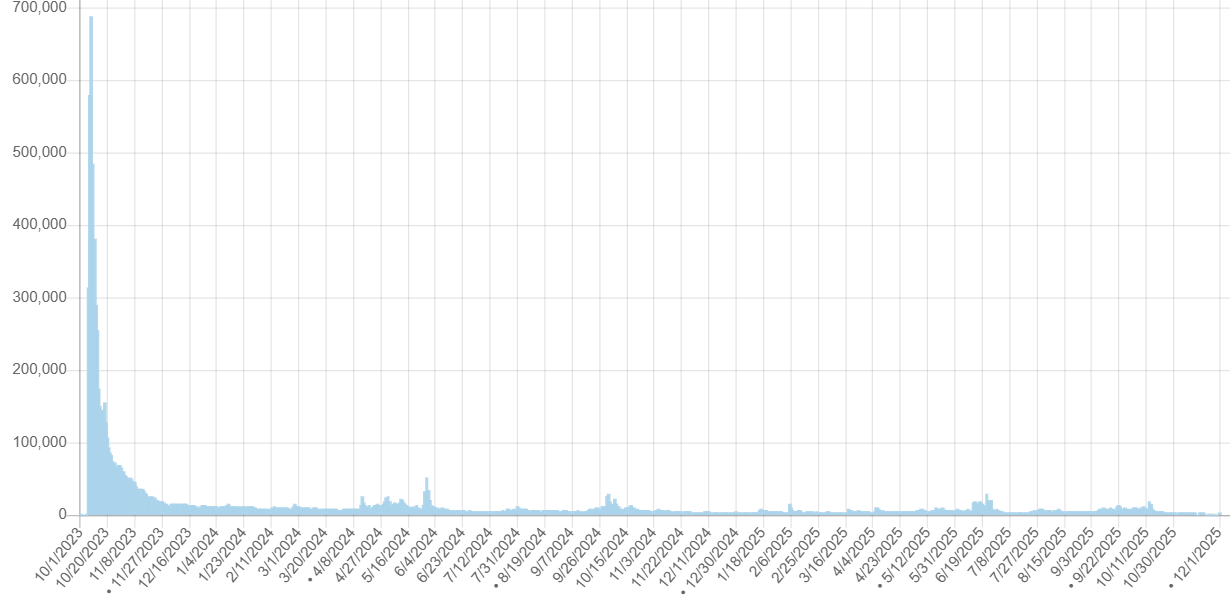}
    \caption{\textbf{Daily Wikipedia pageviews for ``Israeli–Palestinian conflict'' (October 1, 2023–December 1, 2025).} The data show a dramatic spike in information-seeking immediately after October 7, 2023, followed by a sustained decline and stabilization at lower levels.
    }
    \label{suppfig:Daily_Wiki_PageViews}
\end{figure}
\pagebreak

\begin{figure}[htp]
    \centering    \includegraphics[width=1\textwidth,height=0.9\textheight, keepaspectratio]{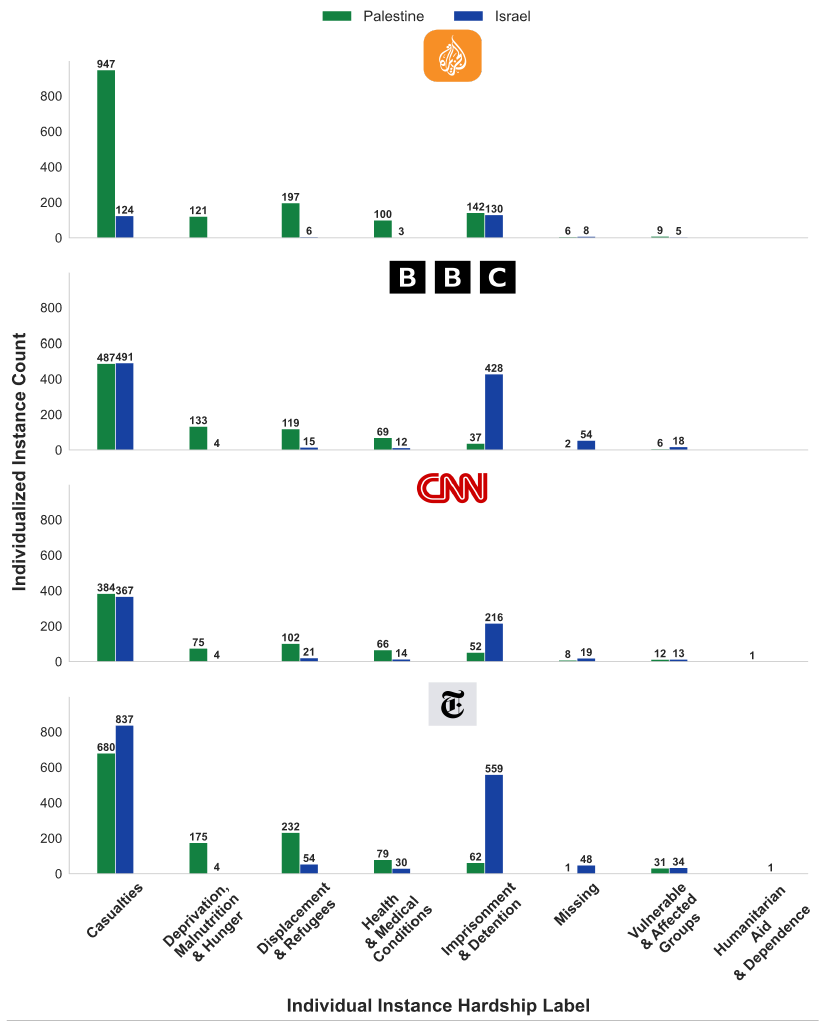}
    \caption{\textbf{Individualized instance counts by hardship category.} Bar plots show, for each media source, the total number of individualized story mentions, grouped by the most severe hardship detected within each instance.
    }
    \label{suppfig:Indiv_Hardship}
\end{figure}
\pagebreak

\begin{figure}[h!]
    \centering
    \includegraphics[width=0.9\linewidth]{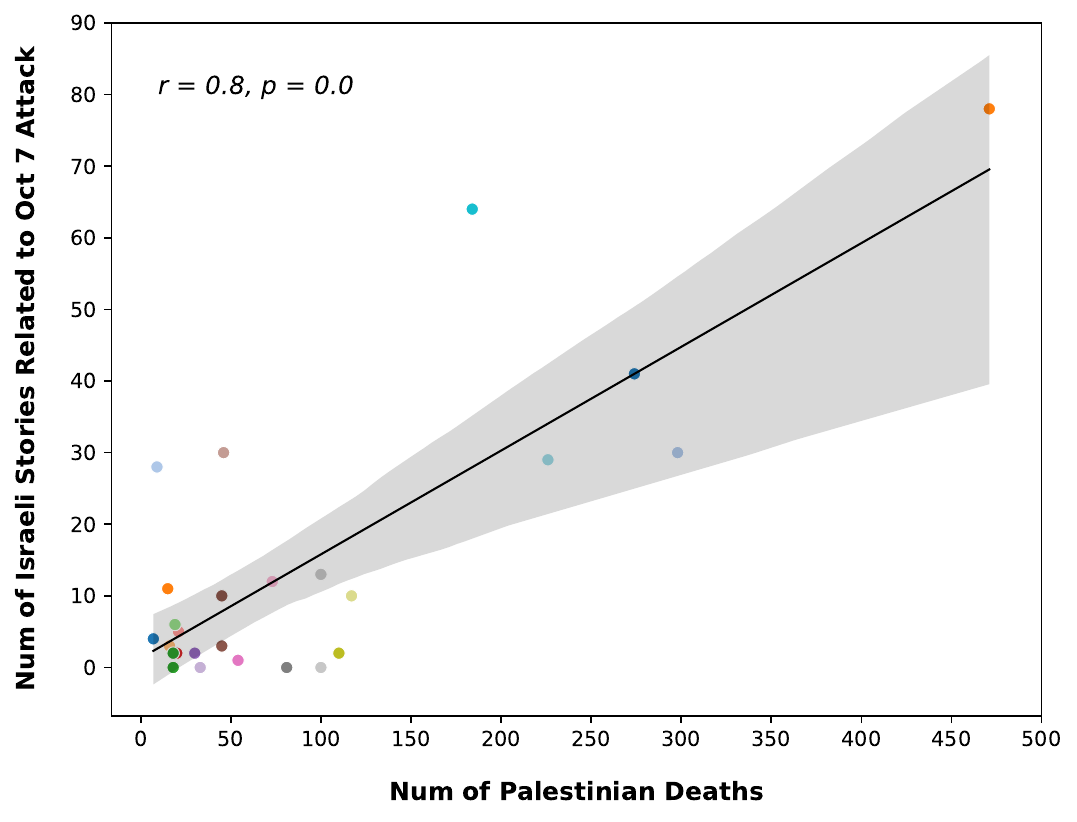}
    \caption{\textbf{Correlation between Palestinian civilian deaths and Israeli story coverage.} Shown is the relationship between daily Palestinian civilian deaths and the number of October 7–related Israeli stories published on the same or following day. Each point corresponds to a day with high Palestinian casualties.}
    \label{supp_fig:correlation_IsrStories_PalDeaths}
\end{figure}
\pagebreak

\begin{figure}[htp]
    \centering    \includegraphics[width=1\textwidth,height=0.9\textheight, keepaspectratio]{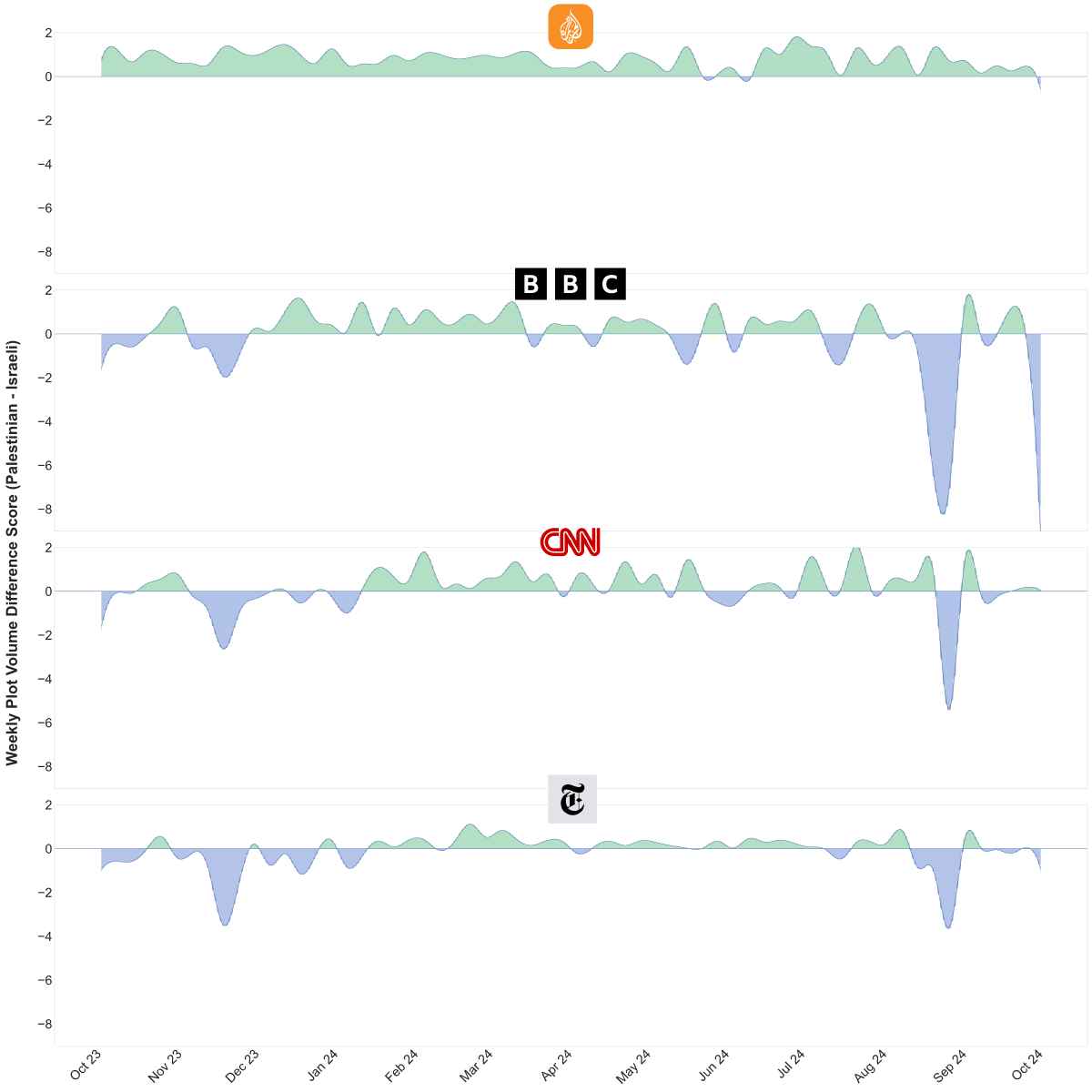}
    \caption{\textbf{Plot Volume difference score across all media sources.} Line plots show the difference in weekly average Plot Volume scores between the Palestinian and Israeli stories for all media sources over the first 12 months of the conflict.
    }
    \label{suppfig:IG_Empathy_Plot_Volume}
\end{figure}

\pagebreak
%\section*{Supplementary Figures}
\begin{figure}[htp]
    \centering    \includegraphics[width=1\textwidth,height=0.9\textheight, keepaspectratio]{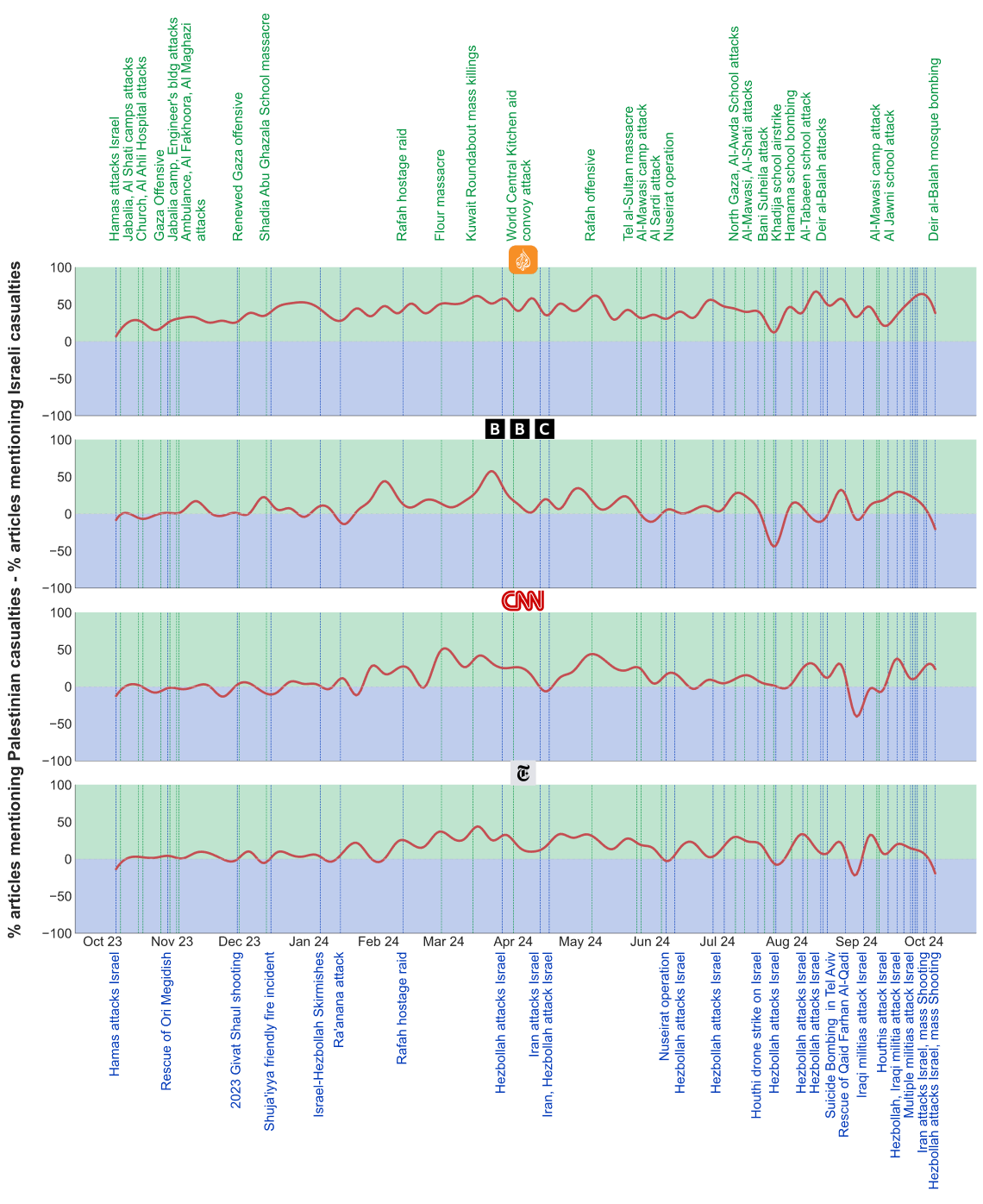}
    \caption{\textbf{Disparity in casualty reporting across media sources.} Line plots show the percentage difference in articles referencing Palestinian and Israeli casualties over the first 12 months, with major events impacting each side marked in green (Palestinian) and blue (Israeli).
    }
    \label{suppfig:CS_Percent_Difference}
\end{figure}

\pagebreak

\begin{sidewaysfigure}[htp]
    \centering    \includegraphics[width=1\textwidth,height=1\textheight, keepaspectratio]{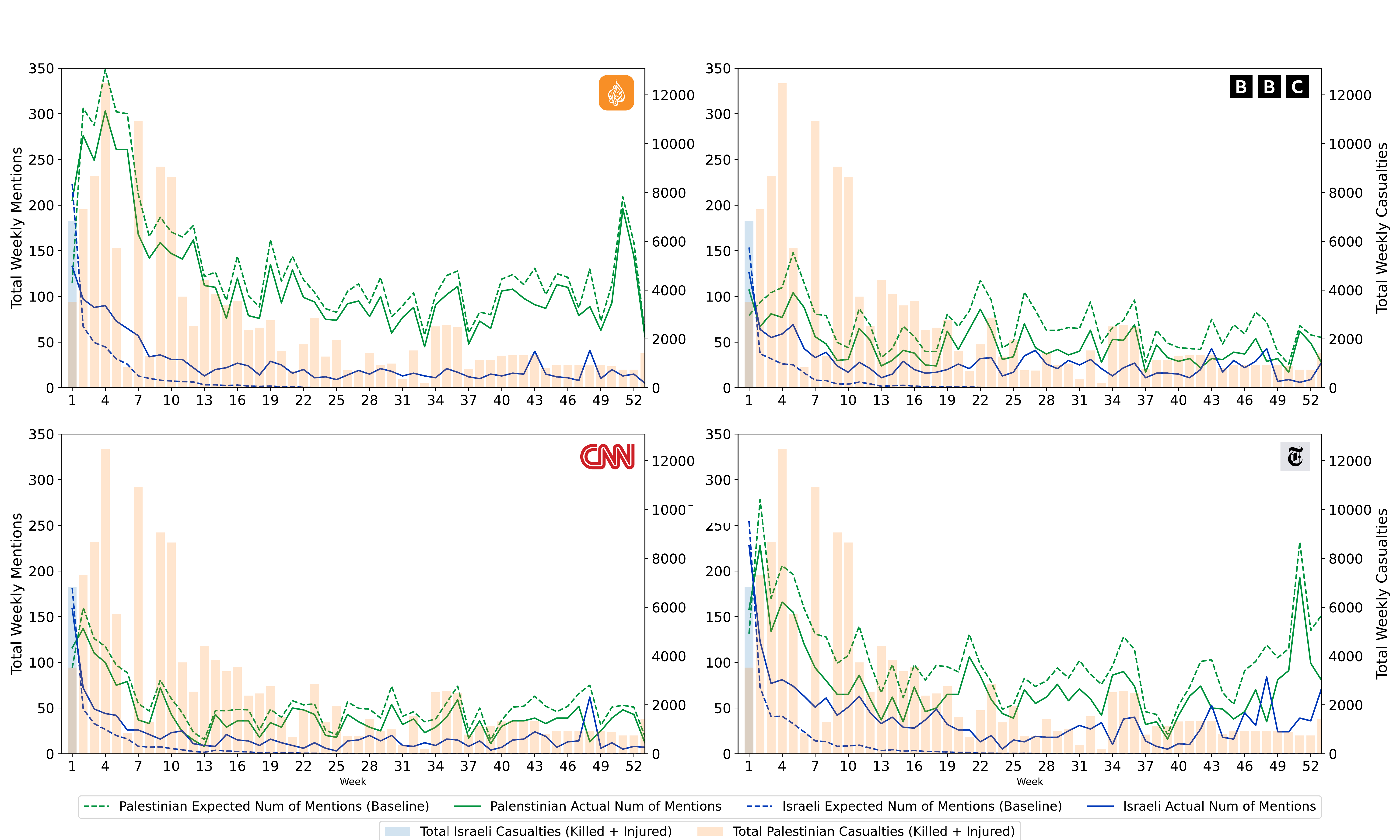}
    \caption{\textbf{Baseline comparison of the expected versus actual number of casualty-related CVN mentions for each news outlet over time}. The graph illustrates the percentage difference between the expected (baseline) and actual mentions for Palestinians and Israelis in each outlet.}
    \label{suppfig:Baseline}
\end{sidewaysfigure}

\pagebreak

\begin{figure}[htp]
    \centering    \includegraphics[width=.95\textwidth,height=\textheight,keepaspectratio]{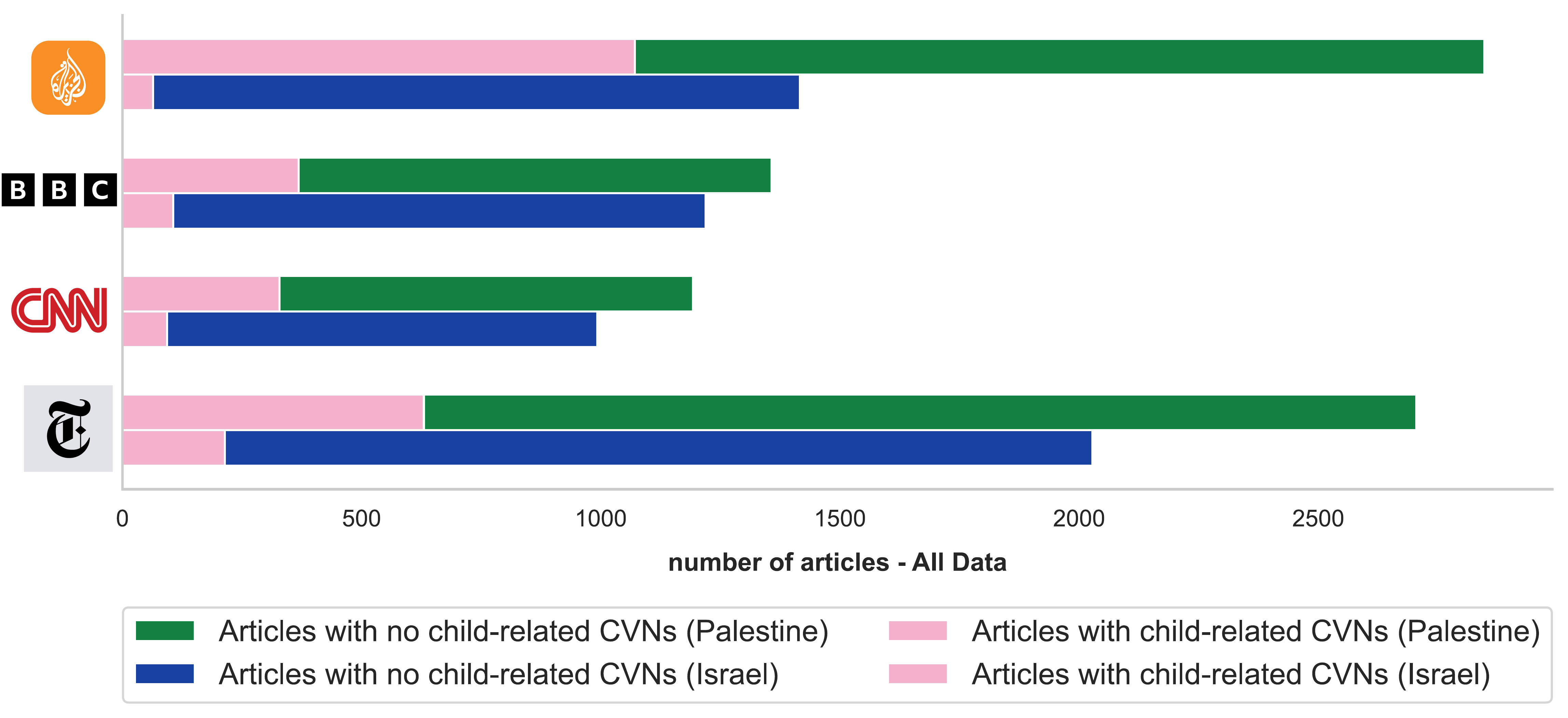}
    \caption{
    \textbf{Proportion of Articles Reporting Child-Related CVNs by Side and Media Source.} Share of articles containing at least one child-related CVN compared to articles with no mention of child-related CVNs, categorized by side (Palestine and Israel) across four media sources. The bars represent the proportion of articles within each category for each media source, highlighting differences in reporting on child-related CVNs for both sides.}
    \label{supp_fig:children_CVN}
\end{figure}

\pagebreak

\begin{figure}[htp]
    \centering    \includegraphics[width=.95\textwidth,height=\textheight,keepaspectratio]{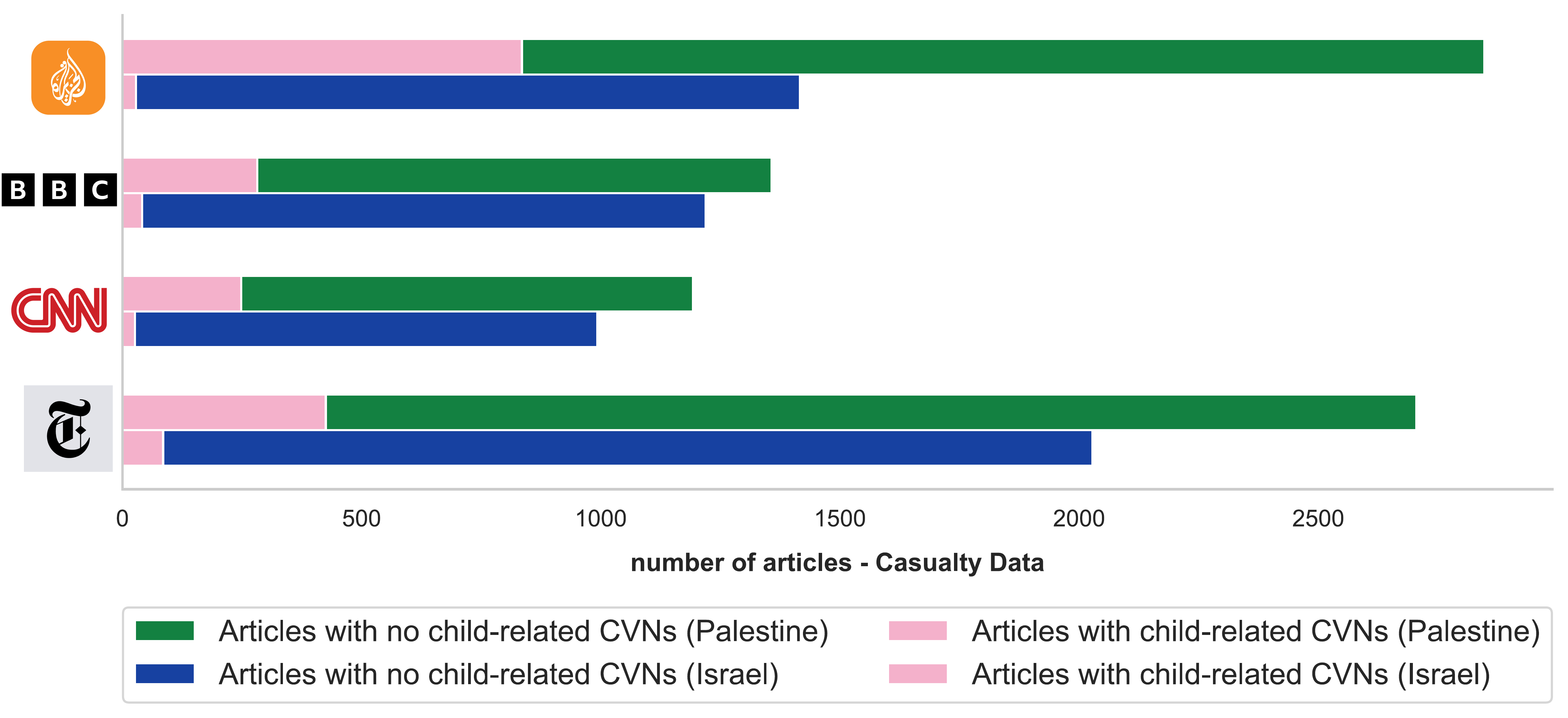}
    \caption{
    \textbf{Proportion of Articles Reporting Child-Related CVNs by Side and Media Source for Casualties-related CVNs only.} Share of articles containing at least one child-related CVN compared to articles with no mention of child-related CVNs, categorized by side (Palestine and Israel) across four media sources. The bars represent the proportion of articles within each category for each media source, highlighting differences in reporting on child-related CVNs for both sides.}
    \label{supp_fig:children_CVN_casualty}
\end{figure}

\pagebreak

%\section*{Supplementary Tables}
\pagebreak
\FloatBarrier 
\begin{table}[H]
{\fontsize{9}{9.2}\selectfont{
\caption{\textbf{Major Events Impacting Palestinian Civilians.} This table provides a comprehensive list of significant events during the first year of the conflict where Palestinian civilians were harmed by military operations, detailing the date and a brief description of each event. The table represents a compilation of the \textit{Non-Battles Attacks} \cite{enwiki_1310903603} and the \textit{Massacres in Palestine} \cite{enwiki_1311468219} lists, into a single comprehensive representation.} 
\label{supp_tab:Major_Events_Palestine} 
\begin{center}
\begin{tabular}{ll}
\toprule
\multicolumn{1}{c}{\textbf{Date}} & \multicolumn{1}{c}{\textbf{Event}}\\
\midrule
7-Oct-23
&Hamas attacks Israel
\\
9-Oct-23
&Jabalia camp market airstrike
\\
 9-Oct-23
&Al-Shati refugee camp airstrike
\\
 17-Oct-23
&Al-Ahli Arab Hospital explosion
\\
 19-Oct-23
&Church of Saint Porphyrius airstrike
\\
27-Oct-23
&Gaza Offensive
\\
31-Oct-23
& Jabalia refugee camp airstrike
\\
31-Oct-23
&Engineer's Building airstrike
\\
3-Nov-23
&Al-Shifa ambulance airstrike
\\
4-Nov-23
&Al-Fakhoora school airstrike
\\
4-Nov-23
&Al-Maghazi camp airstrike
\\
1-Dec-23
&Renewed Gaza offensive
\\
13-Dec-23
&Shadia Abu Ghazala School massacre
\\
 12-Feb-24
&Rafah hostage raid
\\
 29-Feb-24
&Flour massacre
\\
 14-Mar-24
&Kuwait Roundabout mass killings
\\
 1-Apr-24
&World Central Kitchen aid convoy attack
\\
 6-May-24
&Rafah offensive
\\
 26-May-24
&Tel al-Sultan massacre
\\
 28-May-24
&Al-Mawasi refugee camp attack
\\
 6-Jun-24
&Al-Sardi school attack
\\
 8-Jun-24
&Nuseirat operation (Hostage release massacre)
\\
 9-Jul-24
&Attacks over North Gaza
\\
 9-Jul-24
&Al-Awda School massacre
\\
 13-Jul-24
&Al-Mawasi airstrikes
\\
 13-Jul-24
&Al-Shati attack
\\
 22-Jul-24
&Bani Suheila attack
\\
 26-Jul-24
&Khadija school airstrike
\\
 3-Aug-24
&Hamama school bombing
\\
 10-Aug-24
&Al-Tabaeen school attack
\\
17-Aug-24
&August 2024 Deir al-Balah attacks
\\
10-Sep-24
&Al-Mawasi refugee camp attack
\\
 11-Sep-24
&Al-Jawni school attack
\\
 6-Oct-24&Deir al-Balah mosque bombing\\
\bottomrule
\end{tabular}
\end{center}
}}

\vspace{10pt} % Adds a small vertical space
    
    \begin{tablenotes}
\footnotesize
\item \emph{\textbf{Note:}} In addition to the 2 lists used to compile the table, we added 3 critical events that reflect days of major military offensives conducted in Gaza that resulted in dozens of civilian casualties. First, the Gaza Offensive event on 27 October 2023 - the day the military started its ground operations on the enclave. Second, the Renewal of the Gaza Offensive event on 1 December 2023 - the day the military resumed its offensive after the breakdown of the first truce. Third, the Rafah Offensive event on 6 May 2024 - the day the military started its operations against Rafah city in the south of Gaza, where much of the population have sought refugee in.  
\end{tablenotes}

\end{table}

\begin{table}[H]
{\fontsize{9}{9.2}\selectfont{
\caption{\textbf{Major Events Impacting Israeli Civilians.} This table provides a comprehensive list of significant events when Israeli civilians were affected, detailing the date, and a brief description of each event.}
\label{supp_tab:Major_Events_Israel}
\begin{center}
\begin{tabular}{ll}
\toprule
\multicolumn{1}{c}{\textbf{Date}} & \multicolumn{1}{c}{\textbf{Event}}\\
\midrule
7-Oct-23
&Hamas attacks Israel
\\
 30-Oct-23
&Rescue of Ori Megidish
\\
 30-Nov-23
&2023 Givat Shaul shooting
\\
 15-Dec-23
&Shuja'iyya friendly fire incident
\\
 6-Jan-24
&Israel-Hezbollah Skirmishes
\\
 15-Jan-24
&2024 Ra'anana attack (vehicle ramming and stabbing event)\\
 12-Feb-24
&Rafah hostage raid
\\
27-Mar-24
&Hezbollah attacks Israel
\\
13-Apr-24
&Iran attacks Israel 
\\
17-Apr-24
&Hezbollah attacks Israel
\\
8-Jun-24
&Nuseirat operation (Hostage release massacre)
\\
12-Jun-24
&Hezbollah attacks Israel
\\
 29-Jun-24
&Hezbollah attacks Israel
\\
 4-Jul-24
&Hezbollah attacks Israel
\\
 19-Jul-24
&Houthi drone strike on Israel
\\
 27-Jul-24
&Hezbollah attacks Israel
\\
 8-Aug-24
&Hezbollah, Israel Exchange fire
\\
 16-Aug-24
&Hezbollah attacks Israel
\\
 17-Aug-24
&Hezbollah attacks Israel
\\
 19-Aug-24
&Suicide Bombing  in Tel Aviv
\\
27-Aug-24
&Rescue of Qaid Farhan Al-Qadi
\\
4-Sep-24
&Iraqi militias attack Israel
\\
 15-Sep-24
&Houthis attack Israel
\\
 19-Sep-24
&Hezbollah attacks Israel
\\
 22-Sep-24
&Hezbollah, Iraqi militia attack Israel
\\
 25-Sep-24
&Hezbollah, Iraqi militia attack Israel
\\
26-Sep-24
&Hezbollah attacks Israel
\\
27-Sep-24
&Multiple militias attack Israel
\\
28-Sep-24
&Hezbollah, Houthis attack Israel
\\
1-Oct-24
&Iran attacks Israel
\\
1-Oct-24
&Israel - Mass shooting
\\
2-Oct-24
&Hezbollah attacks Israel
\\
6-Oct-24&Israel - Mass shooting\\
\bottomrule
\end{tabular}
\end{center}
}}

\vspace{10pt} % Adds a small vertical space

    \begin{tablenotes}
\footnotesize
\item \emph{\textbf{Note:}} We could not find a single comprehensive list of events affecting Israeli citizens. The few sources we found had short lists which overlooked many significant incidents. As such, we decided to compile our own list by reviewing the conflict events day by day, consulting official news reports, and selecting - as much as possible - those events that significantly impacted Israeli civilians. 
\end{tablenotes}

\end{table}

\begin{table}[H]
{\fontsize{12}{14}\selectfont{
\caption{\textbf{Count of Individualized and Grouped Instances.} This table provides the instance counts obtained using the LLM and the prompts in Supplementary Note~\ref{supp_note:IG_Prompts} and the Individual to Group count ratios per side for each media source.}
\label{supp_tab:IG_counts_ratios}
\begin{center}
\begin{tabular}{clccc}
\toprule
&  &  &  & \textbf{Individual to} \\
\textbf{Source} & \textbf{Side} & \textbf{Individual} & \textbf{Group} & 
\textbf{Group Ratio} \\
\midrule
AJE&Palestine& 1540& 17773&0.09
\\
&Israel& 299& 4257&0.07
\\
 & & & &\\
BBC&Palestine& 863& 7414&0.12
\\
& Israel& 1042& 5118&0.20
\\
 & & & &\\
CNN&Palestine& 718& 8075&0.09
\\
&Israel& 669& 5178&0.13
\\
 & & & &\\
NYT&Palestine& 1291& 16413&0.08
\\
&Israel& 1628& 10488&0.16\\

\bottomrule
\end{tabular}
\end{center}
}}
\end{table}

\pagebreak

\begin{table}[H]
{\fontsize{12}{22}\selectfont{
\caption{\textbf{Count of Western Media's Individualized Hardship-related Stories per Side per Source for four major events.} This table provides the count of stories published by the Western media outlets (BBC, CNN and NYT) for major events witnessing a large number of Palestinian casualties. The counts reflect the number of stories published on the day of the event as well as the day after. Israeli counts represent stories related to the October 7 attacks.}
\label{supp_tab:Conc_reporting}
\begin{center}
\begin{tabular}{llcccccc}
\toprule
\ & & \multicolumn{2}{c}{\textbf{BBC}} & \multicolumn{2}{c}{\textbf{CNN}} & \multicolumn{2}{c}{\textbf{NYT}}\\

\textbf{Event} & \textbf{Date} & \textbf{Pal} & \textbf{Isr} & \textbf{Pal} & \textbf{Isr} & \textbf{Pal} & \textbf{Isr} \\
\midrule
\multicolumn{8}{l}{\textbf{Dates with Palestinian only major event:}}\\
Renewed Gaza Offensive& 01 Dec 2023& 6& 19& 5& 21& 6&24\\
Flour massacre& 29 Feb 2024& 3& 1& 2& 7& 11&2\\
Al-Mawasi refugee camp attack&  10 Sep 2024&  3&  0&  8&  0&  7&  5\\
 & & & & & & &\\
\multicolumn{8}{l}{\textbf{Dates with both Palestinian and Israeli major events:}}\\
Al-Thalatheni Street Strike&  15 Jan 2024&  2&  13&  4&  10&  0&  12\\
Nuseirat operation&  08 Jun 2024&  4&  8&  3&  13&  12&  19\\
Al Safad School Attack& 01 Sep 2024& 4& 86& 0& 38& 1&57\\

\bottomrule
\end{tabular}
\end{center}
}}
\end{table}

\begin{table}[H]
\centering
\caption{\textbf{Child-related Individualized Stories.} 
(a) Counts and percentages of individualized child references compared to total number of individualized stories across the four media outlets, compared with the actual proportion of children killed on each side. 
(b) Distribution of individualized child stories by side within each media outlet, compared with the distribution of child deaths.}
\label{supp_tab:Child_reference_counts}

% ---------- Table (a) ----------
\noindent\textbf{(a)} \\
\vspace{0.3em}
\resizebox{\textwidth}{!}{%
\begin{tabular}{llcccccc}
\toprule
    &  
    & \multicolumn{3}{c}{\textbf{Individualized Stories}} & 
    \multicolumn{3}{c}{\textbf{Deaths in 1$^{st}$ Year}}  \\
    \textbf{Source} &  \textbf{Side}
    & \textbf{Child-related} 
    & \textbf{Total} 
    & \textbf{\% Child-related} 
    & \textbf{Children}
    & \textbf{Total} 
    & \textbf{\% Children}  \\
\midrule
AJE   & Palestine & 784 & 1540 & 51\% & 14000 & 42000 &34\% \\
      & Israel    &  80 &  299 & 27\% & 38 & 1400 & 3\% \\
      & \textbf{Both} & \textbf{864} & \textbf{1839} & \textbf{47\%} & \textbf{14038} & \textbf{43400} & \textbf{32\%}\\
&&&&&\\
BBC   & Palestine & 474 &  863 & 55\% &14000 & 42000&34\% \\
      & Israel    & 380 & 1042 & 36\% & 38 & 1400 &3\% \\
    & \textbf{Both} & \textbf{854} & \textbf{1905} & \textbf{45\%} & \textbf{14038} & \textbf{43400} & \textbf{32\%}\\
&&&&&\\
CNN   & Palestine & 403 &  718 & 56\% & 14000 & 42000 &34\% \\
      & Israel    & 297 &  669 & 44\% & 38 & 1400 & 3\% \\
    & \textbf{Both} & \textbf{700} & \textbf{1387} & \textbf{50\%} & \textbf{14038} & \textbf{43400} & \textbf{32\%}\\
&&&&&\\
NYT   & Palestine & 630 & 1291 & 49\% & 14000 & 42000 &34\% \\
      & Israel    & 654 & 1628 & 40\% & 38 & 1400 & 3\% \\
    & \textbf{Both} & \textbf{1284} & \textbf{2919} & \textbf{44\%} & \textbf{14038} & \textbf{43400} & \textbf{32\%}\\
\bottomrule
\end{tabular}
}

\vspace{1em}

% ---------- Table (b) ----------
\noindent\textbf{(b)} \\
\vspace{0.3em}
\resizebox{0.77\textwidth}{!}{%
\begin{tabular}{llccccc}
\toprule
    &  
    & \multicolumn{2}{c}{\textbf{Child-related}} &
    & \multicolumn{2}{c}{\textbf{Children}} \\
    &  
    & \multicolumn{2}{c}{\textbf{Individualized Stories}} &
    & \multicolumn{2}{c}{\textbf{Deaths in 1$^{st}$ Year}} \\
    
    \textbf{Source} & \textbf{Side} 
    & Count 
    & Percent &
    & Count 
    & Percent \\
\midrule
AJE   & Palestine & 784 & 91\% &   & 14000 & $>$99.7\% \\
      & Israel    &  80 &  9\% & &38 &  $<$0.3\% \\
&&&&&\\
BBC   & Palestine & 474 & 56\% & &14000 & $>$99.7\% \\
      & Israel    & 380 & 44\% & &38 &  $<$0.3\%  \\
&&&&&\\
CNN   & Palestine & 403 & 57\% & &14000 & $>$99.7\%  \\
      & Israel    & 297 & 43\% & &38 &  $<$0.3\%  \\
&&&&&\\
NYT   & Palestine & 630 & 49\% & &14000 & $>$99.7\% \\
      & Israel    & 654 & 51\% & &38 &  $<$0.3\%  \\
\bottomrule
\end{tabular}
}
\end{table}

\begingroup
\setlength{\tabcolsep}{5pt}         % tighter columns (default ~6pt)
\renewcommand{\arraystretch}{1.05}  % slightly tighter rows
\footnotesize                           % or \small \footnotesize / \scriptsize if needed

\begin{longtable}{@{}cclcc@{}}
\caption{\textbf{Distribution of Civilian Victim Numbers (CVNs) by Type for Palestine and Israel across the four Media Outlets.} This table presents the distribution of CVNs reported by AJE, BBC, CNN, and NYT, separated by victim types for Palestine and Israel. Although presented sequentially by source, the table allows for direct comparison of coverage patterns across outlets and sides.}
\label{supp_tab:Statistics_Types_Label_Distribution}\\
\toprule
\textbf{Source}&\textbf{Side}&\textbf{Label}&\textbf{Number of Mentions}&\textbf{Percent}\\
\midrule
\endfirsthead

\multicolumn{5}{r}{\emph{Continued on next page}}\\
\endfoot

\bottomrule
\endlastfoot

% -------- AJE --------
AJE & Palestine & \textbf{\color{red}Casualties} & \textbf{\color{red}6346} & \textbf{\color{red}61.1} \\
 &  & Displacement and Refugees & 1370 & 13.2 \\
 &  & Vulnerable and Affected Groups & 1182 & 11.4 \\
 &  & Other & 631 & 6.1 \\
 &  & Imprisonment and Detention & 439 & 4.2 \\
 &  & Health and Medical Conditions & 203 & 2.0 \\
 &  & Deprivation, Malnutrition and Hunger & 119 & 1.1 \\
 &  & Humanitarian Aid and Dependence & 51 & 0.5 \\
 &  & Missing and Non-Casualty Victims & 41 & 0.4 \\
 &  &  &  &  \\
 & Israel & \textbf{\color{red}Casualties} & \textbf{\color{red}1452} & \textbf{\color{red}53.1} \\
 &  & Imprisonment and Detention & 945 & 34.6 \\
 &  & Other & 205 & 7.5 \\
 &  & Vulnerable and Affected Groups & 57 & 2.1 \\
 &  & Displacement and Refugees & 41 & 1.5 \\
 &  & Missing and Non-Casualty Victims & 29 & 1.1 \\
 &  & Health and Medical Conditions & 5 & 0.2 \\
\midrule

% -------- BBC --------
BBC & Palestine & \textbf{\color{red}Casualties} & \textbf{\color{red}2497} & \textbf{\color{red}58.8} \\
 &  & Displacement and Refugees & 636 & 15.0 \\
 &  & Vulnerable and Affected Groups & 460 & 10.8 \\
 &  & Other & 274 & 6.5 \\
 &  & Imprisonment and Detention & 200 & 4.7 \\
 &  & Health and Medical Conditions & 108 & 2.5 \\
 &  & Deprivation, Malnutrition and Hunger & 33 & 0.8 \\
 &  & Humanitarian Aid and Dependence & 26 & 0.6 \\
 &  & Missing and Non-Casualty Victims & 10 & 0.2 \\
 &  &  &  &  \\
 & Israel & \textbf{\color{red}Casualties} & \textbf{\color{red}1472} & \textbf{\color{red}46.3} \\
 &  & Imprisonment and Detention & 1299 & 40.9 \\
 &  & Other & 200 & 6.3 \\
 &  & Vulnerable and Affected Groups & 93 & 2.9 \\
 &  & Displacement and Refugees & 58 & 1.8 \\
 &  & Missing and Non-Casualty Victims & 50 & 1.6 \\
 &  & Health and Medical Conditions & 5 & 0.2 \\
\midrule\\

% -------- CNN --------
\addlinespace
\midrule
CNN & Palestine & \textbf{\color{red}Casualties} & \textbf{\color{red}2219} & \textbf{\color{red}54.7} \\
 &  & Vulnerable and Affected Groups & 593 & 14.6 \\
 &  & Displacement and Refugees & 580 & 14.3 \\
 &  & Other & 325 & 8.0 \\
 &  & Imprisonment and Detention & 175 & 4.3 \\
 &  & Health and Medical Conditions & 84 & 2.1 \\
 &  & Deprivation, Malnutrition and Hunger & 40 & 1.0 \\
 &  & Humanitarian Aid and Dependence & 34 & 0.8 \\
 &  & Missing and Non-Casualty Victims & 6 & 0.1 \\
 &  &  &  &  \\
 & Israel & \textbf{\color{red}Casualties} & \textbf{\color{red}1053} & \textbf{\color{red}42.0} \\
 &  & Imprisonment and Detention & 1008 & 40.2 \\
 &  & Other & 220 & 8.8 \\
 &  & Vulnerable and Affected Groups & 91 & 3.6 \\
 &  & Missing and Non-Casualty Victims & 89 & 3.5 \\
 &  & Displacement and Refugees & 43 & 1.7 \\
 &  & Health and Medical Conditions & 4 & 0.2 \\
 &  & Deprivation, Malnutrition and Hunger & 2 & 0.1 \\
\midrule

% -------- NYT --------
NYT & Palestine & \textbf{\color{red}Casualties} & \textbf{\color{red}4048} & \textbf{\color{red}51.3} \\
 &  & Displacement and Refugees & 1290 & 16.4 \\
 &  & Vulnerable and Affected Groups & 1068 & 13.5 \\
 &  & Other & 772 & 9.8 \\
 &  & Imprisonment and Detention & 386 & 4.9 \\
 &  & Health and Medical Conditions & 166 & 2.1 \\
 &  & Humanitarian Aid and Dependence & 81 & 1.0 \\
 &  & Deprivation, Malnutrition and Hunger & 67 & 0.8 \\
 &  & Missing and Non-Casualty Victims & 8 & 0.1 \\
 &  &  &  &  \\
 & Israel & \textbf{\color{red}Casualties} & \textbf{\color{red}2113} & \textbf{\color{red}44.5} \\
 &  & Imprisonment and Detention & 1572 & 33.1 \\
 &  & Other & 485 & 10.2 \\
 &  & Vulnerable and Affected Groups & 239 & 5.0 \\
 &  & Displacement and Refugees & 195 & 4.1 \\
 &  & Missing and Non-Casualty Victims & 132 & 2.8 \\
 &  & Health and Medical Conditions & 9 & 0.2 \\

\end{longtable}
\endgroup

\pagebreak

\begin{table}[H]
\fontsize{12}{14}\selectfont
\centering
\caption{\textbf{Most commonly used source citation phrases when reporting CVNs about Israel}}
\label{tab:cvn_coverage_1}
\begin{tabular}{ll}
    \toprule
    \multicolumn{1}{c}{\textbf{Style}} & \multicolumn{1}{c}{\textbf{Phrases}} \\
    \midrule
  Israeli officials say&  ``Israeli officials say”\\
  &  ``Israeli official told''\\
 &``Israeli authorities say”\\
 &``Israeli government says''\\
 &``Israeli government announced''\\
 &``officials said''\\
 &``Authorities said''\\
 &``Emergency responders said''\\
 & ``according to Israeli officials''\\
  &  ``Israel says''\\
  & ``Israel believes''\\
 &\\
 the Israeli military says& ``the IDF says''\\
  & ``Israel Defense Forces said''\\
 & ``an IDF spokesperson said''\\
 & ``Israeli military says''\\
 & ``according to Israeli military''\\
 &\\
  American officials say& ``American officials said''\\
  & ``U.S. officials said''\\
  & ``President Biden said''\\
   &\\
  Hamas says& ``Hamas says''\\
  & ``Hamas claims''\\
  \bottomrule
\end{tabular}
\end{table}

\begin{table}[H]
{\fontsize{12}{14}\selectfont{
    \centering
    \caption{\textbf{Most commonly used source citation phrases when reporting CVNs about Palestine}}
    {%
    \begin{tabular}{ll}
        \toprule
        \multicolumn{1}{c}{\textbf{Style}} & \multicolumn{1}{c}{\textbf{Phrases}}  \\

        \midrule
  Palestinian Health Ministry says&  ``The Palestinian Health Ministry in Gaza said''\\
  &  ``Gaza’s Health Ministry said''\\
 &\\
 Medical / Health officials say&``health officials say''\\
 &``health officials in Gaza said''\\
 &``medical sources said''\\
 &``medics reported''\\
 &\\
  Palestinian authorities say& ``Palestinian authorities say''\\
  & ``Palestinian officials say''\\
 &``authorities in Gaza said''\\
 &``authorities reporting''\\
 &``local officials say''\\
 &``local authorities say''\\
 &\\
  Aid agencies / NGO says& ``aid workers say''\\
  & ``The United Nations said''\\
  & ``The U.N. humanitarian agency said''\\
  & ``UNRWA said''\\
  & ``According to UNRWA''\\
  & ``the U.N. said on Monday''\\
  &``Human Rights Watch said''\\
 &Mentions of specific NGOs or charities\\
 &\\
  Hamas says& ``Hamas says''\\
  & ``a Hamas spokesman said''\\
 &\\
  Hezbollah says& ``Hezbollah said''\\
  & ``Hezbollah confirmed''\\
 &\\
  Israeli officials say& ``Israeli officials said''\\
  & ``the Israeli military said''\\
 
 &\\
  Other & ``Rights groups say''\\
  & ``media networks say''\\
 & ``press freedom monitors have said''\\
 \bottomrule
    \end{tabular}
    \label{tab:cvn_coverage_2}
    }
}}

\end{table}

\begin{table}[H]
    \centering
    \caption{\textbf{Percentage of CVNs Reported with References Across Media Sources and Sides}}
    \begin{tabular}{llcccc}
        \toprule
        \makecell{\textbf{Media} \\ \textbf{source}} & \textbf{Side} & \makecell{\textbf{\# of CVNs} \\ \textbf{with a Reference}} & \makecell{\textbf{Total \#} \\ \textbf{of CVNs}} & \textbf{Percentage} & \textbf{Ratio} \\
        \midrule
        AJE  & Palestine & 2,549 & 8,436 & 30.0\% &  \multirow{2}{*}{1.1}\\
        & Israel & 583 & 2,152 & 27.0\% & \\
        \\
        BBC & Palestine & 1,829 & 3,616 & 50.6\% & \multirow{2}{*}{3.0}\\
        & Israel & 382 & 2262 & 16.9\% &\\
        \\
        CNN & Palestine & 1,681 & 3,356 & 50.1\% & \multirow{2}{*}{1.9}\\
        & Israel & 509 & 1,921 & 26.5\% &\\
        \\
        NYT & Palestine & 2,873 & 6,752 & 42.6\% & \multirow{2}{*}{1.6}\\
        & Israel & 1,035 & 3,920 & 26.4\% & \\
        \bottomrule
    \end{tabular}
    \label{tab:cvn_percent_coverage}
\end{table}

\begin{table}[H]
\centering
\caption{\textbf{\textcolor{black}{Final List of Doubt-Casting Phrases and Their Categories.}} This table presents the finalized list of doubt-casting phrases identified in the analysis, grouped by their respective categories. While the phrases listed here primarily concern the Palestinian side, our initial framework assumed that doubt-casting could occur toward both Palestinian and Israeli accounts. The asymmetry in the final list reflects what emerged empirically from the data rather than an a priori focus on one side.}
\label{supp_tab:Casting_Doubt_Phrases_Categories}

\resizebox{0.685\textwidth}{!}{%
\begin{tabular}{lc}

 \textbf{Doubt-Casting Phrase}& \textbf{Category}\\
            \midrule
            ..., an organization controlled by Hamas
& Source Doubting
\\
 ..., which is controlled by Hamas
&Source Doubting
\\
 Hamas Health Ministry
&Source Doubting
\\
 Hamas and medical officials 
&Source Doubting
\\
 Hamas authorities
&Source Doubting
\\
 Hamas government
&Source Doubting
\\
 Hamas health officials
&Source Doubting
\\
 Hamas officials
&Source Doubting
\\
 Hamas said
&Source Doubting
\\
 Hamas security officials
&Source Doubting
\\
 Hamas sources in Gaza 
&Source Doubting
\\
 Hamas-controlled Gaza
&Source Doubting
\\
 Hamas-controlled Gaza Strip
&Source Doubting
\\
 Hamas-controlled government
&Source Doubting
\\
 Hamas-controlled health authorities
&Source Doubting
\\
 Hamas-led authorities
&Source Doubting
\\
 Hamas-ruled Gaza
&Source Doubting
\\
 Hamas-ruled territory
&Source Doubting
\\
 Hamas-run Gaza Health Ministry
&Source Doubting
\\
 Hamas-run Gaza Ministry of Health
&Source Doubting
\\
 Hamas-run Ministry of Health
&Source Doubting
\\
 Hamas-run Palestinian territory
&Source Doubting
\\
            Hamas-run al-Aqsa TV reported
& Source Doubting
\\
 Hamas-run health ministry
& Source Doubting
\\
 Hamas-run territory
& Source Doubting
\\
 according to Hamas' military wing
& Source Doubting
\\
 according to a Hamas website
& Source Doubting
\\
 according to health officials in Hamas-controlled Gaza
& Source Doubting
\\
            according to officials in the Hamas-run territory
& Source Doubting
\\
            controlled by Hamas
& Source Doubting
\\
            drawing from sources in Hamas-controlled Gaza
& Source Doubting
\\
            sources in Hamas-run Gaza
& Source Doubting
\\
            sources in the Hamas-controlled enclave
& Source Doubting
\\
            who were not on the scene
& Source Doubting
\\
            …, which is part of the political arm of Hamas
& Source Doubting
\\
            allegedly
& Uncertainty in Numbers
\\
            are thought to have been killed
& Uncertainty in Numbers
\\
            claim
& Uncertainty in Numbers
\\
            difficult to know how many
& Uncertainty in Numbers
\\
            no reliable and current figures
& Uncertainty in Numbers
\\
            reportedly
& Uncertainty in Numbers
\\
            said to be
& Uncertainty in Numbers
\\
            said to have been killed
& Uncertainty in Numbers
\\
 said to have been wounded
&Uncertainty in Numbers
\\
            was reported to have& Uncertainty in Numbers\\
            \bottomrule
        \end{tabular}
    }

\end{table}

\begin{table}[H]
\centering
\caption{\textbf{\textcolor{black}{Breakdown of Doubt-Casting Phrases in Casualty-Reporting Sentences in AJE}} This table presents the total frequency of each doubt-casting phrase identified in AJE's casualty-reporting sentences. It also shows how these frequencies are distributed across Palestinian and Israeli CVN mentions, providing insights into the framing dynamics for each side within AJE’s reporting.}
\label{supp_tab:Casting_Doubt_AJE_Phrase_Count}
    
    \resizebox{0.77\textwidth}{!}{%
        \begin{tabular}{lccc}

            & \multicolumn{3}{c}{\textbf{Al Jazeera English}} \\
            & \multicolumn{3}{c}{(Total of \textbf{10,601} CVNs)} \\
            \cmidrule(r){2-4}
            Doubt-casting phrases & Total mentions & Mentions& Mentions\\
 & that include& referencing&referencing\\
 & the phrase& Palestinian&Israeli\\
 & & victims&victims\\
            \midrule
            ..., an organization controlled by Hamas
& 0& 0& 0
\\
            ..., which is controlled by Hamas
& 0& 0& 0
\\
 Hamas Health Ministry
& 1& 1&0
\\
 Hamas and medical officials 
& 0& 0&0
\\
            Hamas authorities
& 2& 2& 0
\\
            Hamas government
& 1& 1& 0
\\
            Hamas health officials
& 0& 0& 0
\\
            Hamas officials
& 4& 4& 0
\\
            Hamas said
& 3& 3& 0
\\
            Hamas security officials
& 0& 0& 0
\\
            Hamas sources in Gaza 
& 0& 0& 0
\\
 Hamas-controlled Gaza
& 0& 0&0
\\
 Hamas-controlled Gaza Strip
& 0& 0&0
\\
 Hamas-controlled government
& 1& 1&0
\\
 Hamas-controlled health authorities
& 0& 0&0
\\
 Hamas-led authorities
& 0& 0&0
\\
 Hamas-ruled Gaza
& 2& 2&0
\\
 Hamas-ruled territory
& 1& 1&0
\\
 Hamas-run Gaza Health Ministry
& 0& 0&0
\\
 Hamas-run Gaza Ministry of Health
& 0& 0&0
\\
 Hamas-run Ministry of Health
& 3& 3&0
\\
 Hamas-run Palestinian territory
& 1& 1&0
\\
 Hamas-run al-Aqsa TV reported
& 0& 0&0
\\
 Hamas-run health ministry
& 4& 4&0
\\
 Hamas-run territory
& 2& 2&0
\\
 according to Hamas' military wing
& 0& 0&0
\\
 according to a Hamas website
& 0& 0&0
\\
 according to health officials in Hamas-controlled Gaza
& 0& 0&0
\\
 according to officials in the Hamas-run territory
& 0& 0&0
\\
 controlled by Hamas
& 1& 1&0
\\
 drawing from sources in Hamas-controlled Gaza
& 0& 0&0
\\
 sources in Hamas-run Gaza
& 0& 0&0
\\
 sources in the Hamas-controlled enclave
& 0& 0&0
\\
 who were not on the scene
& 0& 0&0
\\
 …, which is part of the political arm of Hamas
& 0& 0&0
\\
            allegedly
& 11& 7& 4
\\
            are thought to have been killed
& 0& 0& 0
\\
            claim
& 10& 5& 5
\\
            difficult to know how many
& 0& 0& 0
\\
            no reliable and current figures
& 0& 0& 0
\\
            reportedly
& 44& 32& 12
\\
            said to be
& 2& 2& 0
\\
            said to have been killed
& 0& 0& 0
\\
 said to have been wounded
& 0& 0&0
\\
            was reported to have& 0& 0& 0\\
\midrule
\textbf{Total}&93&72&21\\
\bottomrule
\end{tabular}
}

\end{table}

\begin{table}[H]
\centering
\caption{\textbf{\textcolor{black}{Breakdown of Doubt-Casting Phrases in Casualty-Reporting Sentences in BBC}} This table presents the total frequency of each doubt-casting phrase identified in BBC's casualty-reporting sentences. It also shows how these frequencies are distributed across Palestinian and Israeli CVN mentions, providing insights into the framing dynamics for each side within BBC’s reporting.}
\label{supp_tab:Casting_Doubt_BBC_Phrase_Count}
    
    \resizebox{0.77\textwidth}{!}{%
        \begin{tabular}{lccc}

            & \multicolumn{3}{c}{\textbf{BBC}} \\
            & \multicolumn{3}{c}{(Total of \textbf{5,898} CVNs)} \\
            \cmidrule(r){2-4}
            Doubt-casting phrases & Total mentions & Mentions& Mentions\\
 & that include& referencing&referencing\\
 & the phrase& Palestinian&Israeli\\
 & & victims&victims\\
            \midrule
            ..., an organization controlled by Hamas& 0& 0& 0
\\
            ..., which is controlled by Hamas& 2& 2& 0
\\
 Hamas Health Ministry& 0& 0&0
\\
 Hamas and medical officials & 0& 0&0
\\
            Hamas authorities& 1& 1& 0
\\
            Hamas government& 5& 5& 0
\\
            Hamas health officials& 3& 3& 0
\\
            Hamas officials& 5& 5& 0
\\
            Hamas said& 12& 12& 0
\\
            Hamas security officials& 0& 0& 0
\\
            Hamas sources in Gaza & 0& 0& 0
\\
 Hamas-controlled Gaza& 2& 2&0
\\
 Hamas-controlled Gaza Strip& 0& 0&0
\\
 Hamas-controlled government& 0& 0&0
\\
 Hamas-controlled health authorities& 0& 0&0
\\
 Hamas-led authorities& 0& 0&0
\\
 Hamas-ruled Gaza& 0& 0&0
\\
 Hamas-ruled territory& 1& 1&0
\\
 Hamas-run Gaza Health Ministry& 4& 4&0
\\
 Hamas-run Gaza Ministry of Health& 0& 0&0
\\
 Hamas-run Ministry of Health& 3& 3&0
\\
 Hamas-run Palestinian territory& 2& 2&0
\\
 Hamas-run al-Aqsa TV reported& 0& 0&0
\\
 Hamas-run health ministry& 800& 800&0
\\
 Hamas-run territory& 3& 3&0
\\
 according to Hamas' military wing& 0& 0&0
\\
 according to a Hamas website& 0& 0&0
\\
 according to health officials in Hamas-controlled Gaza& 0& 0&0
\\
 according to officials in the Hamas-run territory& 1& 1&0
\\
 controlled by Hamas& 2& 2&0
\\
 drawing from sources in Hamas-controlled Gaza& 0& 0&0
\\
 sources in Hamas-run Gaza& 0& 0&0
\\
 sources in the Hamas-controlled enclave& 0& 0&0
\\
 who were not on the scene& 0& 0&0
\\
 …, which is part of the political arm of Hamas& 0& 0&0
\\
            allegedly& 1& 1& 0
\\
            are thought to have been killed& 1& 0& 1
\\
            claim& 3& 3& 0
\\
            difficult to know how many& 0& 0& 0
\\
            no reliable and current figures& 0& 0& 0
\\
            reportedly& 88& 72& 16
\\
            said to be& 8& 6& 2
\\
            said to have been killed& 4& 4& 0
\\
 said to have been wounded& 0& 0&0
\\
            was reported to have& 0& 0& 0\\
\midrule
\textbf{Total}&951&932&19\\
\bottomrule
\end{tabular}
}

\end{table}

\begin{table}[H]
\centering
    \caption{\textbf{\textcolor{black}{Breakdown of Doubt-Casting Phrases in Casualty-Reporting Sentences in CNN}} This table presents the total frequency of each doubt-casting phrase identified in CNN's casualty-reporting sentences. It also shows how these frequencies are distributed across Palestinian and Israeli CVN mentions, providing insights into the framing dynamics for each side within CNN’s reporting.}
\label{supp_tab:Casting_Doubt_CNN_Phrase_Count}
    
    \resizebox{0.77\textwidth}{!}{%
        \begin{tabular}{lccc}

            & \multicolumn{3}{c}{\textbf{CNN}} \\
            & \multicolumn{3}{c}{(Total of \textbf{6,569} CVNs)} \\
            \cmidrule(r){2-4}
            Doubt-casting phrases & Total mentions & Mentions& Mentions\\
 & that include& referencing&referencing\\
 & the phrase& Palestinian&Israeli\\
 & & victims&victims\\
            \midrule
            ..., an organization controlled by Hamas& 1& 1& 0
\\
            ..., which is controlled by Hamas& 6& 6& 0
\\
 Hamas Health Ministry& 0& 0&0
\\
 Hamas and medical officials & 0& 0&0
\\
            Hamas authorities& 2& 2& 0
\\
            Hamas government& 1& 1& 0
\\
            Hamas health officials& 0& 0& 0
\\
            Hamas officials& 1& 1& 0
\\
            Hamas said& 5& 5& 0
\\
            Hamas security officials& 0& 0& 0
\\
            Hamas sources in Gaza & 0& 0& 0
\\
 Hamas-controlled Gaza& 20& 20&0
\\
 Hamas-controlled Gaza Strip& 2& 2&0
\\
 Hamas-controlled government& 2& 2&0
\\
 Hamas-controlled health authorities& 5& 5&0
\\
 Hamas-led authorities& 0& 0&0
\\
 Hamas-ruled Gaza& 1& 1&0
\\
 Hamas-ruled territory& 0& 0&0
\\
 Hamas-run Gaza Health Ministry& 3& 3&0
\\
 Hamas-run Gaza Ministry of Health& 4& 4&0
\\
 Hamas-run Ministry of Health& 32& 32&0
\\
 Hamas-run Palestinian territory& 0& 0&0
\\
 Hamas-run al-Aqsa TV reported& 0& 0&0
\\
 Hamas-run health ministry& 18& 18&0
\\
 Hamas-run territory& 0& 0&0
\\
 according to Hamas' military wing& 0& 0&0
\\
 according to a Hamas website& 0& 0&0
\\
 according to health officials in Hamas-controlled Gaza& 0& 0&0
\\
 according to officials in the Hamas-run territory& 0& 0&0
\\
 controlled by Hamas& 8& 8&0
\\
 drawing from sources in Hamas-controlled Gaza& 2& 2&0
\\
 sources in Hamas-run Gaza& 5& 5&0
\\
 sources in the Hamas-controlled enclave& 35& 35&0
\\
 who were not on the scene& 0& 0&0
\\
 …, which is part of the political arm of Hamas& 0& 0&0
\\
            allegedly& 5& 4& 1
\\
            are thought to have been killed& 1& 0& 1
\\
            claim& 6& 5& 1
\\
            difficult to know how many& 0& 0& 0
\\
            no reliable and current figures& 0& 0& 0
\\
            reportedly& 9& 8& 1
\\
            said to be& 0& 0& 0
\\
            said to have been killed& 2& 2& 0
\\
 said to have been wounded& 0& 0&0
\\
            was reported to have& 0& 0& 0\\
\midrule
\textbf{Total}&176&172&4\\
\bottomrule
\end{tabular}
}

\end{table}

\begin{table}[H]
\centering
    \caption{\textbf{\textcolor{black}{Breakdown of Doubt-Casting Phrases in Casualty-Reporting Sentences in NYT}} This table presents the total frequency of each doubt-casting phrase identified in NYT's casualty-reporting sentences. It also shows how these frequencies are distributed across Palestinian and Israeli CVN mentions, providing insights into the framing dynamics for each side within NYT’s reporting.}
\label{supp_tab:Casting_Doubt_NYT_Phrase_Count}
    
    \resizebox{0.77\textwidth}{!}{%
        \begin{tabular}{lccc}

            & \multicolumn{3}{c}{\textbf{The New York Times}} \\
            & \multicolumn{3}{c}{(Total of \textbf{10,728} CVNs)} \\
            \cmidrule(r){2-4}
            Doubt-casting phrases & Total mentions & Mentions& Mentions\\
 & that include& referencing&referencing\\
 & the phrase& Palestinian&Israeli\\
 & & victims&victims\\
            \midrule
            ..., an organization controlled by Hamas& 0& 0& 0
\\
            ..., which is controlled by Hamas& 12& 12& 0
\\
 Hamas Health Ministry& 0& 0&0
\\
 Hamas and medical officials & 0& 0&0
\\
            Hamas authorities& 1& 1& 0
\\
            Hamas government& 5& 5& 0
\\
            Hamas health officials& 0& 0& 0
\\
            Hamas officials& 10& 10& 0
\\
            Hamas said& 12& 12& 0
\\
            Hamas security officials& 0& 0& 0
\\
            Hamas sources in Gaza & 0& 0& 0
\\
 Hamas-controlled Gaza& 12& 12&0
\\
 Hamas-controlled Gaza Strip& 0& 0&0
\\
 Hamas-controlled government& 3& 3&0
\\
 Hamas-controlled health authorities& 0& 0&0
\\
 Hamas-led authorities& 1& 1&0
\\
 Hamas-ruled Gaza& 0& 0&0
\\
 Hamas-ruled territory& 0& 0&0
\\
 Hamas-run Gaza Health Ministry& 12& 12&0
\\
 Hamas-run Gaza Ministry of Health& 1& 1&0
\\
 Hamas-run Ministry of Health& 5& 5&0
\\
 Hamas-run Palestinian territory& 0& 0&0
\\
 Hamas-run al-Aqsa TV reported& 0& 0&0
\\
 Hamas-run health ministry& 32& 32&0
\\
 Hamas-run territory& 6& 6&0
\\
 according to Hamas' military wing& 0& 0&0
\\
 according to a Hamas website& 0& 0&0
\\
 according to health officials in Hamas-controlled Gaza& 1& 1&0
\\
 according to officials in the Hamas-run territory& 1& 1&0
\\
 controlled by Hamas& 12& 12&0
\\
 drawing from sources in Hamas-controlled Gaza& 0& 0&0
\\
 sources in Hamas-run Gaza& 0& 0&0
\\
 sources in the Hamas-controlled enclave& 0& 0&0
\\
 who were not on the scene& 0& 0&0
\\
 …, which is part of the political arm of Hamas& 1& 1&0
\\
            allegedly& 0& 0& 0
\\
            are thought to have been killed& 0& 0& 0
\\
            claim& 10& 9& 1
\\
            difficult to know how many& 0& 0& 0
\\
            no reliable and current figures& 0& 0& 0
\\
            reportedly& 22& 21& 1
\\
            said to be& 6& 4& 2
\\
            said to have been killed& 0& 0& 0
\\
 said to have been wounded& 0& 0&0
\\
            was reported to have& 1& 1& 0\\
\midrule
\textbf{Total}&166&162&4\\
\bottomrule
\end{tabular}
}

\end{table}

\begin{comment}
\begin{table}[H]
{\fontsize{7}{7}\selectfont{
\caption{\textbf{Child-related Mentions in Media Coverage of Palestine and Israel: Counts and Percentages Across Four Outlets.} Counts and percentages of articles with Child-related CVN for Palestine and Israel across four media sources. The table presents the number of articles mentioning numbers of children, the total number of articles, and the corresponding percentage for each side (Palestine and Israel) in AJE, BBC, CNN, and NYT.}
\label{supp_tab:children_CVN}
\begin{center}
\resizebox{\textwidth}{!}{%
\begin{tabular}{llccccc}
\toprule
& & \textbf{Articles with} & \textbf{Articles} & \textbf{Articles} & \textbf{Percent} & \\

\textbf{Media} & & \textbf{Casualty-} & \textbf{with} & \textbf{with} & \textbf{Casualty-} & \textbf{Percent}\\

\textbf{source} & \textbf{Side} & \textbf{related} & \textbf{Child} & \textbf{CVNs} & \textbf{related} & \textbf{Child}\\

 & & \textbf{Child CVNs} & \textbf{CVNs} & & \textbf{Child CVNs} & \textbf{CVNs}\\
\midrule
\multirow{2}{*}{AJE} & Palestine & 835 & 1,071 & 2,847 & 29.3\% & 37.6\% \\ 
 & Israel& 28 & 64 & 1,416 & 2.0\% & 4.5\% \\ \\
\multirow{2}{*}{BBC}& Palestine & 282 & 368 & 1,357 & 20.8\% & 27.1\%\\ 
& Israel& 41& 106  & 1,219 & 3.4\%& 8.7\% \\ \\
\multirow{2}{*}{CNN}& Palestine & 248 & 328 & 1,193 & 20.8\%& 27.5\% \\ 
& Israel& 26& 93  & 993 & 2.6\% & 9.4\%  \\ \\
\multirow{2}{*}{NYT}& Palestine  & 425 & 630 & 2,705 & 15.7\% & 23.3\% \\ 
& Israel& 85& 214 & 2,028 & 4.2\%& 10.6\% \\
\bottomrule
\end{tabular}
}
\end{center}
}}
\end{table}
 
\end{comment}

\begin{table}[H]
{\fontsize{7}{7}\selectfont{
\caption{\textbf{Child-related Mentions in Media Coverage of Palestine and Israel: Counts and Percentages Across Four Outlets (with baseline reference).} Counts and percentages of articles with Child-related CVNs for Palestine and Israel across four media sources. The table presents the number of articles mentioning numbers of children, the total number of articles, and the corresponding percentage for each side (Palestine and Israel) in AJE, BBC, CNN, and NYT. The rightmost column provides a baseline reference: the approximate share of children among civilian casualties for each side.}
\label{supp_tab:children_CVN}
\begin{center}
\resizebox{\textwidth}{!}{%
\begin{tabular}{llcccccc}
\toprule
 & & \textbf{Articles with} & \textbf{Articles} & \textbf{Articles} & \textbf{Percent} & \textbf{Percent} & \textbf{Baseline: children}\\
\textbf{Media} & \textbf{Side} & \textbf{Casualty-} & \textbf{with} & \textbf{with} & \textbf{Casualty-} & \textbf{Child} & \textbf{among civilian}\\
\textbf{source} &  & \textbf{related} & \textbf{Child} & \textbf{CVNs} & \textbf{related} & \textbf{CVNs} & \textbf{casualties}\\
 &  & \textbf{Child CVNs} & \textbf{CVNs} &  & \textbf{Child CVNs} &  & \textbf{(reference)}\\
\midrule
\multirow{2}{*}{AJE} & Palestine & 835 & 1{,}071 & 2{,}847 & 29.3\% & 37.6\% & \multirow{2}{*}{\(\approx 40\text{–}44\%\)}\\
 & Israel & 28 & 64 & 1{,}416 & 2.0\% & 4.5\% & \multirow{2}{*}{\(\approx 5\%\) (civ.)}\\
\\[-4pt]
\multirow{2}{*}{BBC} & Palestine & 282 & 368 & 1{,}357 & 20.8\% & 27.1\% & \multirow{2}{*}{\(\approx 40\text{–}44\%\)}\\
 & Israel & 41 & 106 & 1{,}219 & 3.4\% & 8.7\% & \multirow{2}{*}{\(\approx 5\%\) (civ.)}\\
\\[-4pt]
\multirow{2}{*}{CNN} & Palestine & 248 & 328 & 1{,}193 & 20.8\% & 27.5\% & \multirow{2}{*}{\(\approx 40\text{–}44\%\)}\\
 & Israel & 26 & 93 & 993 & 2.6\% & 9.4\% & \multirow{2}{*}{\(\approx 5\%\) (civ.)}\\
\\[-4pt]
\multirow{2}{*}{NYT} & Palestine & 425 & 630 & 2{,}705 & 15.7\% & 23.3\% & \multirow{2}{*}{\(\approx 40\text{–}44\%\)}\\
 & Israel & 85 & 214 & 2{,}028 & 4.2\% & 10.6\% & \multirow{2}{*}{\(\approx 5\%\) (civ.)}\\
\bottomrule
\end{tabular}
}
\end{center}
\par\vspace{4pt}\footnotesize
\emph{Note.} Baseline provides context for interpreting media shares. For \textbf{Israel}, children were \(\approx 5\%\) of \emph{civilian} deaths on Oct.~7 (and \(\approx 3\%\) of all deaths including soldiers). For \textbf{Palestine}, children comprise \(\approx 40\text{–}44\%\) of \emph{civilian} deaths over the first year (Oct.~7, 2023–Oct.~7, 2024).

}}
\end{table}

\pagebreak
\section{Extracting, Labeling, and Validating Instances for the Individualized vs. Category-Based Reporting Analysis}
\label{supp_note:IG_Prompts}

{\singlespacing
In this analysis, we classified how articles report instances of civilian hardships, distinguishing between two styles:
\begin{enumerate}
    \item \textbf{Individualization}: The report focuses on the hardship of a \textit{single} identified person, family, or couple. The individual(s) may be identified by details such as name, age, role, or family background. The key feature is that the person is clearly singled out, and their hardship is highlighted in a way that invites reader sympathy. For convenience, we refer to these as ``individualized instances'' or ``stories''.
    \item \textbf{Grouping}: The report describes the hardship of \textit{multiple} individuals collectively, without singling out any one person. The focus is on the group as a whole, rather than on an individual’s experience. We refer to these as ``grouped instances'' or ``grouped mentions.''
\end{enumerate}

\vspace{0.5\baselineskip}

To conduct this analysis, we followed a two-step process using two distinct LLM prompts:

\vspace{0.5\baselineskip}

\noindent \textbf{Step 1: Prompt 1}

\vspace{0.5\baselineskip}

\noindent First, we extracted all instances of civilian hardships from the media articles. Prompt 1, instructed the LLM to identify both individualized and grouped instances within an article, regardless of how many appeared. The model was also tasked with recording:
\begin{itemize}
    \item the human entity experiencing the hardship (individual or group),
    \item the location of the hardship,
    \item the side of the entity,
    \item whether the entity was a civilian,
    \item whether they were the primary figure (for individualized instances) or quoted (for grouped instances), and
    \item the exact phrases in which the instance appeared.
\end{itemize}

After the collection of the data in step 1, the data was carefully filtered to extract individualized stories. These were defined as instances where the human entity was an individual associated with either Palestine or Israel, and where the hardship occurred in Gaza or Israel. Only unique entities within each article were retained, resulting in 10,747 instances distributed as follows: 3,638 and 4,412 primary instances for Israel and Palestine, respectively, and 1,120 and 1,577 secondary instances for Israel and Palestine, respectively. In our analysis, we focused on instances where the entity was the primary figure, as secondary instances were considered stories branching from those of a primary one. This decision was supported by the observation that the secondary-to-primary instance ratios were nearly identical (0.31 for Israel and 0.36 for Palestine) indicating comparable representation of both instance types across sides. To compare the two proportions (0.31 vs. 0.36), we calculated Cohen’s h (Cohen, 1988) effect size to be 0.11, which according to Cohen’s benchmarks (0.20 = small, 0.50 = medium, 0.80 = large) indicates that the difference between the proportions is negligible, with secondary instances making only a minimal contribution relative to primary ones.

\vspace{2\baselineskip}

\noindent \textbf{Step 2: Prompt 2}

\vspace{0.5\baselineskip}

\noindent Next, we analyzed the extracted individualized stories in greater detail. The goal was twofold: (1) to identify the most significant hardship described, and (2) to determine whether it was directly related to the October 7 attack or hostages. Prompt 2 took as input the phrases containing the story, the name or identifier of the main individual, and the full article. The LLM then:

\begin{enumerate}

\item Located all sentences in the article where the individual or their hardship was mentioned (since a single story could span multiple, non-consecutive sentences).

\item Assigned a single hardship label from the following controlled set: ``\textit{Casualties}'', ``\textit{Displacement and Refugees}'', ``\textit{Imprisonment and Detention}'', ``\textit{Health and Medical Conditions}'', ``\textit{Deprivation, Malnutrition and Hunger}'', ``\textit{Humanitarian Aid and Dependence}'', ``\textit{Missing}'', ``\textit{Vulnerable and Affected Groups}'', or ``\textit{Other Hardship}.''

\item Indicated whether the hardship was directly connected to the October 7 attack or hostages.
\end{enumerate}

Finally, given the prominence of reported child casualties in this context, we conducted a third analysis (Prompt 3) to determine whether an individualized story or grouped instance mentioned children. Prompt 3 instructed the LLM to review the entity name/identifier, the instance phrase, and the full article, and to return whether the instance contained a reference to a child (e.g., “her child,” “many children,” “9-year-old Sarah”). This analysis used the UN definition of a child, i.e. a person below the age of 18, unless relevant laws recognize an earlier age of majority~\cite{UN_Children}, to ensure accuracy. 

Below, we present the full text of each prompt along with examples of the input provided and the corresponding output returned by the LLM.

}

\begin{tcolorbox}[enhanced,fit to height=20cm,colback=teal!25!black!2!white,colframe=teal!90!black,title= Prompt 1: Extracting Individualized and Grouped Instances]
  
\# \textbf{Role and Objective}

\begin{itemize}
    \item You are a helpful autonomous extraction agent with expertise in Psychology and Advanced Textual Analysis.
    \item Your objective is to thoroughly analyze a user-provided news article and extract all occurrences of the “Individualized” and “Grouped” instances.
    \item You are an agent - please keep going until the user’s query is completely resolved, before ending your turn and yielding back to the user. Only terminate your turn when you are confident that **all valid instances** have been correctly identified. You do not stop after providing only a few examples and you pay particular attention to extracting instances from long sentences.
    \item You must strictly follow the specific set of instructions defined below.
    \item Your responses are only in JSON.
\end{itemize}

\# \textbf{Instructions}\\

Use the criteria below to identify both the individualized and grouped instances.

\begin{itemize}
    \item An instance is a description of an Entity experiencing Hardship.
    \item In case of an Individualized instance, the Entity is a singled-out Individual which may be a single person or a single family or a single couple only. Example: “Tom”, “a mother”, “a child”, “a boy”, “a family”, “a couple”, “a refugee”, “a hostage”, “a man”, “a soldier”, “an NGO operator”, “a volunteer”, or “an Israeli”.
    \item In case of a Grouped instance, the Entity is a Group where a Group refers to multiple individuals described collectively, referred to as one collective unit without singling out any person by name or treating them as distinct individuals, where their shared role, status or condition is emphasized. Example: mothers, children, boys, families (plural), couples (plural), refugees, hostages, victims, men, Jews, Muslims, Palestinians, Israelis, civilians, captives, people, injured, senior officers, aid workers, rescue teams, soldiers, NGO operators, volunteers, wounded…etc. A Group may be combined with Numbers to indicate a definite, exact quantity, such as: “Ten children”, “1500 men”, “3,400 people”, or with Quantifiers to indicate an indefinite or approximate quantity, such as “several men”, “many Israelis”, “few women”, “all hostages”, or “dozens injured”.
    \item A Hardship refers to the Entity’s exclusive experience of suffering or difficulty, it reflects a case of severe emotional or physical pain. Example: “Mary was killed in the bombing” or “Ten civilians were killed”. The hardship account must be relevant to the Israel-Palestine conflict. Keep in mind that an account does not have to take place in Palestine or Israel to be considered relevant to the conflict.
\end{itemize}

\end{tcolorbox}

\begin{tcolorbox}[enhanced,fit to height=20cm,colback=teal!25!black!2!white,colframe=teal!90!black,title= Prompt 1: Extracting Individualized and Grouped Instances (continued)]
  
\begin{itemize}
    \item A hardship, may have the following types:
    \begin{itemize}
        \item \textbf{Emotional (severe)}: such as suffering, struggle, distress, worry or concern about self or close relations only (such as family, friends, loved ones), unease, desperation, sadness, anxiety, loss, grief, anguish, exposure to difficult experiences, severe psychological, deep emotional, traumatic or painful experiences/strain, or feeling overwhelmed by a difficult situation.
        \item \textbf{War-related}: such as shortages of food, water, fuel, medicine; fear of death or arrest, feeling threatened, exposure to horrible scenes or bloodshed, being unsheltered or exposed to bad weather, trapped in rubble of collapsed building, waiting in lines, feeling an existential threat, sacrificing to help self or others, getting assassinated, being targeted by strikes or tortured.
        \item \textbf{Physical}: getting killed, injured, wounded, attacked.
    \end{itemize}
    \item In case of a Grouped instance, the hardship may take any of the following forms:
    \begin{itemize}
        \item \textbf{Generalized Hardship}, this refers to a collective experience of suffering or difficulty that affects the Group, such as “the civilians were traumatized by the attacks”.
        \item \textbf{Generalized Statistics}, such as “90 women and children were killed in the strike”.
        \item \textbf{Broad Descriptions of Events}, such as “the civilians are suffering from a widespread famine in the city”.
    \end{itemize}
    \item A hardship has the following properties:
    \begin{itemize}
        \item A hardship may be mentioned either explicitly/directly (“citizens were killed”, “Tom was attacked”) or implicitly/indirectly (“citizens were released” implying previous captivity or “David was in tears” implying deep sadness) in the article.
        \item A hardship may be Primary such as “David was killed”, or Secondary such as “David’s wife was murdered.”
        \item Important - A hardship does not have to be the main focus of the article context. Meaning, it may be mentioned in the background to the sentence’s main message.
    \end{itemize}
    \item A hardship must reflect a case of severe emotional or physical pain. As such, the following hardship cases do not qualify as valid ones:
    \begin{itemize}
        \item Worry/concern/emotional distress about those other than self and close relations like family, friends or loved ones.
       \item Political or emotional dissatisfaction reactions: they reflect perceptions, frustrations, or unmet expectations, such as: frustration, dismissal of concerns, being dismissed, political pressure, political anxiety, disappointment about anything, being neglected, dismissal of views, or feeling unheard.
    \end{itemize}
\end{itemize}

\end{tcolorbox}

\begin{tcolorbox}[enhanced,fit to height=20cm,colback=teal!25!black!2!white,colframe=teal!90!black,title= Prompt 1: Extracting Individualized and Grouped Instances (continued)]

\begin{itemize}[label={}]
    \item
    \begin{itemize}
        \item Potential hardship that did not happen yet, such as “if the authorities cut the power, this will put the civilians at risk”.
        \item Important: Inflicting harm on others is not a valid hardship, such as a man killing or injuring others must not be considered to be experiencing a hardship, rather, he is inflicting one on others.
    \end{itemize}
\end{itemize}

\# \textbf{Special Notes}

\begin{itemize}
    \item If two or more clearly identified, unconnected individuals are reported to have experienced the same hardship, see example in this case, you should not treat them as a Group since the individuals constituting the group have not been collectively referenced, rather they were individually identified. As such, you should treat them all (example “Mary and Tom”) as a special case of a single individual and return them all as one Individual (Entity) of an Individualized instance.
    \item Important - In some cases, a Group identifier on its own could imply an implicit hardship, resulting in the Grouped instance being just one word long. Example: just the word “hostages” implies that a group of civilians were taken against their will (the grouped hardship). Similarly, “prisoners”, “the wounded”, “the detained”, “the starving”, “victims” all follow the same logic.
    \item Analyze image or photo captions as part of the article text.
    \item Important: An individualized instance may have an unnamed Entity, example: “a 70-year old woman” or “a child”, the most important thing is that the singled out Entity is singular.
    \item Important! A sentence may contain more than one Grouped instance, example: ``They killed 1100 civilians and abducted 300 others'', so here this sentence should yield 2 instances; “killed 1100 civilians” and “abducted 300 others”.
    \item An article may contain neither Individualized nor Grouped instances at all, so it is fine to report as such provided you read the article carefully to make sure that indeed no instance was missed.
    \item An article may have a large number of Individualized or Grouped instances, so there is no upper limit on the number of instances to report back. Feel free to report as many instances as needed to cover all of the article’s instances.
\end{itemize}

\# \textbf{Response Fields}

\begin{itemize}
    \item \textbf{Type}: Return “Individualized” or “Grouped” based on the type of the detected instance as per the instructions provided above. Important: pay special attention to the criteria outlined above before returning a response.
\end{itemize}
\end{tcolorbox}

\begin{tcolorbox}[enhanced,fit to height=20cm,colback=teal!25!black!2!white,colframe=teal!90!black,title= Prompt 1: Extracting Individualized and Grouped Instances (continued)]

\begin{itemize}
    \item \textbf{Entity}: return the instance Entity’s identifier exactly as it appears in the article. The identifier must be explicitly mentioned in the article (“many civilians in Gaza had to go to refugee camps” - in this case, the identifier is “many civilians”) and not inferred from the text (“the refugee camps are full”).
    \item \textbf{Side}: provide an educated best guess regarding the Side to which the Entity supports, belongs to or is sympathetic with based on the article and the extra context provided below. Return only “Palestine”, “Israel”, “Both” or “Other”.
    \item \textbf{Civilian\_Status}: Specify if the Entity’s status is “Military”, “Government” (any non-military related government position), or “Civilian” (any other affiliation such as hostage(s), teacher(s), doctor(s), nurse(s), patient(s), settler(s), NGO worker(s), UN worker(s), UN official(s), etc.). Note for the purposes of this analysis, any hostage/captive is considered a “Civilian”, regardless of their pre-abduction role. Important: Any individual or group belonging to Hamas or IDF must not be regarded as a Civilian.
    \item \textbf{Location}: return the physical location where the instance’s hardship is taking place. Return only either “Israel” (this includes cities like Tel Aviv, Ashdod, Haifa), “Gaza”, “West Bank” or “Other” (this includes any other location).
    \item \textbf{Primary}: Return "Yes" if the reported Entity is the main central figure of the instance narrative and the one that is set up as the entry point of the instance’s narrative/story events, otherwise return “No”. Note that the Primary entity’s hardship narrative does not branch from another entity’s narrative. Note: the Primary entity may be experiencing a secondary hardship. See example 9. In the case of a Grouped Prompt, return only “N\_A”.
    \item \textbf{Quoted}: In the case of a Grouped Prompt, return “Yes” only if the instance is mentioned within a direct quotation (enclosed in either single or double quotation marks - important: a quotation may include an ellipsis (...) ). Otherwise, return “No”. For an Individualized Prompt, always return “N\_A”.
\end{itemize}

\begin{itemize}
    \item \textbf{Phrases}: return the part of the sentence from the article (verbatim) in which the instance’s Entity and the hardship they experienced are both explicitly and clearly mentioned.
\end{itemize}

\# \textbf{JSON Response format:}\\

Your response MUST adhere to the following JSON format:\\
\{“instances”:[\{“Type”:“...”, “Entity”:“...”, “Side”:“...”, “Civilian\_Status”:“...”, “Location”:“...”, “Primary”:“...”, “Quoted”:“...”, “Phrases”:“...”\}, \{“Type”:“...”, “Entity”:“...”, “Side”:“...”, “Civilian\_Status”:“...”, “Location”:“...”, “Primary”:“...”, “Quoted”:“...”, “Phrases”:“...”\}...]\}.\\

If you cannot find any instances within the article to report, return:\\ 
\{“instances”:[\{“Type”:“N\_A”, “Entity”:“N\_A”, “Side”:“N\_A”, “Civilian\_Status”:“N\_A”, “Location”:“N\_A”, “Primary”:“N\_A”, “Quoted”:“N\_A”, “Phrases”:“N\_A”\} ]\}. 
\end{tcolorbox}

\begin{tcolorbox}[colback=blue!5!white,colframe=teal!75!black,title=Example of Sentences with Individualized and Grouped Instances, breakable]

  \textbf{Example 1:} ``A man whose sister and two nieces were killed in the Hamas attacks on Israel is campaigning to free his brother-in-law from captivity.
  
  Steve Brisley's sister, Lianne Sharabi, nieces Noiya, 16, Yahel, 13, and their family dog were murdered on 7 October last year, when Hamas launched a series of cross-border attacks from the Gaza Strip on southern Israel.''\\

  \textbf{Example 2:} ``Ghada Elnajjar, a Palestinian American who has lost more than 80 members of her extended family to the Israel-Hamas war, which she calls `Israel's ongoing genocide.' ''\\

  \textbf{Example 3:} ``Non-stop Israeli bombardment of the Gaza Strip destroyed residential buildings and killed more than 3,400 Palestinians.''\\

  \textbf{Example 4:} ``They have treated more than 400 patients because of the surprise attack by Hamas.''\\
  
\end{tcolorbox}

\begin{tcolorbox}[title=JSON Output, width=\textwidth, breakable]
\begin{lstlisting}[language=json, numbers=none, breaklines=true]
{ { Type: Individualzied,
    Entity: Steve Brisley,
    Side: Israel,
    Civilian_Status: Civilian,
    Location: Israel,
    Primary: Yes,
    Quoted: N_A,
    Phrases: A man whose sister and two nieces were killed in the Hamas attacks on Israel is campaigning to free his brother-in-law from captivity.
  },  
  { Type: Individualzied,
    Entity: Ghada Elnajjar,
    Side: Palestine,
    Civilian_Status: Civilian,
    Location: Other,
    Primary: Yes,
    Quoted: N_A,
    Phrases: Ghada Elnajjar, a Palestinian American who has lost more than 80 members of her extended family to the Israel-Hamas war, which she calls `Israel's ongoing genocide.'.
  },
  { Type: Grouped,
    Entity: more than 3,400 Palestinians,
    Side: Palestine,
    Civilian_Status: Civilian,
    Location: Gaza,
    Primary: N_A,
    Quoted: No,
    Phrases: non-stop Israeli bombardment of the Gaza Strip destroyed residential buildings and killed more than 3,400 Palestinians.
  },
  { Type: Grouped,
    Entity: more than 400 patients,
    Side: Israel,
    Civilian_Status: Civilian,
    Location: Israel,
    Primary: N_A,
    Quoted: No,
    Phrases: They have treated more than 400 patients because of the surprise attack by Hamas.
   } }
\end{lstlisting}
\end{tcolorbox}

\begin{tcolorbox}[enhanced,fit to height=20cm,colback=teal!25!black!2!white,colframe=teal!90!black,title= Prompt 2: Labeling Individualized Instances]
\# \textbf{Role and Objective}

\begin{itemize}
    \item You are a helpful autonomous textual extraction \& labeling agent with expertise in Psychology and Advanced Textual Analysis.
    \item The user shall provide you with 3 inputs:
    \begin{itemize}
        \item A news article.
        \item An Individual’s Identifier: the identity of a single person or human entity mentioned in the article (e.g. “David”, “the woman”, “a teacher”, “a family”...etc.).
        \item A Phrase: an excerpt phrase from the news article where the individual was mentioned.
    \end{itemize}
    \item Your objective is to use the identifier and the phrase to correctly identify the individual and the individual’s “story” within the article. The story refers to all the sentences in the article related to the Individual or where the Individual or their hardship(s) was mentioned.You shall then use the story to return the requested fields.
    \item You are an agent - please keep going until the user’s query is completely resolved, before ending your turn and yielding back to the user. Only terminate your turn when you are confident that **all the individual’s sentences** have been correctly identified and all the fields have been properly answered. You do not stop after checking only a few article sentences.
    \item Your responses are only in JSON.
\end{itemize}

\# \textbf{Instructions}\\

\#\# \textbf{Hardship Labels}

Return only one of those hardship labels as per the instructions below: “Casualties”, “Displacement and Refugees”, “Imprisonment and Detention”, “Health and Medical Conditions”, “Deprivation, Malnutrition and Hunger”, “Humanitarian Aid and Dependence”, “Missing”, “Vulnerable and Affected Groups”, or “Other Hardship”.\\

\#\# \textbf{Thought Process}

\begin{enumerate}
    \item Read the Individual’s identifier and the Phrase.
    \item Read the news article carefully.
    \item Use information in 1 to Identify the individual within the article.
    \item Create an empty list to store the individual’s story sentences.
    \item Go over the article one sentence at a time and decide if the sentence is related to the Individual or where the Individual or their hardship(s) was mentioned in any way.
\end{enumerate}
\end{tcolorbox}

\begin{tcolorbox}[enhanced,fit to height=20cm,colback=teal!25!black!2!white,colframe=teal!90!black,title= Prompt 2: Labeling Individualized Instances]

\begin{enumerate}[start=6]
    \item If the sentence is relevant, add it to the list created in 4.
    \item Repeat steps 5 and 6 till all article sentences are covered and the individual’s story has been entirely collected.
    \item Stop and scan all the story sentences then assign a single label to describe the most severe hardship described in the individual’s story, whoever may have experienced it. Meaning, do not limit the hardship labeling to only the individual’s experiences but expand it to include any hardship that was experienced by any character mentioned in the individual’s story. Example “Tom’s daughter was killed in the air strike”, here we have 2 hardships: the killing of the daughter (Casualties) and Tom’s emotional pain of losing his daughter (Other Hardship). Since in this case the “being killed” hardship is more severe than the “emotional pain” hardship, then you must return the “Casualties” label as the most severe detected hardship even though this hardship was not experienced by Tom, but since it was mentioned in his story then it must be returned. Again, this is very important: the most severe hardship may not be experienced directly by the individual but by others mentioned in his story/narrative.
\end{enumerate}
  
\#\# \textbf{Special Note}

An exception: if the most severe detected hardship is inflicted by the user-provided individual upon others, then in this case you should return a special label - “Inflicted upon others”. Example: if the user-provided individual is “Tom” and if the story of Tom is only “Tom killed 7 people” then in this case Tom is causing hardship upon others and so this is a valid exception case.\\

\#\# \textbf{Requested Fields}

\begin{itemize}
    \item \textbf{Hardship}: return the label that describes the most severe hardship mentioned in the individual’s story, as explained in the instructions above. It must be just one of the labels provided in the Instructions above.
    \item \textbf{Justify}: provide a reason for the selection of the hardship label. Keep it as brief and concise as possible.
    \item \textbf{Oct\_7\_Attack}: is the Individual or any of those mentioned in the Individual’s story experiencing a hardship directly related to the Oct 7 attack? Example: a person recounting the kidnapping of a loved one on Oct 7, or a person recounting the death or injury or any hardship faced by an Oct 7 hostage later after the attack. If so, return “Yes”, otherwise, return “No”. Consider only Israeli hostages and victims of the Oct 7 attack.
\end{itemize}

\# \textbf{Response Format}

Your response MUST adhere to the following JSON format: \{ “instances”: [ \{“Hardship”: “...”, “Justify”: “...”, “Oct\_7\_Attack”: “...” \}] \}.\\
\end{tcolorbox}

\begin{tcolorbox}[colback=blue!5!white,colframe=teal!75!black,title=Example of Sentences with Individualized Instances, breakable]

  \textbf{Sentence 1:} Judith Raanan and her daughter Natalie, were abducted by Hamas during its October 7 attack on southern Israel.\\

  \textbf{Sentence 2:} Amiram Cooper ... who were brutally abducted to the Gaza Strip on October 7, that they are no longer alive and that their bodies are held by the Hamas terrorist organization.\\

  \textbf{Sentence 3:} A similar situation has taken hold for Marwa Talbani, 32, and her family as displacement affects all these mothers and their children the same way\\

  \textbf{Sentence 4:} 'She will be treated at a hospital in Dayton, Ohio, for third-degree burns on her face.', 'Saja Bilal Junaid, 3, the youngest of the children. Saja’s head was covered entirely in bandages except for a small opening for her eyes, nose and mouth. Her face was charred with third-degree burns after an Israeli airstrike hit her home at Jabalia refugee camp, the aid group said.'\\

  \textbf{Sentence 5:} Ahmed says his father was the kindest man he had ever known. He worked hard driving a taxi and in construction to build a home for his children and educate them well.\\
  
\end{tcolorbox}

%The table below shows part of the LLM output for the instances taken from the examples above

\begin{tcolorbox}[title=JSON Output, width=\textwidth, breakable]
\begin{lstlisting}[language=json, numbers=none, breaklines=true]
{ { Hardship: Imprisonment and Detention,
    Justify: Judith Raanan and her daughter Natalie were abducted by Hamas and held captive in Gaza, which constitutes imprisonment and detention. Although they were later released, the most severe hardship in their story is their abduction and captivity.,
    Oct_7_Atack: Yes,
  },
  { Hardship: Casualties,
    Justify: Amiram Cooper was abducted on October 7, held hostage in Gaza, and has now been declared dead, with his body still held by Hamas. The story also mentions the murder of his brother-in-law on October 7. The most severe hardship is the killing of Amiram Cooper and others.,
    Oct_7_Atack: Yes,
  },
  { Hardship: Deprivation, Malnutrition and Hunger,
    Justify: tearful couple,
    Oct_7_Atack: ,
  },
  { Hardship: Deprivation, Malnutrition and Hunger,
    Justify: Marwa Talbani and her family are described as suffering from severe hunger, food scarcity, and the inability to provide adequate nutrition for their children, with children crying from hunger and parents rationing minimal food. This is the most severe hardship in her story, surpassing displacement and economic hardship.,
    Oct_7_Atack: No,
  },
  { Hardship: Casualties,
    Justify: Ahmed's father was killed in an air strike along with many other family members, making 'Casualties' the most severe hardship in his story.,
    Oct_7_Atack: No,
  } }
\end{lstlisting}
\end{tcolorbox}

\begin{tcolorbox}[enhanced,fit to height=20cm,colback=teal!25!black!2!white,colframe=teal!90!black,title= Prompt 3: Identifying Child Mentions in Individualized and Grouped Instances]
\# \textbf{Role and Objective}

\begin{itemize}
    \item You are a helpful autonomous textual extraction \& labeling agent with expertise in Psychology and Advanced Textual Analysis.
    \item The user shall provide you with the following inputs:
    \begin{itemize}
        \item A news article.
        \item An Entity’s Identifier: the identifier of a human entity within the article. Can refer to a “single” individual such as “David”, “the woman”, “a teacher”, “a family”...etc, or multiple individuals described collectively (i.e. a group) such as “refugees”, “20 children”, “many women”...etc.
        \item An Entity’s Type: “Individualized” indicates that the entity refers to a singular individual/person while “Grouped” indicates that the entity refers to multiple individuals described collectively.
        \item A Phrase: an excerpt phrase from the news article where the Entity was mentioned.
    \end{itemize}
    \item Your objective is to use the Identifier and the Phrase to correctly identify the Entity within the article. If the Entity Type is “Individual” , you must identify that Individual Entity’s story which is defined as all the sentences in the article related to the Entity or where the Entity or their hardship(s) was mentioned.You shall then use the story to answer the requested field below. If the Entity Type is “Group”, then simply answer the requested field based on the Entity’s identifier and there is no need to identify the story.
    \item You are an agent - please keep going until the user’s query is completely resolved, before ending your turn and yielding back to the user. Only terminate your turn when you are confident that **all the Entity’s sentences** have been correctly identified and the requested field has been properly answered. You do not stop after checking only a few article sentences, rather you must check the entire article.
    \item Your responses are only in JSON.
\end{itemize}

\#\# \textbf{Requested Fields}\\

\textbf{Is\_Child}: If the Entity type is Individualized, then if the Entity or any of those mentioned in the Entity’s story is a child(ren), return the term in the article text where they are mentioned. If the Entity type is Grouped, then return the Entity Identifier if it explicitly refers to a child(ren). Otherwise, return “N\_A”. Note that in both cases, the child mention must be explicit, if however it is not clear as in “she lost many family members”, then do not assume a child(ren) was mentioned.\\

\#\# \textbf{Special Instructions:}

\begin{itemize}
    \item If the Entity type is Individualized and multiple children are mentioned in an Individual's Entity’s story, like “Sara lost her 2 children, Larry and Tom”, then just one of any of the multiple terms referring to the children, in this case it can be “children”, “Larry” or “Tom”.
    \item It is important to note that a child is defined as any person under the age of 18 years, so if there is a mention of age in the article, make sure to cross-reference it against 18 years to determine if the detected person/group is indeed referring to a child(ren) or not.
\end{itemize}

\# \textbf{Response Format}

Your response MUST adhere to the following JSON format: \{ “instances”: [ \{“Is\_Child”: “...”\}] \}.

\end{tcolorbox}

\begin{tcolorbox}[colback=blue!5!white,colframe=teal!75!black,title=Example of Sentences with Individualized and Grouped Instances, breakable]

  \textbf{Sentence 1:} ...including Kfir Bibas, who was 9 months old at the time. He is still unaccounted for..\\

  \textbf{Sentence 2:} Yaffa Adar was a resident of Kibbutz Nir Oz, she has three children, eight grandchildren and seven great-grandchildren. She was abducted by Hamas on Oct. 7 and remained in captivity for 49 days.\\

  \textbf{Sentence 3:} Hamas released 86 women and children it was holding\\

  \textbf{Sentence 4:} Israel's relentless aerial bombardment and ground operations have killed 34,183 people, mostly women and children, according to the Hamas-run health ministry.\\

  \textbf{Sentence 5:} Ms. Al Shaikh’s two granddaughters ran toward her, crying, “Where’s Mom?”.\\

  \textbf{Sentence 6:} the release of Palestinian women and minors held in Israeli prisons — a group that has grown in size since the Oct. 7 attacks.\\

  \textbf{Sentence 7:} On Friday, hundreds of people carrying what is left of their personal belongings poured onto the street on foot.\\

\end{tcolorbox}

%The table below shows part of the LLM output for the instances taken from the examples above

\begin{tcolorbox}[title=JSON Output, width=\textwidth]
\begin{lstlisting}[language=json, numbers=none, breaklines=true]
{   {   Is_Child: Kfir Bibas},
    {   Is_Child: children},
    {   Is_Child: 86 women and children},
    {   Is_Child: 34,183 people, mostly women and children},
    {   Is_Child: two granddaughters},
    {   Is_Child: minors},
    {   Is_Child: Not Applicable},
}
\end{lstlisting}
\end{tcolorbox}

\newpage
\subsection*{Validation Procedure}
{\singlespacing
%We focused on validating the results of the first main prompt as this covered the hardest task, semantically speaking, as opposed to the other 2 prompts which included direct, simpler labeling tasks.\\

\noindent \textbf{Rater procedure and blinding.} Two trained raters from the author team independently validated model outputs while blinded to the source outlet: article files were exported with outlet metadata removed and assigned neutral IDs. Initial disagreements were resolved through discussion to a consensus label; when needed, a third author adjudicated. All performance statistics below are computed against the consensus labels.\\

\noindent \textbf{Validation of the main (most semantically demanding) prompt.} We focused on validating the results of the first main prompt because it required the most semantic judgment, unlike the other two, which involved more direct labeling. We randomly selected 25 articles from each of the four media sources (100 articles total), yielding 1,071 instances for evaluation. We computed accuracy = 89\%, precision = 96\%, and recall = 85\%. Here, true negatives include instances where (i) the hardship did not occur in Gaza or Israel, (ii) the human entity is not a civilian, (iii) the entity does not belong to either the Palestinian or Israeli side, or (iv) the entity is not the primary figure in the narrative.

}

\pagebreak
\clearpage
{\color{black}
\section{Experimental Evidence: Methods and Results}
\label{note:experimental_evidence}
{\singlespacing
\subsection*{Participants}
Participants' answers were collected using Prolific Academic through a Qualtrics link. Participation in the study was paid £1.03 and lasted 8 minutes. The final sample consisted of 262 participants (130 women, 130 men and 2 individuals who did not provide information about their gender, $M_{age} = 41.30$, $SD_{age}= 12.38$). A post-hoc sensitivity power analysis, with $\beta - 1 = .80$ and $\alpha = .05$, showed that this sample size was sufficient to detect effects $r>.17$ for Pearson's correlations. Moreover, a sensitivity power analysis ($\alpha = .05$, $1-\beta=.80$) indicated that, with three between-subjects groups (N = 262; ns = 92, 88, and 82), the design was powered to detect effects of $\eta^2 \ge .04$, corresponding to a small-to-medium effect size. The study received IRB approval from New York University Abu Dhabi.

\subsection*{Procedure}
Participants completed an online survey administered via Qualtrics. After accessing the study, they first read an informed consent form describing the purpose of the research. Only participants who provided informed consent proceeded with the study.

Participants were then randomly assigned to one of three experimental conditions, which manipulated the journalistic identity they were asked to adopt: independent journalist, journalist from the stronger country (Setaria), or journalist from the weaker country (Arding). In all conditions, participants read a detailed description of a fictional geopolitical conflict between Setaria and Arding. The scenario emphasized that, although both sides had suffered casualties, Setaria was militarily and economically stronger, whereas Arding was more vulnerable. The scenario also specified the final civilian death toll of a recent struggle between the two countries (10 civilians from Setaria and 40 civilians from Arding).

Participants were instructed to imagine themselves as ``journalists'' covering the conflict for a major news outlet aligned with their assigned condition (or a neutral outlet in the independent condition). They were asked to consider the ethical responsibility of journalism and the real-world consequences that reporting choices may have on public understanding of the conflict.
Next, participants completed a victim selection task. They were presented with a visual display of 50 civilian victims, each represented by a circle: blue circles denoted victims from Setaria and red circles denoted victims from Arding (see Suppementary Figure~\ref{supp_fig:surveyscreenshot} for a screenshot of the visual display). By clicking on each circle, participants could read a brief biographical description of the victim. Participants were instructed to select exactly 10 victims whom they would choose to focus on in a news article about the conflict. Selections could be freely made and revised until the required number of victims was reached.\\

\begin{figure}[h!]
    \centering
    \setlength{\fboxrule}{0.5pt}   % border thickness
    \setlength{\fboxsep}{2pt}      % padding between image and border
    \fbox{\includegraphics[width=0.8\linewidth]{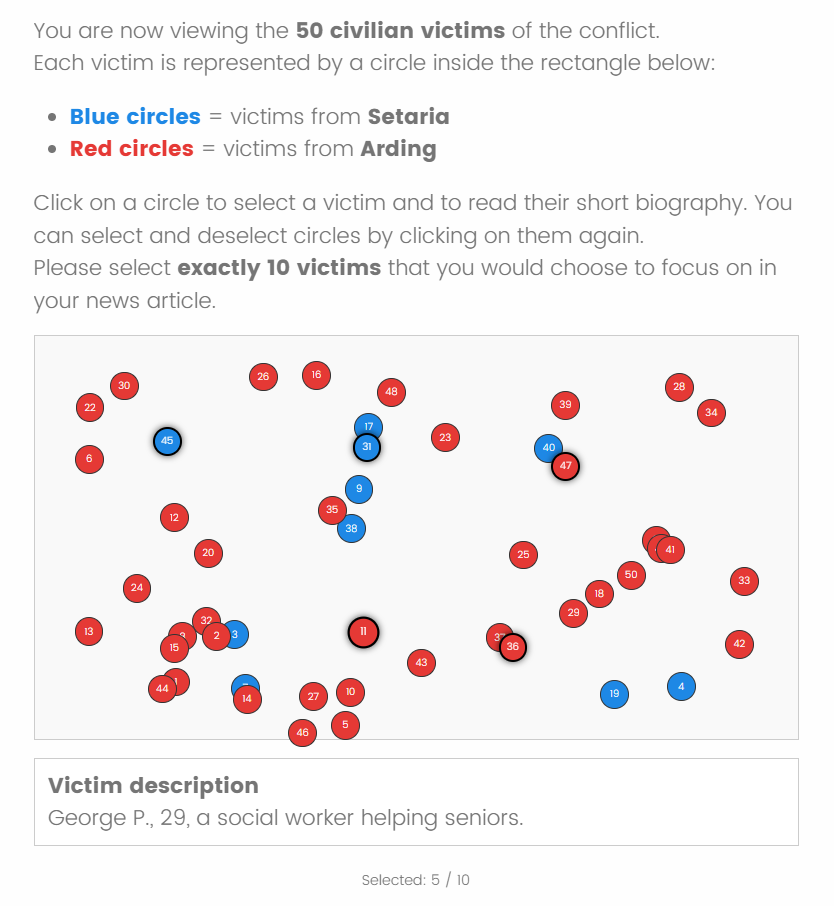}}
    \caption{\textbf{Screenshot of the victim selection task used in the experiment}.}
    \label{supp_fig:surveyscreenshot}
\end{figure}

Following the victim selection task, participants completed a manipulation check assessing their understanding of the assigned journalistic identity, the relative power of the two countries, and the distribution of civilian casualties. Participants then reported how strongly they identified with the role of a journalist, as well as with Setaria and Arding.

Participants also completed a series of items assessing normative beliefs about journalism, including views on objectivity, balance, accountability, and whether journalists should support or prioritize the interests of their own group during times of conflict. Finally, participants provided demographic information (age, gender, political orientation, education, subjective economic status).

All study materials, including the consent form, full instructions, experimental stimuli, and survey items, are publicly available on the Open Science Framework (OSF): \url{https://osf.io/n2s53/}.

\subsection*{Measures}
\textbf{Strong-country victim selection.} During the victim selection task, participants were presented with 50 civilian victims of the conflict, represented as circles, and were instructed to select exactly 10 victims to focus on in a news article. Victims from the militarily and economically stronger country (Setaria) were represented by blue circles, whereas victims from the weaker country (Arding) were represented by red circles. The position of the circles in the designed area and the short-bio describing victims were randomized. The primary dependent measure was the strong-country count, operationalized as the number of victims selected from the stronger country, with possible values ranging from 0 to 10. Higher scores indicated a greater focus on victims from the stronger country relative to the weaker country.\\

\noindent \textbf{Identification.} Participants indicated how strongly they identified with (a) the role of journalist, (b) Setaria, and (c) Arding. Responses were recorded using three separate 0–100 sliders, anchored at 0 (none at all) and 100 (a great deal), with higher values indicating stronger identification.\\

\noindent \textbf{Journalistic norms.} Participants’ beliefs about journalistic norms were assessed using a set of statements tapping professional principles and responsibilities of journalists. Items were organized into four norm dimensions, using three items for each dimension:
\begin{itemize}
    \item \textit{Balance norms} captured the extent to which journalists should remain impartial and avoid taking sides when reporting on conflicts (e.g., ``Journalists should avoid taking sides when reporting on conflicts'', $\alpha = .79$). 
    \item \textit{Accountability norms} reflected the belief that journalists should scrutinize and investigate those in positions of power (e.g., ``Journalists should investigate wrongdoing by those in power'',  $\alpha = .74$). 
    \item \textit{Public service norms} assessed the extent to which journalists should help the public understand complex political issues and provide broader context beyond events (e.g., ``Journalists should help the public understand complex political issues'',  $\alpha = .74$). 
    \item \textit{Advocacy norms} captured the belief that journalists should prioritize or promote the interests of their own group during conflict (e.g., ``Journalists should give more visibility to voices from their side in a conflict'',  $\alpha = .85$). 
\end{itemize}
Participants indicated their agreement with each statement using 0–100 slider scales, anchored at 0 (none at all) and 100 (a great deal), with higher scores indicating stronger endorsement of the respective norm.

\subsection*{Results}
\textbf{Victims selection.} An ANOVA tested whether the number of victims selected from the stronger country differed across conditions (independent-identity, strong country-identity, weak country-identity). The effect of condition was significant, $F(2, 259) = 24.68, p < .001, \eta^2 = .16$. Descriptively, participants selected more victims from the strong country in the strong country-identity condition ($M = 3.71$, $SD = 2.32$, $n = 88$) than in the independent-identity condition ($M = 2.96$, $SD = 1.43$, $n = 92$), and fewer in the weak-identity condition ($M = 1.76$, $SD = 1.59$, $n = 82$).

Given that victims from the strong country constituted 10 of the 50 victims (20\%), a neutral choice pattern corresponds to selecting 2 victims from the strong country out of 10. In the independent-identity condition, the 95\% CI was [2.66, 3.25], indicating that 2 was not included and selections were reliably above the neutral benchmark. In the strong country-identity condition, the 95\% CI was [3.21, 4.20], again indicating that 2 was not included. In contrast, in the weak country-identity condition, the 95\% CI was [1.41, 2.11], meaning that 2 was included, and selections did not clearly differ from the neutral benchmark.

Tukey-adjusted post hoc comparisons showed that the strong country-identity condition selected more victims from the strong country than the independent-identity condition ($M_{diff} = 0.75$), $p = .017$. Both the independent-identity condition ($M_{diff} = 1.20$), $p < .001$, and the strong-identity condition ($M_{diff} = 1.95$), $p < .001$, selected more Setaria victims than the weak-identity condition.\\

\noindent \textbf{Identification.} An ANOVA tested whether identification with the strong country differed across conditions (independent-identity, strong country-identity, weak country-identity). The effect of condition was significant, $F(2, 260) = 51.93$, $p < .001$, $\eta^2 = .285$. Descriptively, participants reported higher identification with the strong country in the strong country-identity condition ($M = 60.40$, $SD = 22.34$) than in the independent-identity condition ($M = 41.30$, $SD = 20.50$) and the weak country-identity condition ($M = 25.95$, $SD = 23.56$).

Similarly, the effect of condition on identification with the weak country was significant, $F(2, 260) = 46.14$, $p < .001$, $\eta^2 = .262$. Descriptively, participants reported higher identification with the weak country in the weak country-identity condition ($M = 79.51$, $SD = 19.56$) than in the independent-identity condition ($M = 55.95$, $SD = 23.69$) and the strong country-identity condition ($M = 47.55$, $SD = 23.47$).

Finally, the effect of condition on identification with the journalist was not significant, $F(2, 260) = 1.03$, $p = .359$, $\eta^2 = .008$. Descriptively, identification with the journalist was comparable across the neutral condition ($M = 77.51$, $SD = 18.73$), the strong country-identity condition ($M = 73.13$, $SD = 22.62$), and the weak country-identity condition ($M = 75.71$, $SD = 20.41$). This result is plausible with the fact that in all conditions participants have to identify with the role of journalist.\\

\noindent \textbf{Identity as predictor of victims selection.} Pearson correlation analyses examined the associations between the number of victims selected from the strong country and participants’ identification with different targets. The number of strong-country victims selected was positively associated with identification with the strong country, $r = .43$, $p<.001$, indicating that stronger identification with the strong country was linked to greater attention to its victims. Conversely, strong-country victim selection was negatively associated with identification with the weak country, $r = -.45$, $p<.001$. Identification with the journalist was not significantly related to strong-country victim selection, $r = -.06$, $p=.32$.\\

\noindent \textbf{Journalistic norms beliefs as predictor of victims selection.} Pearson correlation analyses examined the associations between the number of victims selected from the strong country and participants’ endorsement of different journalistic norms. The number of strong-country victims selected was not significantly associated with neutrality norms, $r = .07$, $p = .280$, accountability norms, $r = .00$, $p = .951$, public service norms, $r = -.04$, $p = .563$, or advocacy norms, $r = -.03$, $p = .671$, indicating that endorsement of journalistic norms was unrelated to the selection of strong-country victims.\\

\noindent \textbf{Mediation analysis.}
Mediation analyses indicated that the effect of experimental condition on strong-country victim selection was explained by shifts in social identification (standardized coefficients). Identification with the strong country positively predicted the number of strong-country victims selected ($\beta = 0.29$, $SE = 0.06$, $p < .001$), whereas identification with the weak country negatively predicted it ($\beta = -0.31$, $SE = 0.06$, $p < .001$).

Relative to the reference category, the strong-country condition increased identification with the strong country ($\beta = 0.73$, $SE = 0.13$, $p < .001$) and decreased identification with the weak country ($\beta = -0.32$, $SE = 0.13$, $p = .011$), yielding significant indirect effects via both mediators (via strong-country identification: indirect $\beta = 0.21$, 95\% CI [0.10, 0.33], $p < .001$; via weak-country identification: indirect $\beta = 0.10$, 95\% CI [0.01, 0.19], $p = .023$). The total effect was also significant ($\beta = 0.38$, $SE = 0.14$, $p = .006$).

Conversely, the weak-country condition reduced identification with the strong country ($\beta = -0.59$, $SE = 0.13$, $p < .001$) and increased identification with the weak country ($\beta = 0.91$, $SE = 0.13$, $p < .001$), producing significant indirect effects in the opposite direction (via strong-country identification: indirect $\beta = -0.17$, 95\% CI [$-0.27, -0.07$], $p < .001$; via weak-country identification: indirect $\beta = -0.28$, 95\% CI [$-0.42, -0.15$], $p < .001$). The total effect ($\beta = -0.61$, $SE = 0.14$, $p < .001$).

When identification variables were included, direct effects of condition on victim selection were no longer significant (strong-country condition: $\beta = 0.06$, $p = .64$; weak-country condition: $\beta = -0.16$, $p = .27$), indicating full mediation (see Supplementary Figure~\ref{supp_fig:survey_mediation}).

\begin{figure}[h!]
    \centering
    \setlength{\fboxrule}{0.5pt}   % border thickness
    \setlength{\fboxsep}{2pt}      % padding between image and border
    \fbox{\includegraphics[width=0.8\linewidth]{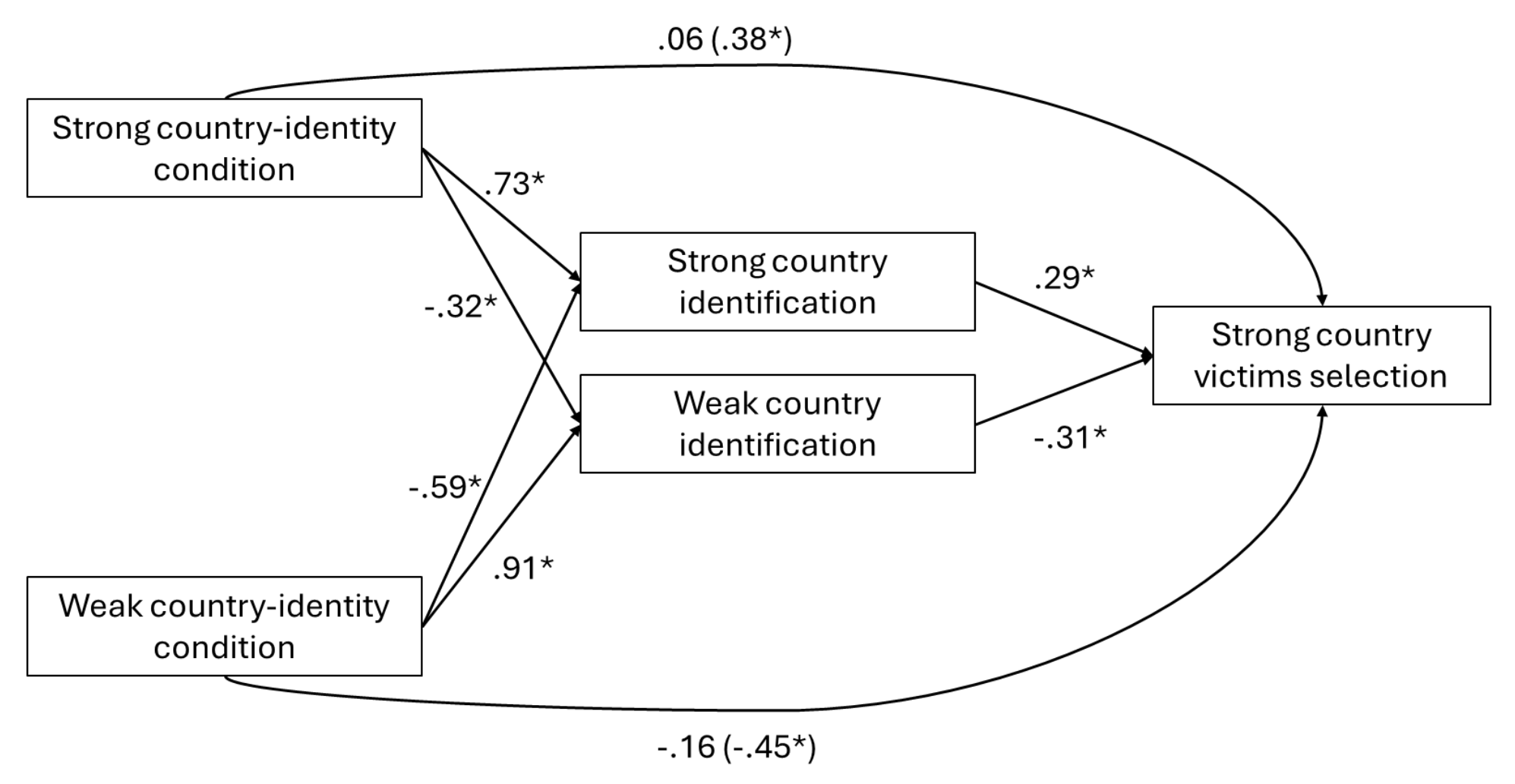}}
    \caption{\textbf{Path diagram of the mediation analysis} testing whether social identification explains the effect of experimental condition on strong-country victim selection. Standardized coefficients are displayed. 
    }
    \label{supp_fig:survey_mediation}
\end{figure}

\subsection*{Discussion}
The experimental study provides critical insight into why equalization bias may occur. When participants were asked to act as journalists in a fictional asymmetric conflict, they systematically over-selected victims from the stronger side, mirroring the equalizing bias observed in real-world Western media. Most importantly, this bias was fully mediated by social identification with the assigned country. Identification with the stronger country increased attention to its victims, while identification with the weaker country reduced it. Once identification was accounted for, experimental conditions no longer predicted victim selection.

This finding directly supports the interpretation that equalizing bias in Western media is not merely a function of professional norms, editorial routines, or information constraints, but is rooted in basic social-psychological processes. Identification subtly but powerfully shapes which lives are perceived as psychologically close, morally salient, and narratively “newsworthy”. Even minimal, experimentally induced identities were sufficient to generate systematic distortions in coverage, suggesting that chronic identification, such as that likely present among Western journalists vis-à-vis Israel, could have even stronger effects.

It is important to highlight that the experimental study demonstrates that equalization bias generalizes beyond the specific historical, political, and institutional context of the Israel–Palestine conflict. Because the conflict was fictitious and stripped of real-world allegiances, this design offers a conservative and robust test of the equalization bias itself.

Finally, by reproducing the equalizing bias in a general population sample, under tightly controlled conditions, the experiment strengthens the claim that this phenomenon reflects a basic social-psychological process rather than a peculiarity of professional journalism, specific media systems, or a particular war.

}
}

\pagebreak
\clearpage
\section{Narrative Asymmetry in Western Media Reporting}
\label{note:recurrent_reporting}
{\singlespacing
In the main manuscript, we show that Western media outlets disproportionately concentrated Israeli casualty coverage on the events of October 7, even months after the attack. To explore this pattern in greater detail, we conducted additional analyses to examine how Israeli stories related to October 7 were distributed over time, particularly in relation to dates marked with high Palestinian civilian casualties, as listed in Supplementary Table \ref{supp_tab:Major_Events_Palestine}. We began by examining the association between Palestinian deaths on a given day and the number of October 7-related Israeli stories published on that day and the following day. As shown in Supplementary Figure~\ref{supp_fig:correlation_IsrStories_PalDeaths}, the results reveal a strong positive linear relationship, with a significant Pearson correlation of 0.8 between Palestinian death counts and the number of October 7-related stories published in the immediate aftermath.

Having established a strong correlation, we next conducted a more detailed investigation by selecting a subsample of four dates on which major Palestinian casualty events occurred and reviewing the Israeli stories published on those dates and the following day. To avoid distortions caused by the natural concentration of coverage immediately surrounding October 7 and its anniversary, we applied two exclusion windows: a two-month period following October 7, 2023, and a one-month period preceding its one-year anniversary. This restricted our analytic window to early December 2023 through early September 2024. Within this span, we aimed to identify four equidistant dates to capture variation across the year. Dividing the window yielded an ideal spacing of about 95 days, but since our analysis was limited to dates of major Palestinian casualties (Supplementary Table~\ref{supp_tab:Major_Events_Palestine}), we allowed for a tolerance of 90–100 days to align with actual events. Based on these criteria, the four dates selected for analysis were:

\begin{itemize}
\item 1 December 2023
\item 29 February 2024 (90-day interval)
\item 8 June 2024 (100-day interval)
\item 11 September 2024 (95-day interval)
\end{itemize}

The articles published on these days and the days after were studied and the findings supported our earlier observations: although Western outlets did cover the Palestinian events as they unfolded, their reporting consistently devoted more narrative depth and emotional detail to Israeli casualties. This emphasis on individualized Israeli stories persisted even on days when there were no reported Israeli casualties, suggesting a systematic pattern in narrative focus. 

We describe the result of our investigation below and show that in the majority of these cases, Israeli-focused stories appeared within hours of the events or by the following day, frequently outnumbering and overshadowing coverage of concurrent Palestinian casualties:

\begin{enumerate}
    \item On 1 December 2023, Israel resumed its assault on Gaza after a brief ceasefire during which hostages and prisoners were exchanged. On that day alone, 184 Palestinians civilians were reported killed as airstrikes flattened residential blocks, and mass evacuation orders forced thousands of elderly, injured, and children alike to march for miles into winter conditions\cite{Jazeera_2023}. Yet alongside reporting these events, Western outlets devoted significant space to Israeli hostage narratives, quoting family members on recovery:

    \begin{quote}
       \textit{``Mentally, she's definitely doing better. She’s more engaged with people and becoming more independent''} (NYT, Dec 1 2023) \cite{Rosman_2023}
    \end{quote}
    and revisiting the trauma of 7 October: 
    \begin{quote}
        \textit{``Around 240 people, from infants to octogenarians, were taken hostage during Hamas’ attack on Israel on October 7. Dozens have been freed but many more remain missing...''} (CNN, Dec 1 2023) \cite{Clarke_2023}.
    \end{quote}
    On that day and the next, the BBC, CNN, and NYT published 19, 21, and 24 Israeli stories respectively, versus just 17 Palestinian stories combined (Supplementary Table \ref{supp_tab:Conc_reporting}).

\item On 29 February 2024, a convoy of aid trucks entered Gaza through Al-Rashid Street in Gaza City. As Palestinians gathered, Israeli forces opened fire, killing 117 and injuring 750 \cite{aljazeera2024un}. The incident, later termed the ``Flour Massacre,'' was so named because victims’ bodies were found covered in both blood and spilled flour. While each side offered competing narratives, the fact remained that scores of starving civilians were killed while trying to receive food aid.

Western media coverage of the event was telling. CNN published two articles within hours: one detailed account of the massacre that presented both sides and described it as the worst tragedy yet, and a second opinion piece titled \textit{``Why have the mothers and fathers of the world forgotten about the child hostages''} \cite{levy2024forgotten}. The latter shifted attention away from the massacre to the plight of Israeli families, with passages such as:

    \begin{quote}
        \textit{    “My four-year-old daughter asks me constantly when her best friend Ariel will come home.”}
    \end{quote}
    
    \begin{quote}
        \textit{ “I’m pregnant, and I fear that my unborn son will never meet his uncle, aunt, and redheaded cousins. I fear that he will be born into a world where taking children hostage is accepted.”}
    \end{quote}

    \begin{quote}
        \textit{“When dropping your children off at school or nursery, when you kiss them goodnight — please think of Kfir and Ariel Bibas. Think of their mother and father.”}
    \end{quote}

    Similarly, NYT reported on the massacre while simultaneously emphasizing Israeli hostages, repeating nearly verbatim in two separate articles \cite{Yazbek_Boxerman_2024,Cumming-Bruce_2024} that their suffering was being marginalized:

    \begin{quote}
    
        \textit{“Turning to acknowledge two former hostages behind her, Aviva Siegel and Raz Ben-Ami, whose husbands are still being held in Gaza, she said that Mr. Türk had reduced them to `a mere footnote' in the council’s discourse.”
        }
    \end{quote}
    
   Across 29 February and 1 March, the BBC, CNN, and NYT collectively published 10 Israeli-focused stories versus 16 Palestinian-focused stories, despite the absence of any Israeli event comparable in scale to the Flour Massacre.

\item The contrast was quite stark on 8 June 2024, when two events unfolded simultaneously: an Israeli mission that led to the rescuing of four hostages from Gaza’s Nusseirat area accompanied by heavy air strikes that killed 274 Palestinians civilians and injured 698 \cite{Amon_2024}. Western outlets reported both, but the framing diverged sharply. The Israeli coverage in CNN \cite{McKenzie2024FourIsraeli} was celebratory:
    
        \begin{quote}
            \textit{``Saturday was an emotional and happy day for the state of Israel and the IDF... }\\
            
            \textit{I couldn’t stop hugging him,'' Orit Meir, mother of Almog Meir Jan, told a news conference Saturday. ``Tomorrow is my birthday so I got my present... ''}
        \end{quote}
        
        while Palestinian casualties in the same article were recounted with skepticism over figures and reminders that medical records did not distinguish between civilians and militants:
    
        \begin{quote}
            \textit{``Al-Awda Hospital director Dr Marwan Abu Nasser told CNN that 142 bodies had been counted at the medical facility by late Saturday, while Al-Aqsa Hospital in Deir al-Balah said 94 bodies had been counted.}
    
            \textit{CNN has no way of verifying casualty numbers reported by Palestinian officials in Gaza. Medical records in the war-torn enclave do not differentiate between civilians and militants killed.}
    
            \textit{An Israeli military spokesperson put the number of casualties from the operation at `under 100,' and had no information on how many of those were civilians.''}
        \end{quote}
        Similary, NYT \cite{Bergman2024IsraelHostages} published the following when reporting on the Israeli side of the event: 
    
        \begin{quote}
            \textit{``The freed hostages — Noa Argamani, 26, Almog Meir Jan, 22, Andrey Kozlov, 27, and Shlomi Ziv, 41 — were kidnapped by Palestinian militants from the Nova music festival during the Hamas-led attack on Oct. 7, when about 1,200 people were killed in Israel and 250 taken hostage, Israel says... All four were in good medical condition and were transferred to a hospital in Israel for further examinations, the Israeli authorities said in a statement... After her rescue, Ms. Argamani spoke with Mr. Netanyahu. `I’m so emotional, it’s been so long since I heard Hebrew,' she said''}.
        \end{quote}
        The BBC reported in another article \cite{Mackintosh_2024} the Palestinian casualties in a grouped faceless manner:
        
        \begin{quote}
            \textit{``One man, who said more than 40 members of his family have been killed since the conflict began in October, described to the BBC being in a house which was hit by a strike. ... This home, which used to house approximately 30 people who then became 50, was bombed... only me, my father, my wife, and a young man survived... we are the only survivors out of 50 people.'' }, 
        \end{quote}
    
        while reported the saved hostages in more detail: 
        
        \begin{quote}
            \textit{``Noa Argamani, 26, Almog Meir Jan, 22, Andrei Kozlov, 27, and Shlomi Ziv, 41, who were abducted from the Nova music festival on 7 October have been returned to Israel... ''}
        \end{quote}
        Yet despite the significant disparity in the casualties reported on that day, BBC, CNN, and NYT ran 8, 13, and 19 Israeli stories respectively, more than double the 19 Palestinian stories combined.
    
\item On 10 September 2024, an Israeli air strike hit an Israeli-designated humanitarian zone in southern Gaza, killing 19 Palestinians and injuring about 60, including sleeping children and entire families buried in the sand \cite{Tawfeeq_Nasser_2024}. This was not an isolated incident; it was one in a series of bombed schools-turned-shelters in the last 30 days. No major Israeli events occurred that day, aside from the military releasing a video about the confinement conditions of six hostages found dead ten days earlier.

Western media coverage reflected this contrast. While all three outlets focused primarily on Palestinian casualties, NYT also published two detailed articles \cite{Livni_2024, Stack_2024} about the hostages, listing their names and ages, describing their living conditions and health before death:

    \begin{quote}
        \textit{“the hostages `suffered from significant malnutrition, severe weight loss and long-term physical neglect,' that some had untreated injuries”} \cite{Livni_2024}
    \end{quote}

    and reporting the distress of their families:
    \begin{quote}
        \textit{“some of the families had found their briefings to be disturbing because `it was very hard for them to see how their loved ones survived and were murdered in those conditions.'”} \cite{Stack_2024}
    \end{quote}

    Encouragingly, towards the end of the year BBC and CNN reported only on Palestinian casualties, and even the New York Times published more Palestinian than Israeli stories (7 versus 4).
\end{enumerate}

While this suggested a shift toward more balanced coverage, we questioned whether such balance only emerged in the absence of major Israeli events. Our earlier analysis of 8 June 2024 raised this possibility: on that day, an Israeli mission rescued four hostages from Gaza’s Nusseirat area, but the operation involved heavy air strikes that killed 274 Palestinians and injured 698. Despite the extraordinary scale of Palestinian casualties, media attention was dominated by hostage-related coverage.\\

To probe this further, we examined two additional dates where Israeli hostage events coincided with daily reported Palestinian casualty events: 15 January 2024 and 1 September 2024. More specifically:

\begin{enumerate}
    \item On 15 January 2024, Israeli airstrikes killed 22 Palestinians and injured many more on Gaza’s Al Thalatheni Street \cite{aljazeera2024_gaza_day101}. For Western media, however, the day’s main marker was ``100 days since October 7.'' Coverage centered on Israeli hostage testimonies and recollections:

    \begin{quote}
        \textit{``They thought they knew death but that didn’t prepare them for Oct 7''} (NYT, Jan 15 2024) \cite{Bergman2024Their},
    \end{quote}
    
    \begin{quote}
        \textit{``Israeli teenager recounts her time as a Hostage in Gaza" } (NYT, Jan 15 2024) \cite{nyt2024hostage},
    \end{quote}
    and moving military testimonials:
    \begin{quote}
        \textit{``They were Israel's eyes on the border but their Hamas warnings went unheard" } (BBC, Jan 15 2024) \cite{Cuddy_2024},
    \end{quote}
    with little new reporting on the Palestinian toll. Again, across two days, BBC, CNN, and NYT ran 13, 10, and 12 Israeli stories, compared to only 6 on Palestinians combined, with NYT having published none.

    \item On 1 September 2024, an Israeli strike on a school-turned-refugee camp killed 11 Palestinians, including some asleep in their tents \cite{aa2024gaza}. Yet coverage centered on the previous day’s recovery of six Israeli hostage bodies\cite{Rasgon_Sobelman_Shankar_Fuller_2024}, with vivid portraits:
    
        \begin{quote}
            \textit{``Hersh was among the innocents brutally attacked while attending a music festival for peace in Israel on 7 October... He lost his arm helping friends and strangers during Hamas’ savage massacre. He had just turned 23. He planned to travel the world.''} (BBC, Sept 1 2024) \cite{bbc2024hostages}
        \end{quote}
        
        and tributes that detailed hobbies, hometowns, and personal histories:
    
        \begin{quote}
            \textit{``Hersh Goldberg-Polin loved soccer and music. He was curious, respectful and passionate about geography and travel, according to his mother. He was born in the Bay Area and moved to Israel when he was 8.''} (NYT, Sept 1 2024) \cite{nyt2024polin}
         \end{quote}
    
        \begin{quote}   
            \textit{``The 23-year-old Israeli American became one of the most recognizable faces of the enduring hostage crisis, after he was taken at gunpoint by Hamas militants from the Nova music festival. Banners and murals demanding his return were often displayed in Jerusalem and around the world.''} (CNN, Sept 2 2024) \cite{CNN2024IsraelGazaHostages} 
        \end{quote}
       
        That day and the next, BBC, CNN, and NYT published 86, 38, and 57 Israeli stories respectively, compared to just five on Palestinians combined, with CNN having published none.

\end{enumerate}

Taken together, these cases reveal a consistent pattern: Western media coverage appeared more balanced only when no major Israeli events occurred. When significant events on both sides coincided, however, Israeli narratives were amplified to such a degree that they overshadowed even the most severe Palestinian tragedies. The June 8 rescue operation, the “100 days” milestone on January 15, and the recovery of hostages on September 1 all illustrate how moments of extraordinary Palestinian suffering were reframed or eclipsed through extensive Israeli-centered reporting. This suggests that the imbalance was not merely a product of ongoing coverage decay but was actively reinforced whenever Israeli stories could be interwoven with Palestinian casualty events, thereby sustaining a structural asymmetry in how loss and victimhood were represented. 

}

\pagebreak
\clearpage
\section{Extracting, Labeling and Validating for the Civilian Victim Numbers Analysis}
\label{supp_note:CS_data_extraction}

{\singlespacing
In this analysis, we aim to investigate the use of numbers and statistics in describing violent actions affecting civilians in news articles. We identify instances where numbers or statistics depict violent actions, such as deaths, injury, displacement, etc. Additionally, we determine the identity or group associated with these numbers, whether the number is associated with a child, and other information needed for our analysis. Finally, we categorize which side the reported civilians belong to: ``Palestinian,'' ``Israeli,'' or ``Other.''

To extract this information reliably, we employed advanced semantic role labeling \cite{palmer2010semantic} to capture key elements such as who performed an action, on whom, and in what context. This process was carried out using a large language model (LLM), which allowed us to detect and structure the relevant details from the text at scale. The prompt provided to the model is reproduced below:

}

\begin{tcolorbox}[enhanced,fit to height=10cm,colback=teal!25!black!2!white,colframe=teal!90!black,title=Prompt for the Civilian Victim Numbers Analysis]
  
Carefully scan the article provided to check for any instances where actions affecting civilians (both adults and children) in anyway such as killing, displacing, kidnapping, wounding, and depriving of food / water / shelter, are described using numbers and statistics. Note that numbers may be mentioned as words or  as outright digits. Also, mentions of casualties could be depicted as an open range, such as “More than 100 were killed”, so be sure to detect and report similar instances. Only report instances where numbers/statistics were mentioned, like “more than a thousand were killed”.\\

Pay special attention to any nuanced or inferred context and provide expert and deep analysis to detect all mentions of statistics within the text covering all scenarios where civilians are affected as explained above. Go slowly and carefully over each and every sentence to detect all instances - it is of extreme importance that no number or statistic is missed, so do not rush to answer quickly. After the initial scan, rescan the article checking for any missed instances. Only return instances where specified numbers are detected. Do not report singular counts like once, twice...etc.\\

To provide you with some context to determine the nationalities in case the article context is not easy to understand, on the 7th of October, 2023, Hamas attacked Israel and took hundreds of hostages back to Gaza. So, if there is a reference to hostages in the article, use this piece of information along with the article context to determine if the hostages are the Israeli ones abducted. Also, note that the Israelis may have dual nationalities, so even if the abducted hostages are said to be americans, they are considered on the sraeli side. Note that Hamas and Gaza are Palestinian.\\

\end{tcolorbox}

\begin{tcolorbox}[enhanced,fit to height=20cm,colback=teal!25!black!2!white,colframe=teal!90!black,title=Prompt for the Civilian Victim Numbers Analysis (continued)]
  
Your response format should be as such:
\begin{enumerate}
    \item \textbf{``Keyword''}: the keyword that triggered the instance detection.
  
     \item \textbf{``Statistics''}:  the detected numbers of affected civilians. If no numbers are detected then do not return the instance.
     
    \item \textbf{``Statistics Type''}: state if the detected statistic is an Age, Date, Duration or Number about Victims, Number about Injured, Statistics about Deaths ...etc. Always state a type even if not listed above. Do not state a generic type like a ``Numeric Statistic", be specific.
    
    \item \textbf{``Association''}: provide an educated guess of the identity of the person or group associated with the detected number.
    
    \item \textbf{``Is Child''}: return Yes if the person or group associated with the detected number is about one or more child or infant, otherwise return No.
    
    \item \textbf{``Is Human''}: return Yes only if the detected statistic type is a count of humans (dead or alive), otherwise return No. Example: 5 men were killed. Here, 5 is the count of killed humans, so return Yes. Example: she stayed for 4 days. 4 is not a count of human beings, it is a duration, so return No. Example: She was 5 years old. here 5 represents the female age, but is not a count of humans, so return No.
    
    \item \textbf{``Group Size''}: if 'Is Human' is Yes, then return the size of the group as such: Singular to indicate a group size of 1, Double to indicate a group size of 2 and Many to indicate a group size of 3 or more. If 'Is Human' is No, return Not\_Applicable.
    
    \item \textbf{``Nationality''}: provide an educated guess of the nationality of the affected person(s) or group(s).
    
    \item \textbf{``Side''}: carefully scan the context to provide an educated guess as to the side to which the detected entity reported in Association supports or belongs to. Your response should be limited to only Palestine, Israel, Both or Other. Note that even though the nationality of the associated entity is not Palestinian or Israeli, but the context may suggest that they are related to either, so in this case you assign them a value of Palestinian or Israeli. Example: “Hundreds of protesters in America came out supporting Palestine”, so the protesters in this case have a nationality $=$ American, and the side they are supporting is Palestine.
    
     \item \textbf{``Sentence''}: return part of the sentence where the keyword was detected.

\end{enumerate}

So each instance is represented by a single dictionary containing items 1 through 10 representing the key\-values of each of the 8 items listed above. So, the json format of the response would be as such: \{ \{instances: [ \{instance 1 \}, \{instance 2 \}, \{instance 3 \},...] \}. Use the names of items 1 through 10 exactly, meaning ``Keyword'' to ``Sentence''.\\

\end{tcolorbox}

{\singlespacing
The returned data was further filtered to retain only instances reporting casualties for three or more humans. This was done by selecting instances with the ``Is Human'' and ``Group Size'' values set to ``Yes'' and ``Many,'' respectively. Given that our research focuses on victims related to the Palestinian or Israeli sides, we further filtered the data to exclude any instances with a different ``Side'' parameter value. Finally, we leveraged the ``Statistics Type'' information to categorize the victims broadly into the following groups: ``Casualties'' (including deaths and injuries), ``Displaced and Refugees,'' ``Imprisonment and Detention,'' ``Missing and Non-Casualty Victims,'' ``Vulnerable and Affected Groups,'' ``Deprivation, Malnutrition and Hunger,'' ``Health and Medical Conditions,'' ``Humanitarian Aid and Dependence,'' and ``Other''. A breakdown of the distribution of these statistics across sides and media outlets can be found in Supplementary Table~\ref{supp_tab:Statistics_Types_Label_Distribution} 

To gain a better understanding, we provide the following article as an example, followed by the LLM output produced from the prompt above:

}

\begin{tcolorbox}[colback=blue!5!white,colframe=teal!75!black,title=Article Example, breakable]
  \textbf{Bruno Mars Tel Aviv concert canceled~\footnote {https://edition.cnn.com/2023/10/07/entertainment/bruno-mars-tel-aviv-concert-canceled/index.html} (CNN)} — Bruno Mars' show scheduled to take place in Tel Aviv, Israel on Saturday night has been canceled, according to concert organizer Live Nation Israel. On the company's verified Instagram earlier on Saturday, a text post to customers read, ``Bruno Mars concert scheduled to take place tonight is cancelled. (sic)''. Mars played in Israel for the first time on Wednesday to a crowd of 60,000, with a second show planned for Saturday, according to his tour schedule. Israel's Prime Minister Benjamin Netanyahu declared his country ``at war,'' after Palestinian militants from Gaza fired a deadly barrage of rockets and sent gunmen into Israeli territory Saturday morning. In response, Israel launched air strikes on Gaza, and fighting has continued throughout the day. The surprise attack and subsequent fighting marks a major escalation of the long-running conflict between the two sides. Israeli authorities say at least 200 people are dead and more than 1,400 wounded in Israel. Meanwhile, Palestinian authorities say at least 232 Palestinians have died, and more than 1,600 are wounded. A new round of rockets fired by Gaza militants on Saturday evening made direct hits on multiple locations inside Israel, including in Tel Aviv, according to Israeli authorities. 
\end{tcolorbox}

\begin{tcolorbox}[title=JSON Output, width=\textwidth]
\begin{lstlisting}[language=json, numbers=none, breaklines=true]
{instances:[
    {   Keyword: at least 200,
        Statistics: 200,
        Statistics Type: Statistics About Deaths,
        Association: Israeli civilians,
        Is Child: No,
        Is Human: Yes,
        Group Size: Many,
        Nationality: Israeli,
        Side: Israel,
        Sentence: Israeli authorities say at least 200 people are dead and more than 1,400 wounded in Israel.},
    
    {   Keyword: more than 1,400,
        Statistics: 1,400,
        Statistics Type: Number About Injured,
        Association: Israeli civilians,
        Is Child: No,
        Is Human: Yes,
        Group Size: Many,
        Nationality: Israeli,
        Side: Israel,
        Sentence: Israeli authorities say at least 200 people are dead and more than 1,400 wounded in Israel.},
        
    {   Keyword: at least 232,
        Statistics: 232,
        Statistics Type: Statistics About Deaths,
        Association: Palestinian civilians,
        Is Child: No,
        Is Human: Yes,
        Group Size: Many,
        Nationality: Palestinian,
        Side: Palestine,   
        Sentence: Meanwhile, Palestinian authorities say at least 232 Palestinians have died, and more than 1,600 are wounded.},
        
    {   Keyword: more than 1,600,
        Statistics: 1,600,
        Statistics Type: Number About Injured,
        Association: Palestinian civilians,
        Is Child: No,
        Is Human: Yes,
        Group Size: Many,
        Nationality: Palestinian,
        Side: Palestine,
        Sentence: Meanwhile, Palestinian authorities say at least 232 Palestinians have died, and more than 1,600 are wounded.}  
]}
\end{lstlisting}
\end{tcolorbox}

\subsection*{Validation Procedure}
\vspace{-1em}
{\singlespacing
\noindent \textbf{Rater procedure and blinding.} Two trained raters from the author team independently validated model outputs while blinded to the source outlet: article files were exported with outlet metadata removed and assigned neutral IDs. Initial disagreements were resolved through discussion to a consensus label; when needed, a third author adjudicated. All performance statistics below are computed against the consensus labels.\\

\noindent \textbf{Validation of statistic extraction and key attributes.} We next evaluated the model’s ability to (1) correctly identify statistics that are actually linked to civilian victims, (2) assign the correct side (Palestinian vs.\ Israeli) to the victim, and (3) indicate whether the statistic involves children. For this validation, 25 articles per outlet (100 total) were randomly sampled and double-coded independently as described above. Across the 100 articles, we obtained precision = 97.1\% (recall = 97.0\%) for identifying relevant statistics, 99.0\% for correct side assignment, and 99.1\% for correctly indicating child-related figures.

}

\pagebreak
\section{Cited Source extraction for the Quantifying Human Cost of War analysis}
\label{supp_note:CD_source_citing}

{\singlespacing
This analysis aims to examine every reported CVN instance to determine whether it was cited by a source within the sentence where it appears. To achieve this, we crafted a prompt to guide the LLM in identifying each CVN within its corresponding sentence in the article. The model was instructed to verify whether the reported CVN was cited, identify the source of the citation, and extract the phrase used in the citation. Below, we present the prompt used, along with examples of analyzed instances and their corresponding LLM outputs.

}

\begin{tcolorbox}[enhanced,fit to height=20cm,colback=teal!25!black!2!white,colframe=teal!90!black,title= Prompt for CVN Source Extraction]
  
You are provided below with a sentence, delimited by double brackets, that reports a number representing a civilian victim. Carefully scan the sentence and identify whether the reported number is cited by a source or not. To help you identify the correct instance within the sentence in case the sentence has multiple numbers, the user shall provide the following 2 fields: The Statistics field (this represents the number to look for) and the Association field (this represents the entity within the sentence that is associated with the number). It is critical that you use these 2 fields to correctly identify the number to analyze within the sentence.\\

\textbf{Your json response should include the following information:}
\begin{enumerate}
    \item \textbf{``Cited'': return whether the source was cited, return Yes or No.}
  
     \item \textbf{``Citing\_Source'':  return the reported source for citing. If the response for Cited is No, return N\_A.}
     
    \item \textbf{``Phrase'': the phrase used for citing the source, VERBATIM. If the response for Cited is No, return N\_A.}
    
\end{enumerate}

In case of a long sentence, carefully scan the sentence to determine the underlying semantic roles to correctly identify if a source is citing the number. Example in this sentence: ``Gaza’s Health Ministry said Israel had struck a hospital in northern Gaza, killing one worker, and struck an ambulance in front of another hospital in the southern Gazan town of Khan Younis, wounding several hospital workers and civilians.'', ``several hospital workers'' here are cited by ``Gaza's Health Ministry''.\\

Here are some examples to help in identifying the source and the phrase:
For the following sentence ``100 citizens were killed in the air strike, according to Palestinian health ministry'', the response should be in JSON format as such exactly: { ``instances'': [ {``Cited'': ``Yes'', ``Citing\_Source'':  ``Palestinian Health Ministry'', ``Phrase'': ``according to Palestinian health ministry''}] }.\\

For the following sentence ``at least 234 Palestinians were killed and more than 1,600 wounded in either gun battles or airstrikes, the Palestinian Health Ministry in Gaza said.'', the response should be in JSON format as such exactly: { ``instances'': [ {``Cited'': ``Yes'', ``Citing\_Source'':  ``the Palestinian Health Ministry in Gaza'', ``Phrase'': ``the Palestinian Health Ministry in Gaza said''}] }.\\

For the following sentence ``Israeli authorities say at least 200 people are dead and more than 1,400 wounded in Israel.'', the response should be in JSON format as such exactly: { ``instances'': [ {``Cited'': ``Yes'', ``Citing\_Source'':  ``Israeli authorities'', ``Phrase'': ``Israeli authorities say'' } ] }\\

For the following sentence ``20 citizens were injured.'', the response should be in JSON format as such exactly: { ``instances'': [ {``Cited'':``No'', ``Citing\_Source'':  ``N\_A'', ``Phrase'': ``N\_A''} ] }\\

For the following sentence "many children were killed, according to news reports.", the response should be in JSON format as such exactly: { "instances": [ {``Cited'': ``Yes'', ``Citing\_Source'':  ``news reports'', ``Phrase'': ``according to news reports''} ] }\\
\end{tcolorbox}

\begin{tcolorbox}[colback=blue!5!white,colframe=teal!75!black,title=Example of Sentences with CVNs, breakable]
  \textbf{Sentence 1:} Israeli authorities say at least 200 people are dead and more than 1,400 wounded in Israel.\\

  \textbf{Sentence 2:} Meanwhile, Palestinian authorities say at least 232 Palestinians have died, and more than 1,600 are wounded.\\

  \textbf{Sentence 3:} Now, as then, Israelis have sustained losses — at least 600 dead, over 2,000 wounded and dozens of hostages taken — that go far beyond anything they suffered in recent years.\\

  \textbf{Sentence 4:} Gaza’s Health Ministry said Israel had struck a hospital in northern Gaza, killing one worker, and struck an ambulance in front of another hospital in the southern Gazan town of Khan Younis, wounding several hospital workers and civilians.\\

   \textbf{Sentence 5:} More than 500 Israelis have been injured, Reuters quoted the country's health ministry as saying.\\

   \textbf{Sentence 6:}  Meanwhile, in Gaza, Israel's deadly attacks continue, with the death toll from Israel's bombardment now  more than 40,000, according to Palestinian health authorities..\\

\end{tcolorbox}

%The table below shows part of the LLM output for the five sentences

\begin{tcolorbox}[title=JSON Output, width=\textwidth]
\begin{lstlisting}[language=json, numbers=none, breaklines=true]
{   
    {   Cited: Yes,
        Citing Source: Israeli authorities,
        Phrase: Israeli authorities say,
        },

    {   Cited: Yes,
        Citing Source: Palestinian authorities,
        Phrase: Palestinian authorities say,
        },

    {   Cited: No,
        Citing Source: N_A,
        Phrase: N_A,
        },
        
    {   Cited: Yes,
        Citing Source: Gaza's Health Ministry,
        Phrase: Gaza's Health Ministry said,
        },

    {   Cited: Yes,
        Citing Source: the country's health ministry,
        Phrase: Reuters quoted the country's health ministry as saying,
        }

    {   Cited: Yes,
        Citing Source: Palestinian health authorities,
        Phrase: according to Palestinian health authorities,
        },
            
    ]}
\end{lstlisting}
\end{tcolorbox}

\newpage
\subsection*{Validation Procedure}

{\singlespacing
\noindent \textbf{Rater procedure and blinding.} Two trained raters from the author team independently validated model outputs while blinded to the source outlet: article files were exported with outlet metadata removed and assigned neutral IDs. Initial disagreements were resolved through discussion to a consensus label; when needed, a third author adjudicated. All performance statistics below are computed against the consensus labels.\\

\noindent \textbf{Validation on CVN instances (pooled).} Finally, to assess performance over the broader pool of CVN instances, we randomly sampled 100 instances per outlet (400 total) and evaluated them against the consensus labels. The model achieved accuracy = 93\%, precision = 100\%, and recall = 85\%.

}
\pagebreak
\section{Casting Doubt Phrase extraction for the Quantifying Human Cost of War analysis}
\label{supp_note:CS_phrase_extraction}

{\singlespacing
This analysis aims to identify phrases that cast doubt on the numbers or statistics reported in the Quantifying Human Cost of War analysis. To compile a list of such doubt-casting phrases, we processed all sentences containing casualty-related CVNs (as detailed in Supplementary Note~\ref{supp_note:CS_data_extraction}) using a specially crafted prompt designed to guide the LLM. The prompt instructed the model to identify and extract phrases that appeared to question the accuracy or reliability of the reported figures. Once the doubt-casting phrases were retrieved, we quantified their frequency across the entire dataset of sentences containing CVNs. Below, we provide the prompt used to identify these phrases, along with examples of sentences analyzed and the corresponding outputs generated by the LLM.

}

\begin{tcolorbox}[enhanced,fit to height=19cm,colback=teal!25!black!2!white,colframe=teal!90!black,title=Prompt to Identify Doubt-Casting Phrases]
  
First, I will provide you with some precontext: for some time now there has been an ongoing conflict between Israel and Hamas in Gaza which have resulted in civilian casualties on both sides. Second, you are provided with a sentence (shown below delimited by double square brackets) from a news article about this ongoing conflict and you are required to carefully scan it and return verbatim phrases used by the author(s) that show bias and cast doubt over numbers of civilian casualties reported for both the Israelis or the Palestinians.\\

As an example ``The Hamas-run health ministry says some 9,000 people have been killed.'', here the phrase `Hamas-run health ministry' suggests potential bias or unreliability in the reported casualty figures, as it implies the source may have a vested interest. We are mainly interested in this kind of bias that casts doubts over civilian casualties only.\\

However, in this other example: ``where Israel has killed more than 34,000 Palestinians'', in this phrase there is no casting doubts over the number of reported casualties, they are merely reporting the number of casualties only that the actual number is not verified so saying `more than' is a way to approximate the right number. Thus, do not detect such instances. Again, we are only interested in instances that cast doubt over the accuracy of the numbers of reported civilian casualties.\\

\textbf{Your json response should be in the following format:}
\begin{enumerate}
    \item \textbf{``Keyword'': return the keyword that casts the doubt.}
  
     \item \textbf{``Justification'':  provide a single sentence brief justification of why you chose this phrase and how it casts doubt.}
     
    \item \textbf{``Side'': You should only return Palestine, Israeli, Other to represent the side you believe it is being doubted. In case of the example provided above, the doubt is casted against hamas which is Palestinian so you should return Palestinian.}
    
    \item \textbf{``Src'': the source reporting the casualties or injuries or deaths - return verbatim.}
    
    \item \textbf{``Sentence'': return the part of the sentence (verbatim) that contains the biased phrase.}
    
\end{enumerate}

The response should be in JSON format like this: { ``instances'': [ {``Keyword'': ``according to Hamas run health ministry'', ``Justification'':  ``the phrase according to casts doubt over the fairness of the source'',``Side'': ``Palestine'', ``Src'': ``Hamas run health ministry'',``Sentence'': ``100 children were killed, according to Hamas run health ministry''}, ... ] }\\

\end{tcolorbox}

\begin{tcolorbox}[colback=blue!5!white,colframe=teal!75!black,title=Example of Sentences with CVNs, breakable]
  \textbf{Sentence 1:} Gaza health officials, who are part of the Hamas government, say more than 11,000 people have been killed in the Palestinian enclave since the war began.\\

  \textbf{Sentence 2:} Israel’s relentless airstrikes have killed more than 8,000 people, according to the Hamas-run health ministry.\\

  \textbf{Sentence 3:} About 250 people were abducted and taken to the Gaza Strip that day, according to Israeli officials.\\

  \textbf{Sentence 4:} The bombs have killed journalists, medical workers, teachers, United Nations staff members as well as 2,700 children, according to the Hamas-led authorities.\\

\end{tcolorbox}

%The table below shows part of the LLM output for this article, where only one instance was detected:
\begin{tcolorbox}[title=JSON Output, width=\textwidth]
\begin{lstlisting}[language=json, numbers=none, breaklines=true]
{instances:[
    {
        Keyword: Gaza health officials, who are part of the Hamas government,
        Justification: The phrase suggests potential bias or unreliability in the reported casualty figures, as it implies the source may have a vested interest.,
        Side: Palestine,
        Src: Gaza health officials, who are part of the Hamas government,
        Sentence: Gaza health officials, who are part of the Hamas government, say more than 11,000 people have been killed in the Palestinian enclave since the war began.},

    {   Keyword: Hamas-run health ministry,
        Justification: The phrase 'Hamas-run health ministry' suggests potential bias or unreliability in the reported casualty figures, as it implies the source may have a vested interest.,
        Side: Palestine,
        Src: Hamas-run health ministry,
        Sentence: Israel's relentless airstrikes have killed more than 8,000 people, according to the Hamas-run health ministry.},
        
    {   Keyword: according to Israeli officials,
        Justification: The phrase 'according to' suggests that the reported number of abducted individuals is based on the claims of Israeli officials, which may imply potential bias or lack of independent verification.,
        Side: Other,
        Src: Israeli officials,
        Sentence: About 250 people were abducted and taken to the Gaza Strip that day, according to Israeli officials.},

    {   Keyword: according to the Hamas-led authorities,
        Justification: The phrase 'according to' suggests potential bias or unreliability in the reported casualty figures, as it implies the source may have a vested interest.,
        Side: Palestine,
        Src: Hamas-led authorities,
        Sentence: The bombs have killed journalists, medical workers, teachers, United Nations staff members as well as 2,700 children, according to the Hamas-led authorities.}   
]}
\end{lstlisting}
\end{tcolorbox}

\pagebreak
\clearpage

\section{Analysis of Media Reporting on the 2008, 2012, and 2014 Wars on Gaza}
\label{supp_note:analysis_of_past_conflicts}
{\singlespacing
\textcolor{black}{
The three prior wars on Gaza share important structural similarities with the 2023 war, while also differing in several consequential respects. As in the most recent conflict, all three earlier wars were characterized by a pronounced asymmetry in military power and, consequently, in civilian victimhood: Israeli civilian casualties accounted for only 3\%, 13\%, and 11\% of total civilian deaths in 2008, 2012, and 2014, respectively. At the same time, these earlier conflicts were substantially more limited in scope than the 2023 war, with shorter durations, lower overall casualty counts, and far less extensive destruction of civilian infrastructure (see Supplementary Table~\ref{supp_tab:past_wars} for comparative information on conflict duration and casualties).}

\begin{table}[H]
{\fontsize{7}{7}\selectfont{
\caption{\textbf{Duration and Civilian Casualties in three Wars prior to 2023.}}
\label{supp_tab:past_wars}
\centering
\resizebox{\textwidth}{!}{%
\begin{tabular}{llccccccc}
\toprule
 & & & \multicolumn{2}{c}{\textbf{Palestine}} & \multicolumn{2}{c}{\textbf{Israel}} & \\
\cmidrule(lr){4-5} \cmidrule(lr){6-7}
\textbf{From} & \textbf{To} & \textbf{Conflict} & \textbf{Deaths} & \textbf{Injuries} & \textbf{Deaths} & \textbf{Injuries} & \textbf{Days}& \textbf{Article Count}\\
\midrule
27 Dec 2008 & 18 Jan 2009 & Cast Lead & 1,440 & 5,380 & 3 & 182 & 22 & 1100\\
&&&&&&&&\\
14 Nov 2012 & 21 Nov 2012 & Pillar of Defense & 101 & 1,399 & 4 & 224 & 7 & 364\\
&&&&&&&&\\
7 Jul 2014 & 26 Aug 2014 & Protective Edge & 1,462 & 11,231 & 6 & 1,600 & 50&1322\\
\\[-4pt]
\bottomrule
\end{tabular}
}
\par\vspace{4pt}\footnotesize

}}
\end{table}

\textcolor{black}{ With respect to international involvement, all three wars received substantial military and diplomatic backing from the United States, albeit to a lesser extent than in the 2023 war~\cite{masters_merrow_2025_us_aid_israel_four_charts}. Journalistic access also varied across conflicts: international journalists were denied entry to the war zone in 2008, whereas access to Gaza was permitted during the 2012 and 2014 wars. To the best of our knowledge, there is little evidence of formal editorial guidelines governing terminology prior to 2023. The main exceptions are the leaked Israel Project guidelines from 2009~\cite{IsraelProject2009} and a terminology list compiled by the International Press Institute in 2013~\cite{EJN_IPI_2013}, neither of which constituted official editorial policies of specific news organizations. By contrast, anecdotal evidence suggests that beginning in 2023, several major newsrooms issued internal editorial guidelines regulating language use~\cite{guardian2024cnn}.}

\textcolor{black}{Another difference between the earlier Gaza wars and the 2023 war concerns the legal characterization of Israeli conduct under international law. Only in the latter case was Israel formally accused of committing genocide. This includes the still-ongoing case filed by South Africa at the International Court of Justice (ICJ) in December 2023, in which the Court found the allegation of genocide to be ``plausible'' in its provisional measures ruling in January 2024, as well as subsequent genocide determinations or conclusions by major international and domestic human rights bodies (including Amnesty International, December 5, 2024~\cite{AmnestyInternational2024}; B’Tselem and Physicians for Human Rights–Israel, July 2025~\cite{BtselemGenocide2025}; the International Association of Genocide Scholars, August 2025~\cite{iags_2025_resolution_gaza}; and the UN Commission of Inquiry, September 16, 2025~\cite{UNCOIGenocideGaza2025}). By contrast, during the 2008 and 2012 wars, Israel was accused by international organizations of war crimes but not genocide. In the case of the 2014 war, some organizations advanced genocide accusations; however, these claims did not culminate in a formal judicial finding or ruling recognizing genocide.}

\section*{Results}

\textcolor{black}{We replicated our main analyses using news coverage of the three preceding wars on Gaza (2008, 2012, and 2014). However, the corpus of articles for these conflicts was substantially smaller due to their relatively short duration (which led us to merge the week-long 2012 data with the comparatively longer 2008 dataset), and the resulting findings should therefore be interpreted with caution. The limited number of articles and critical reporting instances (e.g., reporting of CVNs) also precluded more fine-grained analyses over time or across population subgroups (such as children). Accordingly, we restrict our reporting to three core analyses: (1) the ratio of individualized to collective victim representations, indicative of identifiable victim bias; (2) equalization bias in casualty reporting; and (3) doubt casting.}

\subsubsection*{\textit{Identifiable Victim Bias: Individualized vs Collective Reporting}}

\textcolor{black}{In Supplementary Figure~\ref{suppfig:IG_pastwars}, we replot the ratio of individualized to collective victim representation across the four conflicts (the current war and three prior wars). During the 2008 and 2012 wars, none of the analyzed outlets exhibited a statistically significant bias in individual-to-group mention ratios (GEE regression models). If anything, coverage showed a small and non-significant tendency toward greater individualization of Palestinian victims across all outlets except the BBC.}

\textcolor{black}{A qualitative shift emerges in the 2014 war, where reporting patterns begin to resemble those observed in 2023. During this period, Western outlets increasingly featured individualized portrayals of Israeli relative to Palestinian victims, with statistically significant differences emerging for the BBC and The New York Times. Specifically, the odds of a victim being individualized were 57.3\% lower (BBC) and 38.9\% lower (NYT) for Palestinians compared to Israelis. In contrast, Al Jazeera English showed a tendency in the opposite direction.}

\textcolor{black}{By 2023, these asymmetries become statistically significant across all Western outlets, as reported in the main manuscript.}

\begin{figure}[htbp!]
    \centering    \includegraphics[width=0.99\textwidth, keepaspectratio]{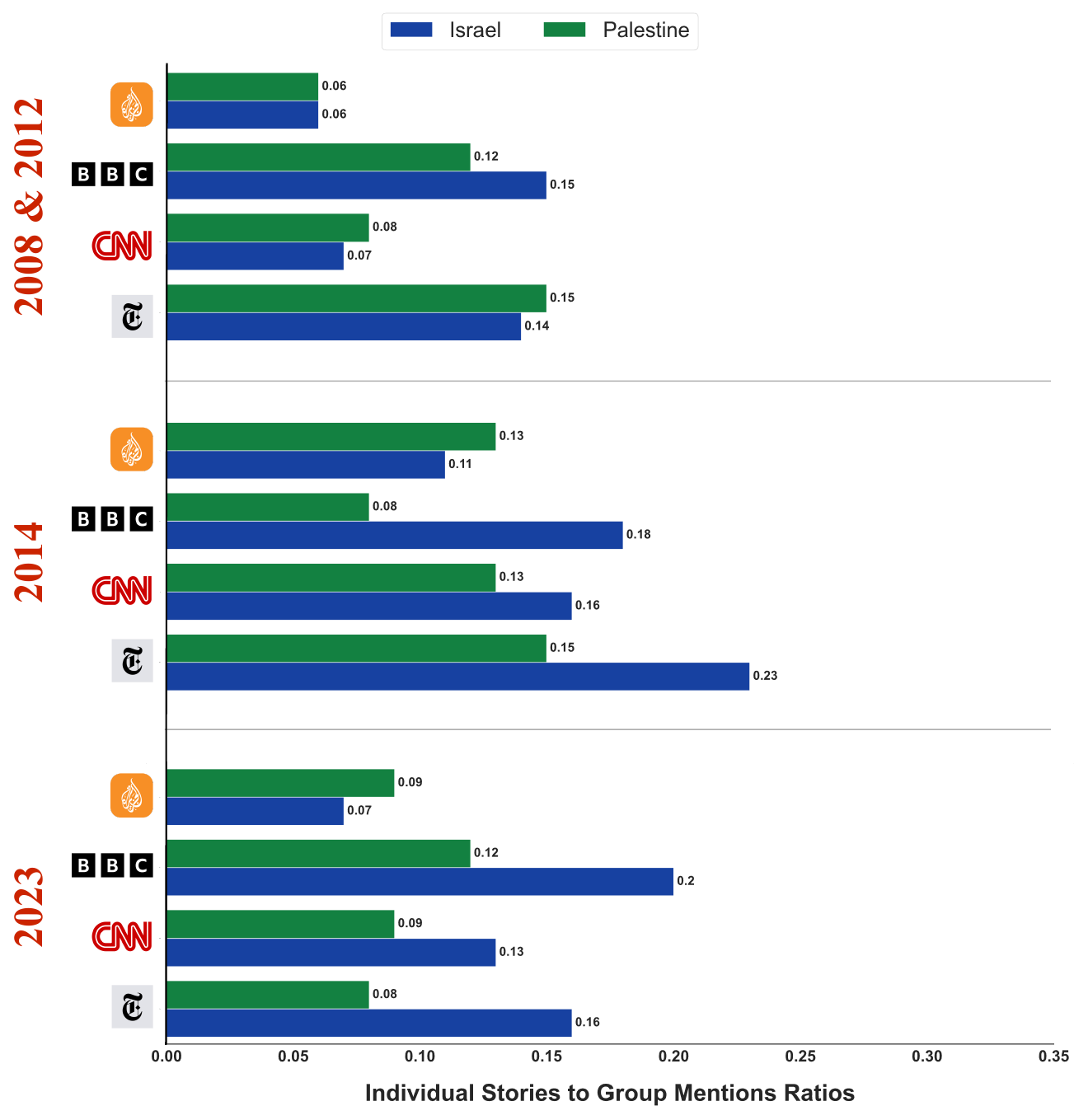}
    \caption{\textbf{Ratio of individualized to grouped mentions per side for each media source across the four wars.} Each bar represents the ratio of individualized to grouped instance counts for each media source and side. }
    \label{suppfig:IG_pastwars}
\end{figure}

\subsection*{\textit{Equalizing bias in casualty reporting}} 

\textcolor{black}{The analysis of equalization bias in casualty reporting was replicated by examining the number of individualized casualty stories for which actual casualty ratios could be established as a point of comparison for each war. As shown in Supplementary Figure~\ref{suppfig:IdvCas_pastwars}, with the exception of AJE, the tendency toward equalization seems to be increasing steadily across successive conflicts, suggesting that an equalizing bias in casualty reporting was already present, albeit in a relatively subtle form, during the earlier wars and is becoming progressively more pronounced over time.}

\begin{figure}[htbp!]
    \centering    \includegraphics[width=0.87\textwidth, keepaspectratio]{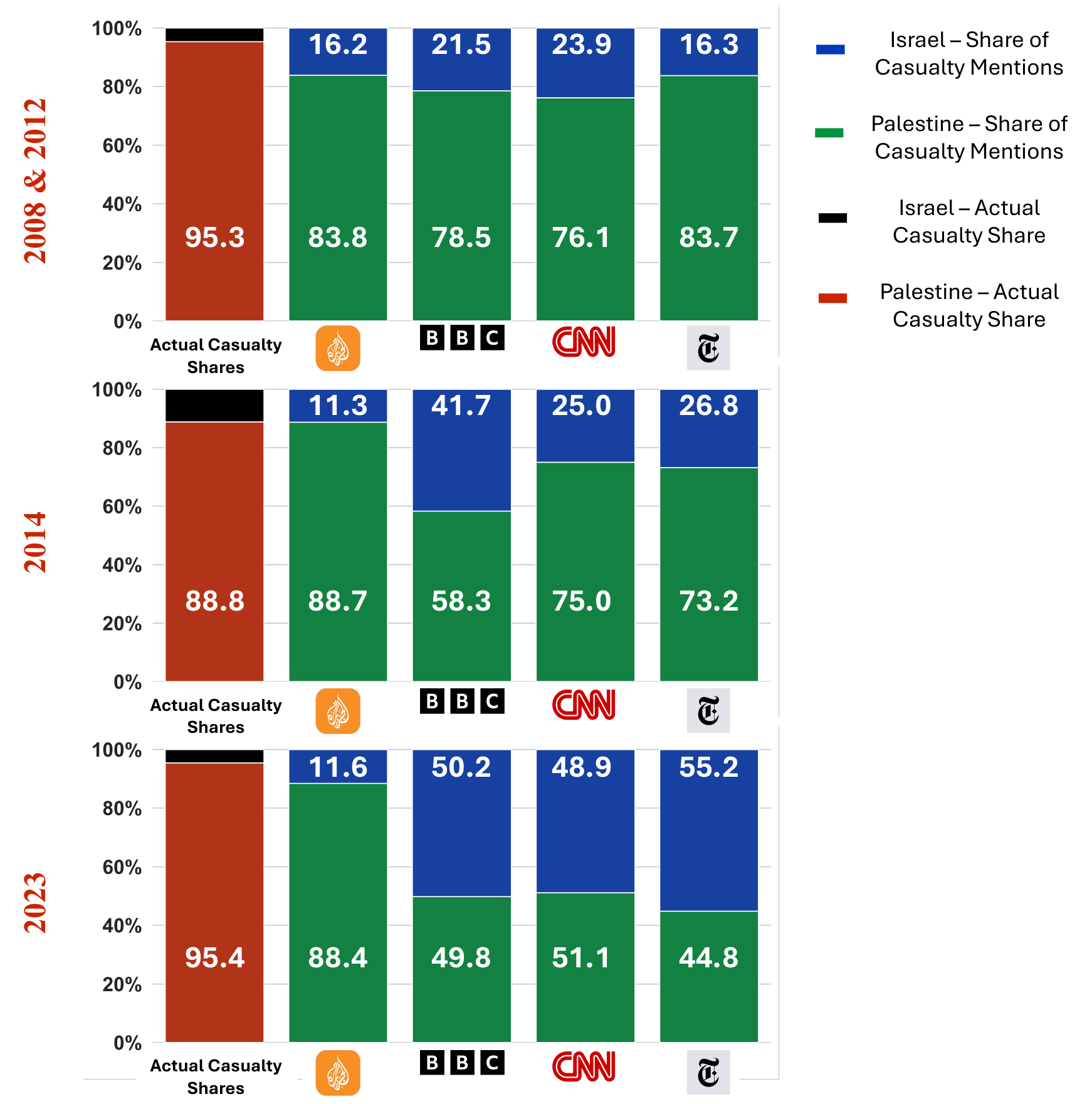}
    \caption{\textbf{Actual casualty counts and individualized casualty-related story percentages per side for each media source.} 
    }
    \label{suppfig:IdvCas_pastwars}
\end{figure}

\subsection*{\textit{Doubt-casting language when quantifying the human cost of war}} 

\textcolor{black}{Prior to 2023, source-doubting expressions such as ``Hamas-run health ministry'' were exceedingly rare across all analyzed media outlets. This pattern is notable given that Hamas had won the Palestinian legislative elections in 2006, had governed the Israeli-occupied Gaza Strip since 2007, and had been designated a ``terrorist organization'' well before the 2008 Gaza war by both the United States (1997) and the European Union (2003).\footnote{By contrast, the United Kingdom designated Hamas as a ``terrorist organization'' only in 2021.} Despite these longstanding political and legal designations, reporting during the 2008, 2012, and 2014 wars shows little evidence of systematic source-doubting or explicit uncertainty framing when civilian victim numbers (CVNs) were cited.}

\textcolor{black}{By contrast, Supplementary Figure 14 reveals a marked shift in 2023. In the 2023 war on Gaza, source-doubting language accompanying CVNs rose sharply and most prominently in BBC coverage, where such expressions appeared in roughly 20\% of CVN containing statements, compared to only 1–2\% in prior wars. CNN also exhibited an increase, though more modest (3.8\% versus 0.7\% previously). Supplementary Figure 15 further shows that this was not merely a short-lived reaction at the onset of hostilities. Although the use of source-doubting language spiked at the beginning of the war, formulations such as ``according to the Hamas-run Ministry of Health'' persisted at elevated levels throughout the year, indicating a sustained shift in reporting practices rather than a temporary adjustment during the conflict’s initial phase. This pattern reflects a new editorial strategy compared to the prior conflicts in 2008, 2012, and 2014, during which casualty figures from Hamas-affiliated institutions were apparently treated with considerably less skepticism. Notably, the origin of casualty statistics was rarely specified when figures were attributed to Israeli or other independent sources.}

\begin{figure}[htbp!]
    \centering    \includegraphics[width=\textwidth]{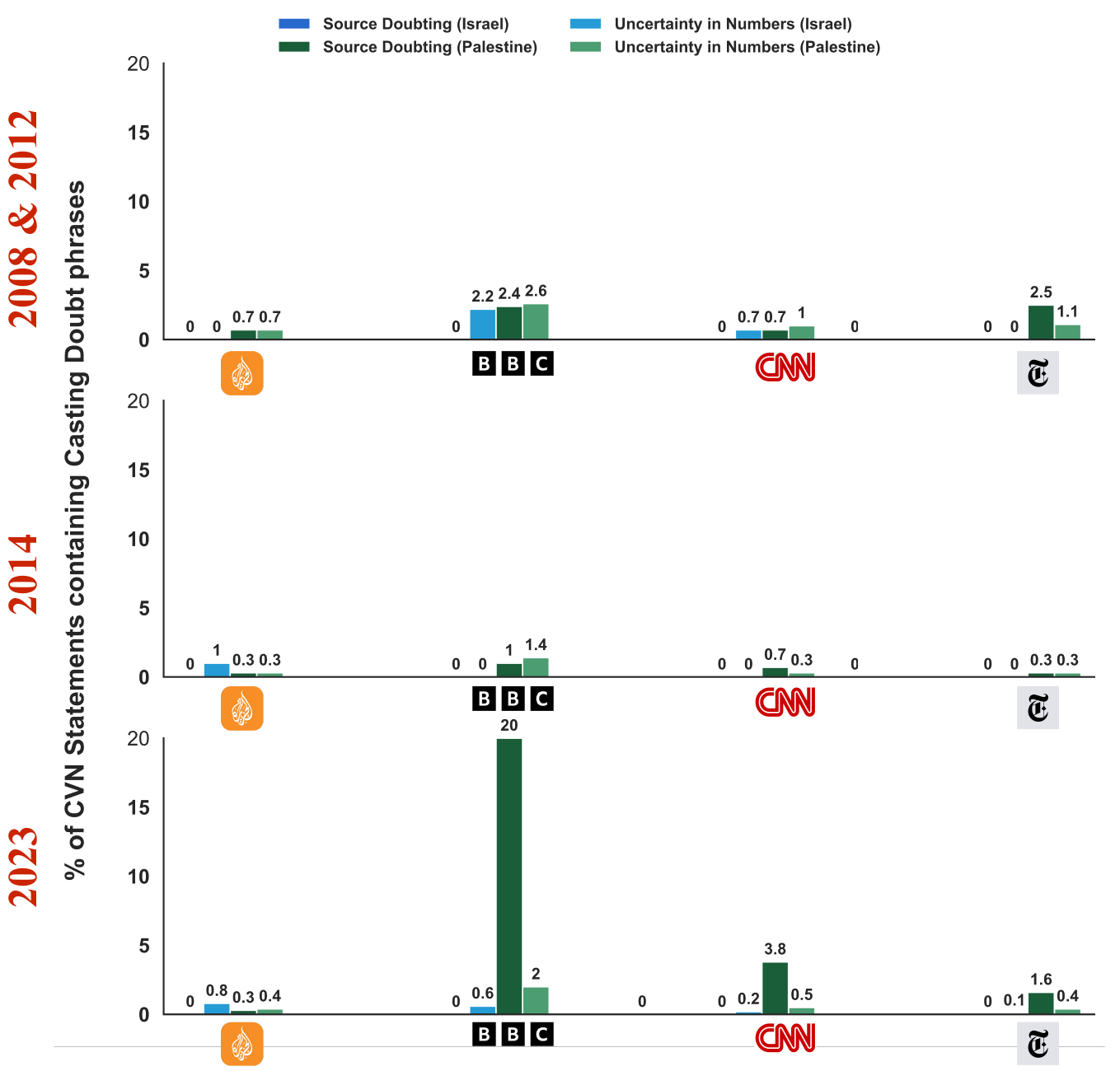}
    \caption{\textbf{Figure shows the overall share of CVN statements that had Casting Doubt phrases in terms of Source Doubting and Uncertainty in Numbers. The Shares are shown per side and media source.} 
    }
    \label{suppfig:past_conflict_overall_cd_shares}
\end{figure}

\begin{figure}[htbp!]
    \centering    \includegraphics[width=\textwidth, height=0.92\textheight,
    keepaspectratio]{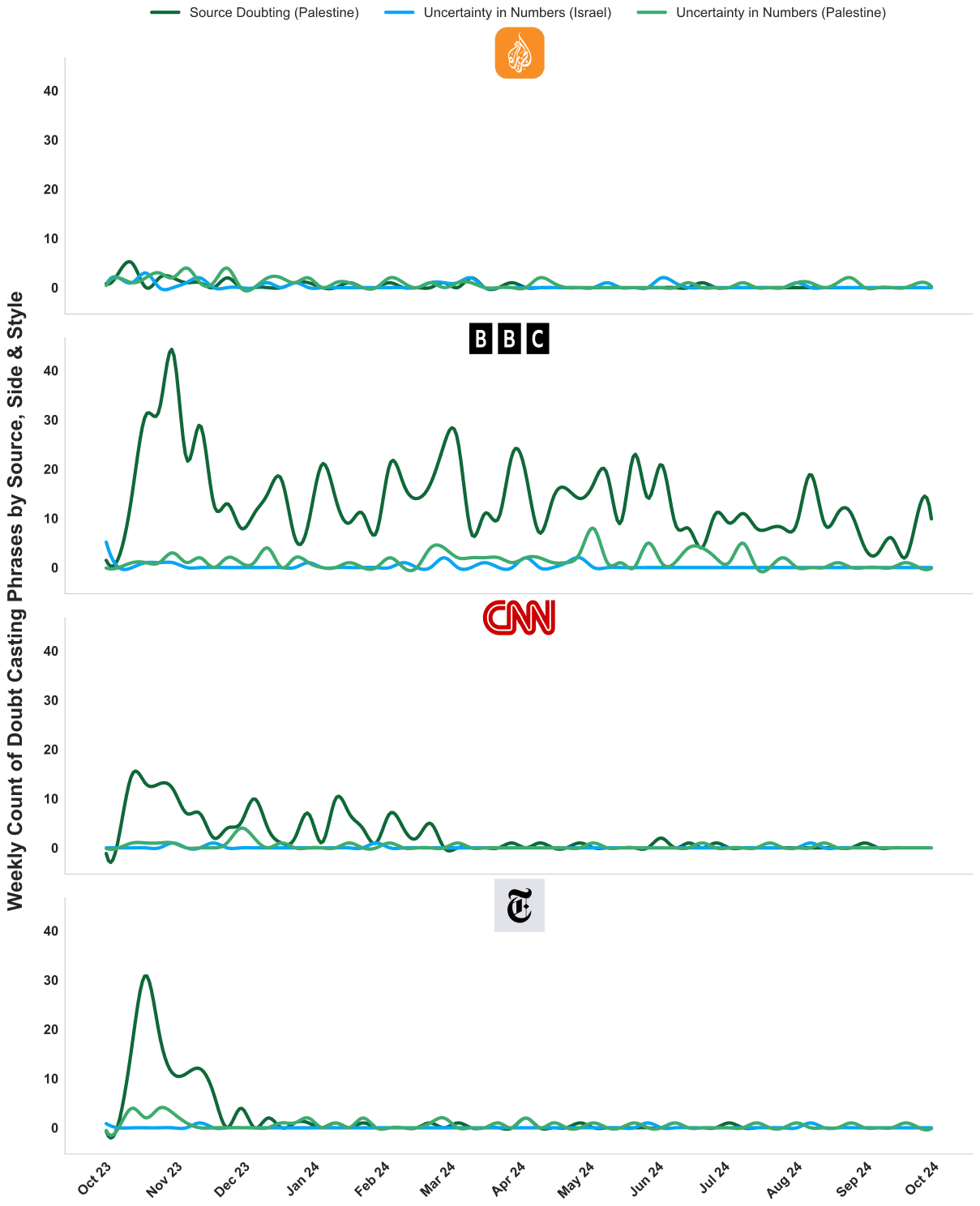 }
    \caption{\textbf{Figure shows the weekly share of CVN statements published in 2023/24 that had Casting Doubt phrases in terms of Source Doubting and Uncertainty in Numbers. The Shares are shown per side and media source.} 
    }
    \label{suppfig:past_conflict_weekly_cd_shares}
\end{figure}

}

\pagebreak

% Bibliography
\clearpage
{\singlespacing
%\bibliography{bib}
%\bibliographystyle{naturemag}

}